\documentclass[sigconf]{acmart}
\AtBeginDocument{%
  }

\setcopyright{acmlicensed}
\copyrightyear{2018}
\acmYear{2018}
\acmDOI{XXXXXXX.XXXXXXX}
\acmConference[Conference acronym 'XX]{Make sure to enter the correct
  conference title from your rights confirmation email}{June 03--05,
  2018}{Woodstock, NY}
\acmISBN{978-1-4503-XXXX-X/18/06}

\usepackage{algorithm}
\usepackage{algorithmicx, algpseudocode}
\usepackage{enumitem}
\usepackage{multirow}
\usepackage{multicol}
\usepackage{mathtools}
\usepackage{amsmath, amsthm}
\usepackage{bbold}
\usepackage{anyfontsize}
\usepackage{placeins}

\usepackage{array}
\usepackage{diagbox} 
\usepackage{booktabs}
\usepackage{tabularx}
\usepackage{makecell}

\usepackage{graphicx}
\graphicspath{{./fig/}}
\usepackage{colortbl}
\usepackage[table]{xcolor}
\usepackage{caption}
\usepackage{subcaption}

\algrenewcommand\algorithmicrequire{\textbf{Input:}}
\algrenewcommand\algorithmicensure{\textbf{Output:}}
\newtheorem{Definition}{definition}

\begin{document}



\title{Scores Know Bob’s Voice: Speaker Impersonation Attack}

\author{Chanwoo Hwang}
\authornote{Department of Mathematics \& Research Institute for Natural Sciences}
\affiliation{%
  \institution{Hanyang University}
  \city{Seoul}
  \country{Republic of Korea}    
}
\email{aa5568@hanyang.ac.kr}

\author{Sunpill Kim}
\authornotemark[1]
\affiliation{%
  \institution{Hanyang University}
  \city{Seoul}
  \country{Republic of Korea}    
}
\email{ksp0352@gmail.com}

\author{Yong Kiam Tan}
\authornote{I$^2$R: Institute for Infocomm Research, A*STAR}
\authornote{NTU: Nanyang Technological University}
\affiliation{%
  \institution{I$^2$R, A*STAR \& NTU}
  \country{Singapore}
}
\email{yongkiam.tan@ntu.edu.sg}

\author{Tianchi Liu}
\affiliation{%
  \institution{National University of Singapore}
  \country{Singapore}    
}
\email{tianchi_liu@u.nus.edu}

\author{Seunghun Paik}
\authornotemark[1]
\affiliation{%
  \institution{Hanyang University}
  \city{Seoul}
  \country{Republic of Korea}  
}
\email{whitesoonguh@hanyang.ac.kr}

\author{Dongsoo Kim}
\authornotemark[1]
\affiliation{%
  \institution{Hanyang University}
  \city{Seoul}
  \country{Republic of Korea}  
}
\email{frds37@hanyang.ac.kr}

\author{Mondal Soumik}
\authornotemark[2] 
\affiliation{%
  \institution{I$^2$R, A*STAR}
  \country{Singapore}    
}
\email{Soumik_Mondal@a-star.edu.sg}

\author{Khin Mi Mi Aung}
\authornotemark[2] 
\affiliation{%
  \institution{I$^2$R, A*STAR}
  \country{Singapore}    
}
\email{Mi_Mi_Aung@a-star.edu.sg}

\author{Jae Hong Seo}
\authornotemark[1]
\authornote{Corresponding Author}
\affiliation{%
  \institution{Hanyang University}
  \city{Seoul}
  \country{Republic of Korea}    
}
\email{jaehongseo@hanyang.ac.kr}

\renewcommand{\shortauthors}{C. Hwang et al.}

\newcommand{\red}{\textcolor{red}}
\newcommand{\blue}{\textcolor{blue}}

\newcommand{\SHP}[1]{ {\color{purple}{Seunghun: #1}}}

\begin{abstract}

\end{abstract}




\received{20 February 2007}
\received[revised]{12 March 2009}
\received[accepted]{5 June 2009}

\begin{abstract}
    Advances in deep learning have enabled the widespread deployment of speaker recognition systems (SRSs), yet they remain vulnerable to score-based impersonation attacks. However, existing attacks operating on raw waveforms require a large number of queries, stemming from the inherent difficulty of performing optimization over high-dimensional audio spaces. Optimization within the latent space of generative models presents a compelling alternative to improve efficiency, yet a fundamental mismatch exists: these latent spaces are typically shaped by data distribution matching, which does not inherently capture the speaker-discriminative geometry. Consequently, optimization trajectories in the latent space often fail to align with the adversarial direction required to maximize the score with the victim.

To resolve this, we propose an inversion-based generative attack framework that explicitly aligns the latent space of the synthesis model with the discriminative feature space. We first analyze the requirements for an \textit{inverse model} to facilitate score-based attacks. Guided by this, we introduce a training strategy enforcing feature-aligned inversion, which geometrically synchronizes the latent inputs with the feature space of SRSs. This synchronization ensures that updates in the latent space translate directly into improvements in attack scores. Furthermore, this capability enables emerging attack paradigms—such as subspace-projection-based attacks—which were previously inapplicable to SRSs due to the lack of a faithful mapping between features and audio.

Experiments demonstrate that our approach achieves superior query efficiency compared to baselines, attaining competitive attack success rates with on average $10\times$ fewer queries. In particular, the enabled subspace-projection-based attack achieves a high success rate of up to 91.65\% using only 50 queries, validating the efficacy of the proposed alignment. These results establish feature-aligned inversion as a crucial instrument for assessing the robustness of modern SRSs against score-based impersonation threats.




\end{abstract}

\maketitle

\section{Introduction}

Speaker recognition systems (SRSs) aim to verify or identify individuals based on their voice characteristics~\cite{BAI202165}.  
Recent advances in deep learning have led to their widespread deployment in security-critical applications such as biometric authentication, forensics, and voice-controlled services~\cite{liu2024disentangling,10760244}. 
In response, their security has attracted growing attention, and a key security threat to SRSs is \emph{speaker impersonation} where an adversary attempts to be accepted by the system as a specific enrolled identity. 

A wide range of attacks against SRSs have been studied under different settings, including human voice mimicry~\cite{kinnunen2019can}, replay attacks~\cite{yoon2020new}, adversarial attacks~\cite{du2020sirenattack, chen2021real, zheng2021black, zuo2024advtts, chen2023qfa2sr}, and speech synthesis-based approaches~\cite{jia2018transfer, lu2021voxstructor}. 
These attacks are typically investigated under different threat models and are not always explicitly framed as impersonation attacks. 
Nevertheless, they share a common functional outcome: generating audio that is accepted by the system as a specific enrolled speaker. 
From this perspective, we collectively refer to such attacks as \emph{speaker impersonation}.

Most existing impersonation attacks, however, assume that the adversary has access to the victim’s voice samples.
While such works help understand security risks under strong adversarial assumptions, access to a victim’s voice samples cannot always be justified in practical deployments due to strict privacy regulations~\cite{voigt2017eu} and standards~\cite{iso2022iso24745}.
They prohibit the storage and distribution of raw biometric data, and enrolled identities are often represented only through templates rather than stored speech signals. 
In such scenarios, an adversary may know that a target identity is enrolled in the system, yet lack access to any corresponding voice recordings.

Under such a constraint, an alternative attack surface arises: the \emph{score-based black-box setting}.
In this setting, the adversary interacts with the SRS solely through similarity scores and has no access to the victim’s voice samples.
This threat model surface is motivated by a typical system's verification process in practical SRSs.
They often provide similarity scores, either explicitly to support flexible decision policies or implicitly as part of their internal decision mechanism.\cite{azure_speaker, aws_transcribe, owasp_api}
These scores quantify how closely a given input matches an enrolled identity and thus constitute a potential source of information for impersonation attempts. 

Despite the real-world relevance of this threat model, speaker impersonation attacks under this setting have received relatively limited attention, in contrast to other biometric domains, e.g., face~\cite{kim2024scores, razzhigaev2020black, vendrow2021realistic}, finger vein~\cite{nguyen2023analysis}, fingerprints~\cite{8698539}.
To our knowledge, FakeBob~\cite{chen2021real} represents one of the few attempts to perform speaker impersonation using score-based feedback only.
While FakeBob proposed a zeroth-order optimization strategy from gradient estimation to iteratively modify the input audio signal, their attack requires a prohibitively large number of score queries. 
This inefficiency arises because direct optimization in the raw audio space introduces a severe mismatch between the search space and the speaker-discriminative representation used by the recognition system.
This observation raises a fundamental research question: 
\emph{Is direct optimization in the audio domain the most appropriate way to formulate score-based speaker impersonation attacks?}

\subsection{Observation: Search Space Misalignment}

To address the question, we revisit score-based attacks through the lens of black-box optimization, identifying the key bottleneck as the misalignment between the search space and the scoring objective.

\subsubsection*{\bf Formulating Score-based Attacks}
Fundamentally, the adversarial goal of score-based impersonation attacks is to find an audio signal whose similarity score with the target identity exceeds a decision threshold.
By viewing the similarity score as the objective, we can reformulate these attacks in terms of a black-box optimization problem to maximize the score.
This formalization generalizes FakeBob as it utilizes an NES-based~\cite{wierstra2014natural} optimizer to solve this problem, leveraging raw audio space as the search space.

While this formulation is general, its practical effectiveness depends critically on the choice of the search space in which optimization is performed. 
As observed by FakeBob, direct optimization in the raw audio space is prohibitively inefficient due to its extremely high dimensionality; the effectiveness of the NES solver is known to degrade significantly as the dimensionality of the input space increases~\cite{ilyas2018black, wierstra2014natural}.
This motivates the adoption of low-dimensional latent spaces through generative models.
In particular, text-to-speech (TTS) models, which convert text to speech while maintaining speaker-specific characteristics, have been extensively studied and widely adopted in practice~\cite{casanova2022yourtts,casanova2024xtts,osman2022emo}.
For this reason, a seemingly natural direction is to leverage the latent space of such models as a search space.

\subsubsection*{\bf Missing Component: Inverse Model}
However, dimensionality reduction strategy alone is insufficient: the latent space must also be aligned with the speaker embedding space induced by the recognition model. 
Without such alignment, score-guided updates lead to poorly conditioned optimization landscapes and slow convergence, even compared to a direct optimization over the raw audio space.
This observation highlights the need for a generative model whose latent space is well-aligned with the template space to facilitate solving the optimization problem.

From this perspective, we identify the inverse model as a key missing component in score-based speaker impersonation. 
By explicitly mapping speaker embeddings back to the signal domain, an inverse model induces a search space that is both low-dimensional and structurally aligned with the impersonation objective, thereby enabling more predictable and query-efficient optimization.
Constructing such an inverse model, however, is non-trivial, as speech signals exhibit substantial variability across content, prosody, and acoustic conditions, and we observe that existing TTS models only provide a loose alignment between their latent spaces and speaker-discriminative representations (See Section~\ref{sec:goodinverse} for details).

\subsection{Our Contributions}

In this work, we study score-based speaker impersonation attacks under a realistic black-box setting without access to victim voice samples.
By identifying inversion as the primary bottleneck in existing approaches, we design an attack framework that enables efficient exploitation of score feedback in SRSs.

Our main contributions are summarized as follows:

\begin{itemize}[leftmargin=*]
    \item \textbf{Optimization-based formulation of score-based speaker impersonation.}  
    We formalize score-based speaker impersonation as a black-box optimization problem and analyze why direct optimization in the raw audio space is inherently inefficient.

    \item \textbf{Identification of inversion as the central bottleneck.}  
    We provide a systematic analysis showing that the absence of effective feature-to-speech inversion is the key factor limiting prior score-based attacks in speaker recognition.

    \item \textbf{Feature-aligned inverse model for speaker embeddings.}  
    We design a feature-to-speech inverse network with identity- and structure-constraint objectives, explicitly aligning the inverse mapping with speaker-discriminative embedding geometry.

    \item \textbf{Query-efficient and inversion-essential impersonation attacks.}  
    The proposed inverse model improves query efficiency by $10\times$ on average (up to $24\times$) over existing score-based attacks, and enables subspace projection–based impersonation~\cite{kim2024scores}, achieving up to $91.65\%$ attack success rate within only 50 score queries.

    \item \textbf{Language-agnostic and potential defenses.}  
    Our attacks are designed in a speaker-discriminative feature space rather than the acoustic–linguistic domain, yielding empirically validated cross-lingual generalization; we also outline potential defense directions against inversion-enabled impersonation attacks.
\end{itemize}

Anonymized source code is available at: \url{https://anonymous.4open.science/r/Scores-Know-Bobs-Voice-5B14}.

\section{Related Work}


We overview various threats and attack scenarios against SRSs from the perspective of speaker impersonation attacks, highlighting their scope and limitations under a score-based black-box threat model.
Table~\ref{tab:rlwork} summarizes representative prior works from this viewpoint.

\begin{table}[t]
\centering
\caption{
Comparison of speaker impersonation attacks against SRSs. The reported number of queries corresponds to the cost of performing the attack task as defined in each original work. \textbf{Ours-NES} and \textbf{Ours-SP} denote two optimization strategies used in our attack: Natural Evolution Strategies (NES) and Subspace Projection (SP), respectively.
Details of these optimization methods are described in Section~\ref{subsec:using_inverse_model}.\\
$\dagger:$ We consider FakeBob applied to speaker verification here. 
}
\resizebox{\linewidth}{!}{
\begin{tabular}{c|c|c|c}
\toprule[1.5pt]
\textbf{Method} &
\begin{tabular}[c]{@{}c@{}}\textbf{Victim Voice}\\ \textbf{Required}\end{tabular} &
\textbf{Query Output} &
\textbf{\#Queries} \\ \hline \hline

\cite{kinnunen2019can}
& \multirow{9}{*}{Yes} & -- & 0 \\ \cline{1-1}\cline{3-4}

\cite{yoon2020new}
& & -- & 0 \\ \cline{1-1}\cline{3-4}

\cite{pani2023voice}
& & -- & 0 \\ \cline{1-1}\cline{3-4}

SV2TTS~\cite{jia2018transfer}
& & -- & 0 \\ \cline{1-1}\cline{3-4}

Voxstructor~\cite{lu2021voxstructor}
& & Feature vector & $\geq 100\mathrm{K}$ \\ \cline{1-1}\cline{3-4}

AdvTTS~\cite{zuo2024advtts}
& & -- & 0 \\ \cline{1-1}\cline{3-4}

QFA2SR~\cite{chen2023qfa2sr}
& & -- & 0 \\ \cline{1-1}\cline{3-4}

SirenAttack~\cite{du2020sirenattack}
& & Score & $\geq 7.5\mathrm{K}$ \\ \cline{1-1}\cline{3-4}

Occam~\cite{zheng2021black}
& & Decision & $\geq 10\mathrm{K}$ \\ \hline \hline

FakeBob$^\dagger$~\cite{chen2021real}
& No & Score & $\geq 10\mathrm{K}$ \\ \hline \hline

\rowcolor{gray!20}
\textbf{Ours-NES}
& 
& 
& \textbf{$\approx$ 500} \\
\cline{1-1}\cline{4-4}

\rowcolor{gray!20}
\textbf{Ours-SP}
& \multirow{-2}{*}{\textbf{No}}
& \multirow{-2}{*}{\textbf{Score}}
& \textbf{50} \\

\bottomrule[1.5pt]
\end{tabular}
}
\label{tab:rlwork}
\vspace{-10pt}
\end{table}

\subsubsection*{\bf Impersonation Attacks via Victim Information.}
Several prior studies consider impersonation attacks under the assumption that the adversary has access to the victim’s voice samples or related information.
This category includes human voice mimicry~\cite{kinnunen2019can}, replay attacks using recorded speech~\cite{yoon2020new}, and voice morphing techniques that combine acoustic characteristics from multiple speakers~\cite{pani2023voice}.
Although these works demonstrate the feasibility of impersonation under strong adversarial assumptions, they are inapplicable to our threat model, where such information is unavailable.

\subsubsection*{\bf Generative and Transfer-based Impersonation Attacks.}
Several studies explore impersonation through generative models that synthesize speech mimicking a target identity.
Early approaches such as SV2TTS~\cite{jia2018transfer} operate in white-box settings and utilize internal model information.
Voxstructor~\cite{lu2021voxstructor} extends this direction to black-box models by recovering victim representations through extensive interaction with the target system, followed by training a separate synthesis model.
However, this approach requires access to feature-level outputs, which are typically treated as sensitive biometric data in deployed systems.
Transfer-based approaches, including AdvTTS~\cite{zuo2024advtts} and
QFA2SR~\cite{chen2023qfa2sr}, further study impersonation by exploiting transferability across models, but they often require victim information during preparation.

\subsubsection*{\bf Adversarial Attacks.}
Another line of work studies adversarial attacks against SRSs, where the attacker generates an audio signal that is accepted as a target identity while remaining close to a given source utterance.
This source-audio constraint plays a central role in shaping the attack objective and has been widely studied in adversarial audio generation (e.g.,~\cite{du2020sirenattack, chen2023qfa2sr, chen2021real}).
While such attacks intersect with impersonation at the decision level, they impose an additional constraint that ties the attack to an existing source signal.
As a result, adversarial attacks and score-based impersonation attacks address related but structurally distinct attack objectives.

\subsubsection*{\bf Query-based Black-box Impersonation Attacks.}
Several works investigate impersonation attacks through black-box interaction with the target SRS, demonstrating that graded feedback—such as similarity scores or decision outcomes—can be exploited to guide an attack.
These studies collectively highlight the risks associated with exposing informative verification responses.
Among them, FakeBob~\cite{chen2021real} is most closely aligned with our setting, as it shows that impersonation is possible using score feedback without access to the victim’s voice.
Notably, FakeBob operates over verification queries evaluated against a fixed enrolled identity, which distinguishes it from attacks that rely on source-conditioned or pairwise querying.
While FakeBob was originally proposed as an adversarial attack that enforces source-audio similarity, its query cost in Tab~\ref{tab:rlwork} reflects this additional constraint; accordingly, we consider unconstrained score-based variants for fair comparison in our impersonation setting.

\subsubsection*{\bf Positioning of Our Work.}
In summary, prior work on speaker impersonation spans diverse threat models, ranging from settings with access to victim-side information to those imposing source-audio constraints or requiring extensive interaction with the target system.
Our work operates under the same score-based black-box threat model as prior query-based impersonation attacks (e.g., FakeBob), where the adversary has no access to the victim’s voice, and focuses on improving efficiency via inversion-aligned optimization.

\section{Problem Setting and Threat Model}

\subsubsection*{\bf Notations.} Distances between speaker representations are measured in a feature space $\mathcal{X}$ via a non-negative distance function $d_{\mathcal{X}}(\cdot,\cdot)$. 
We refer to $(\mathcal{X}, d_{\mathcal{X}})$ as the corresponding feature space equipped with a distance. 
We use $\mathbb{R}^{d}$ to denote the $d$-dimensional real vector space and $\mathbb{S}^{d-1} \subset \mathbb{R}^{d}$ to denote the unit hypersphere.
The standard inner product between two vectors is denoted by $\langle \cdot, \cdot \rangle$.

\subsection{Speaker Recognition Systems (SRSs)}

SRSs process speech signals \( \mathfrak{v} \in \mathcal{V} \) to make decisions based on speaker-related characteristics. 
They rely on a feature extractor \( F : \mathcal{V} \rightarrow \mathbb{R}^{d} \), which maps an input utterance to a speaker feature vector that captures speaker-discriminative information.
In modern deep learning-based SRSs, the extractor is typically implemented using neural networks such as x-vector~\cite{8461375} and ECAPA-TDNN~\cite{desplanques20_interspeech}. 
Commonly, feature vectors are normalized to lie on the unit hypersphere \( \mathbb{S}^{d-1} \subset \mathbb{R}^{d} \), and similarity is measured via cosine similarity. 
Verification decisions are made by thresholding the measured distance at \( \tau \), which is typically selected via standard criteria\footnote{Detailed description of criteria is in Appendix~\ref{subsec:evaluation}} such as Equal Error Rate (EER) or minimum Detection Cost Function (MinDCF), yielding thresholds denoted by
\( \tau_{\mathsf{E}} \) and \( \tau_{\mathsf{M}} \), respectively.

Speaker recognition systems are typically deployed in two modes:
\emph{verification} (1:1) and \emph{identification} (1:$N$).
In this work, we focus exclusively on verification.
Using the extractor \(F\), the system can be described by the following two procedures:
\begin{itemize}[leftmargin=*]
    \item $\mathsf{Enroll}$ takes an enrollment utterance
    \( \mathfrak{v} \in \mathcal{V} \)
    and returns a corresponding speaker feature vector
    \( t := F(\mathfrak{v}) \),
    which is subsequently stored in the system database.
    \item $\mathsf{Ver}$ takes a test utterance
    \( \mathfrak{v}' \in \mathcal{V} \)
    and the stored feature vector \( t \),
    and outputs \( \mathbb{1}\!\left(d_{\mathcal{X}}(F(\mathfrak{v}'), t) < \tau\right) \).
\end{itemize}

\subsection{Threat Model}

We consider a score-based impersonation attack against a speaker verification system. 
The system is accessed through a verification interface, where a client submits a speech signal with a claimed enrolled identity and receives a verification response.

In many modern authentication deployments, verification responses are not restricted to a pure binary accept/reject decision. 
For cloud-based service architectures, producing fine-grained similarity scores is often a functional requirement for flexible policy enforcement (e.g., risk-based authentication), as evidenced by major providers such as Microsoft Azure and AWS~\cite{azure_speaker, aws_transcribe}. 
Even if the user interface displays only a binary decision, backend services frequently transmit verbose JSON responses containing these scores. 
This practice, classified as \textit{Excessive Data Exposure} (API3:2019) by OWASP~\cite{owasp_api}, allows an adversary to extract precise confidence indicators from the network traffic.

In addition, when verification is executed on a client-controlled device (on-device), we assume the adversary has full control over the execution environment. 
In this scenario, standard dynamic analysis techniques, such as Dynamic Binary Instrumentation (DBI), function hooking, or memory inspection, allow the adversary to intercept verification routines and inspect intermediate comparison results prior to thresholding, a capability formally recognized in mobile security testing standards~\cite{owasp_mstg}.

Accordingly, we model the target system as a black-box oracle that provides feedback with respect to the similarity between the submitted speech and the enrolled feature vector. The adversary is characterized as follows:

\begin{itemize}[leftmargin=*]
    \item \textbf{Attacker’s goal.}
    The adversary aims to impersonate a fixed enrolled identity by generating a speech signal \( \mathfrak{v}^\ast \in \mathcal{V} \) that is accepted by the verification system.
    Formally, the goal is to find \( \mathfrak{v}^\ast \) such that
    $d_{\mathcal{X}}\!\left(F(\mathfrak{v}^\ast), t_\mathsf{v}\right) < \tau,$ where \( t_\mathsf{v} = F(\mathfrak{v}_{\text{victim}}) \) denotes the victim's enrolled feature vector and \( \tau \) is the predefined threshold.
    
    \item \textbf{Attacker’s knowledge.}
    The adversary has no access to the victim's audio \( \mathfrak{v}_{\text{victim}} \), the enrolled representation \( t \),
    the training data, or the internal architecture and parameters of the feature extractor \(F \), except for its similarity metric $\langle \cdot,\cdot \rangle$.
    
    \item \textbf{Attacker’s capabilities.}
    The adversary can submit arbitrary speech signals \( \mathfrak{v} \in \mathcal{V} \) for the fixed enrolled identity and obtain a response that reveals a score \( d_{\mathcal{X}}(F(\mathfrak{v}), t_\mathsf{v}) \).
\end{itemize}

\subsubsection*{\bf Score-based Impersonation Attack}
Under the system and threat models described above, a \emph{score-based impersonation attack} considers an adversary who interacts with a speaker verification system through its verification interface and leverages score-level feedback to generate a speech signal impersonating the victim's identity. 
Formally, the adversary submits speech signals \( \mathfrak{v} \in \mathcal{V} \) and observes score feedback $\langle F(\mathfrak{v}, t_{\mathsf{v}}) \rangle$, where $t_\mathsf{v}$ is victim's enrolled feature vector. The attack succeeds if the adversary produces an utterance $\mathfrak{v}^\ast \in \mathcal{V} $ such that $ d_{\mathcal{X}}\!\left(F(\mathfrak{v}^\ast), t\right) < \tau$ for a predefined threshold  \( \tau \).

\section{Why Inverse Models?}

In this section, we show how the inverse model serves as a core building block for score-based impersonation attacks.
We model the score-based attack as a black-box optimization problem, showing the benefit of leveraging inverse models in terms of (mis)alignment between the search space and the objective.
We further experimentally validate this effect, demonstrating that the proposed inverse model substantially reduces the number of optimization steps compared to baseline approaches.
For clarity, we refer to our method as `Ours' when presenting comparisons and figures, although full methodological details are introduced in the following section.

\subsection{Score-based Attack as an Optimization}
A typical approach to instantiate score-based impersonation attacks is to view it as the following optimization problem.
\begin{align}\label{eq:SB}
    \text{minimize }d_{\mathcal{X}}(F(\mathfrak{v}), F(\frak{v}^{\ast})) \text{ s.t. } \mathfrak{v^{\ast}} \in \mathcal{V}
\end{align}
Here, the adversary $\mathcal{A}$ succeeds in the attack if the found solution $\mathfrak{v}^{\ast}$ from Eq.~\eqref{eq:SB} satisfies $d_{\mathcal{X}}(F(\mathfrak{v}), F(\frak{v}^{\ast})) < \tau$.
From this viewpoint, we can classify the existing attack methods in terms of how to solve Eq.~\eqref{eq:SB} via exploiting score queries.
For example, several score-based attacks like Fakebob~\cite{chen2021real} exploit zeroth-order optimization solvers, such as gradient estimation~\cite{chen2017zoo}, genetic algorithm~\cite{alzantot2018did}, or hill-climbing optimization~\cite{guo2019simple}.

\subsubsection*{\bf Reducing Search Space via Latents}
In the above approach, $\mathcal{V}$ is usually high-dimensional, e.g., 48,000 (3sec * 16khz), which renders directly solving Eq.~\eqref{eq:SB} rather cumbersome.
To mitigate this, one can leverage generative models with a latent space to reduce the search space.
That is, for a generative model $G: \mathcal{Z} \rightarrow \mathcal{V}$ with a latent space $\mathcal{Z}$ having much smaller dimensionality than $\mathcal{V}$, we can reformulate Eq.~\eqref{eq:SB} as 
\begin{align}\label{eq:LSB}
    \text{minimize }d_{\mathcal{X}}(F(\mathfrak{v}), F(G(z)) \text{ s.t. } z \in \mathcal{Z}
\end{align}
Note that this approach incurs the trade-off between the size of the search space and the quality of the solution.
In an extreme case, the adversary never succeeds in the attack from solving Eq.~\eqref{eq:LSB} if $G$'s capacity is insufficient, i.e., $\min_{z \in \mathcal{Z}}d_{\mathcal{X}}(F(\mathfrak{v}), F(G(z)) \ge \tau$.

\subsection{Inverse Models for Aligning Search Space}

The effectiveness of latent-space optimization in Eq.~\eqref{eq:LSB} depends critically on how well the latent space $G$ is aligned with the speaker embedding space induced by $F$.
However, for generic generative models, proximity in the latent space does not necessarily correspond to proximity in the embedding space, leading to poorly conditioned optimization landscapes.
To overcome this limitation, we introduce the inverse model as an explicit aligner between the latent space and the embedding space, enabling more predictable and efficient optimization.

\subsubsection*{\bf Inverse Models and Desired Properties}
Given a feature extractor $F$, an \emph{inverse model} $F^{-1}: \mathbb{S}^{d-1} \rightarrow \mathcal{V}$ of $F$ aims to reconstruct a plausible voice $\mathfrak{v}$ that closely matches a given feature template $t$, i.e., $F^{-1}(t) = \mathfrak{v}, F(\mathfrak{v}) \approx t$. 
The existence of such inverse models has been reported in various biometric modalities, including face~\cite{mai2018reconstruction,duong2020vec2face}, fingerprint~\cite{wijewardena2022fingerprint}, finger vein~\cite{kauba2020inverse}, iris~\cite{ahmad2020resist}, and palmprint~\cite{yan2024toward}.

From the perspective of a generative model, the latent representation of the inverse model is explicitly aligned with the feature space.
Consequently, updates in the latent space induce predictable changes in the objective function $d_{\mathcal{X}}(F(\mathfrak{v}), F(G(z)))$ when $G=F^{-1}$, making Eq.~\eqref{eq:LSB} substantially easier to optimize.
These considerations naturally lead us to analyze the essential requirements of an inverse model for score-based impersonation attacks:
\begin{itemize}[leftmargin=*]
    \item \textbf{ID-Constraints (A1).} The output audio from the inverse model must maintain identity constraints of $F$, i.e., $\forall x\in \mathbb{S}^{d-1}, d_\mathcal{X}((F \circ F^{-1})(x),x) < \tau$.
    \item \textbf{Transferability (A2).} The generated inverse audio should be recognized as the same identity by alternative speaker recognition models.
\end{itemize}

The former (\textbf{A1}) implies the alignment between the feature space of a known SRS $F$ and the latent space of its inverse.
On the other hand, the latter (\textbf{A2}) ensures that the alignment between the latent space and other SRS models' feature embeddings is still largely maintained.
Specifically, this property is necessary when launching the score-based attack in the black-box setting; in this case, the adversary cannot access the target SRS and thus exploits the surrogate model and its inverse model.
Here, thanks to (\textbf{A2}), such an inverse model is still effective for conducting the score-based attack based on Eq.~\eqref{eq:LSB} against target SRS.

\subsubsection*{\bf A Candidate Inverse Model: TTS}
\label{subsec:usingttsasinv}
Text-to-speech (TTS) systems generate speech signals from text while allowing certain characteristics of the output to be controlled through auxiliary inputs. Depending on the design objective of the TTS system, these auxiliary inputs may encode different factors, such as speaker identity, speaking style, or prosodic attributes.

In particular, zero-shot text-to-speech (ZS-TTS) models~\cite{nvidia2023zero} synthesize speech for previously unseen speakers by conditioning the generation process on a speaker feature vector extracted from a short reference utterance.
Although ZS-TTS models are developed for general-purpose speech synthesis and voice cloning rather than impersonation attacks, this conditioning mechanism allows them to be functionally interpreted as candidate inverse models that map speaker recognition features back to speech signals. In this work, we focus on TTS models that accept a \emph{speaker feature vector} as an explicit conditioning input.

Based on this criterion, we consider several representative ZS-TTS models as candidate inverse models, including SV2TTS~\cite{jia2018transfer}, Voxstructor~\cite{lu2021voxstructor}, and YourTTS~\cite{casanova2022yourtts}.
These models differ in architectural details but share the common mechanism of conditioning speech generation on speaker feature vectors.
Additional selection criteria and details are provided in Appendix~\ref{appsec:existingtts}.

\subsection{Challenge: Finding Good Inverse Models}\label{sec:goodinverse}

The optimization-based viewpoint provides a principled pipeline for score-based impersonation attack, and the inverse model plays a crucial role.
However, we observe that existing models fail to effectively realize the attack pipeline, revealing a mismatch between current training approaches and the attack objective.
This motivates the design of a new inverse model specifically designed for impersonation attacks.
In this section, we analyze these limitations and summarize the key takeaways. Unless otherwise specified, detailed experimental settings are deferred to Appendix~\ref{app:id_constraints}.

\subsubsection*{\bf Do Existing Candidates Satisfy the Desired Properties?}
\label{subsub:properties}

We evaluate whether existing inverse model candidates satisfy the desired identity-preserving properties under a unified round-trip evaluation protocol.
Specifically, we assess (A1) identity preservation under the local speaker recognition model and (A2) transferability of the preserved identity to an alternative target model.

We consider a local speaker recognition model $F_L$, for which an inverse model $F_L^{-1}$ is instantiated.
Given an input utterance $\mathfrak{v}$, we extract its speaker embedding $x = F_L(\mathfrak{v})$ and synthesize inverse audio $\tilde{\mathfrak{v}} = F_L^{-1}(x)$.
Identity preservation is evaluated by measuring the cosine similarity between speaker embeddings extracted from the original and reconstructed utterances.
When evaluated using the local model $F_L$, the resulting similarity score is defined as
$ s_L(\mathfrak{v}) = \langle F_L(\mathfrak{v}), F_L(\tilde{\mathfrak{v}}) \rangle$, which directly corresponds to property A1.
To evaluate transferability (A2), we compute the same similarity using an alternative target speaker recognition model $F_T$: $s_T(\mathfrak{v}) = \langle F_T(\mathfrak{v}), F_T(\tilde{\mathfrak{v}}) \rangle.$
In our experiments, this protocol is applied to utterances from the VoxCeleb1 test set. For reference, we compare the resulting round-trip similarity distributions against positive (same-speaker) and negative (different-speaker) score distributions under each evaluation model.

\begin{figure}[t]
    \centering
    \includegraphics[width=\linewidth]{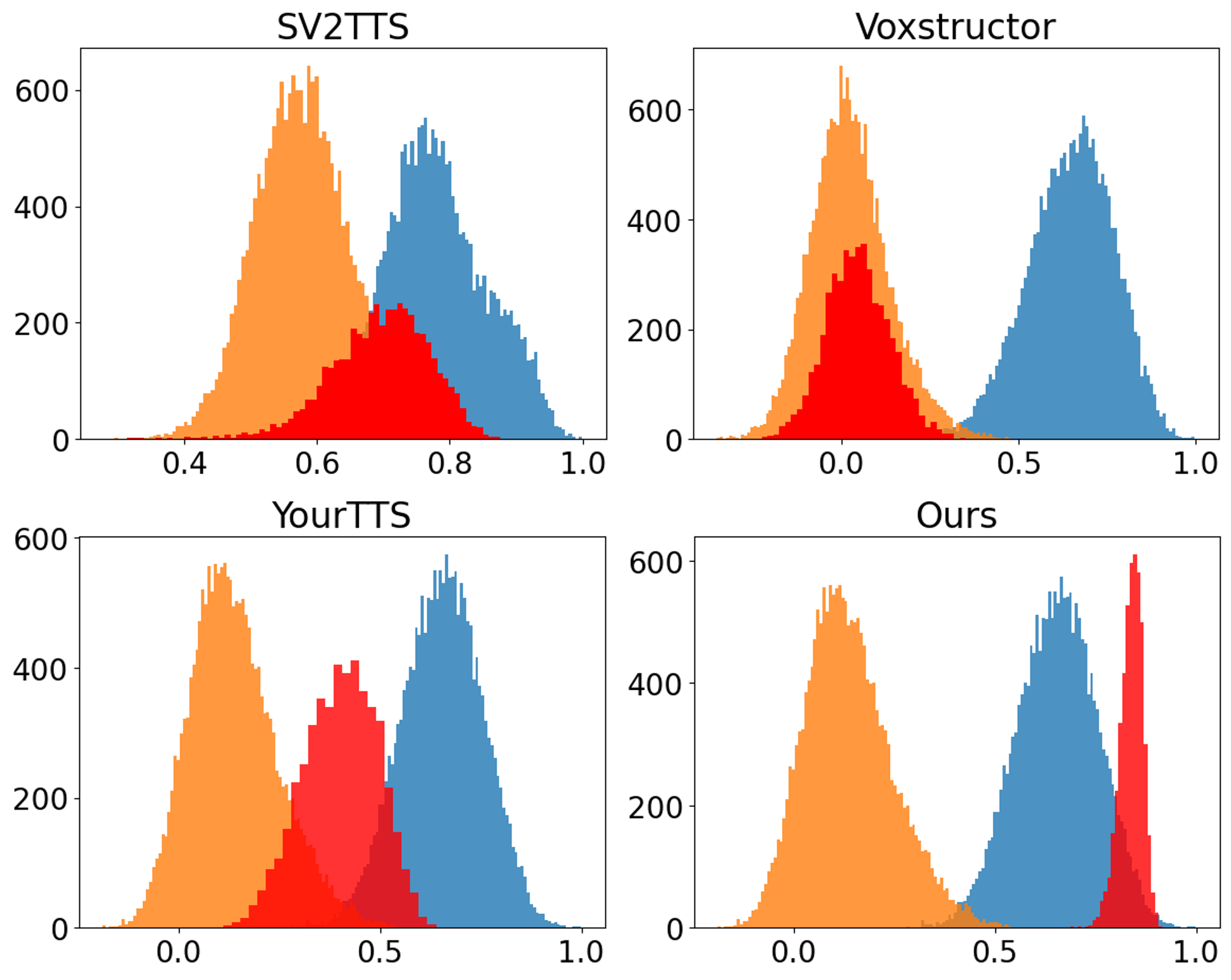}
    \vspace{-20pt}
    \caption{Distributions of round-trip similarity scores $s_L(\mathfrak{v})$ (red) evaluated on the local speaker recognition model $F_L$.
    Positive (blue) and negative (orange) reference distributions are shown for comparison.
    Best viewed in color.}
    \label{fig:id-consL}
\end{figure}

\begin{figure}[t]
    \centering
    \includegraphics[width=\linewidth]{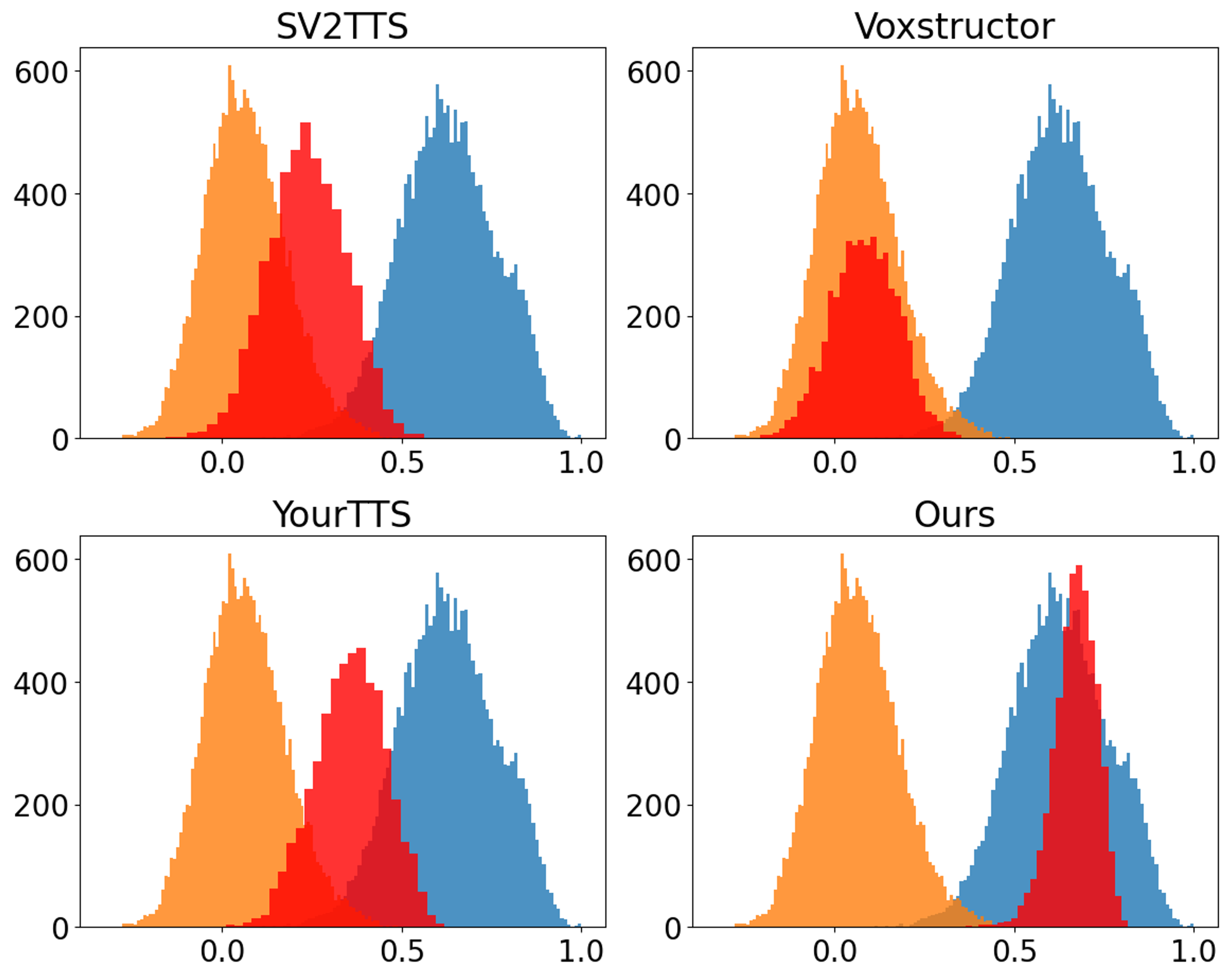}
    \vspace{-20pt}
    \caption{Distributions of round-trip similarity scores $s_T(\mathfrak{v})$ (red) evaluated on the $F_T = T_1$.
    Positive (blue) and negative (orange) reference distributions are shown for comparison.
    Best viewed in color.}
    \vspace{-10pt}
    \label{fig:id-consT}
\end{figure}

Figure~\ref{fig:id-consL} shows the round-trip similarity distributions evaluated by the local model $F_L$, corresponding to property (A1).
For existing TTS-based inverse model candidates, the distributions of $s_L(\mathfrak{v})$ are clearly shifted toward those of negative pairs.
That is, the reconstructed utterances $\tilde{\mathfrak{v}}$ fail to preserve speaker identity even when evaluated by the same model that produced the original embeddings, revealing weak identity preservation in the round-trip mapping.

Figure~\ref{fig:id-consT} further evaluates the same inverse samples under an alternative target model $F_T = T_1$, corresponding to property (A2). (More results on various target models is given in Appendix~\ref{app:id_constraints_transfer})
Compared to the local evaluation, the separation between the round-trip similarity distribution and the positive reference further degrades, and the distributions overlap more closely with negative pairs.
This observation suggests that any residual identity information retained under $F_L$ does not reliably transfer across speaker recognition models, indicating poor cross-model generalization.

Taken together, these results demonstrate that existing inverse model candidates fail to jointly satisfy the desired properties defined in Section~4.2. 
In particular, they exhibit neither strong identity preservation under the local model (A1) nor robust transferability to unseen target models (A2).

\subsubsection*{\bf Trade-Off from Latent Space Reduction}
Recall that the latent space reduction strategy incurs the trade-off between the query complexity and the accuracy of the solution.
To capture this, we introduce the following two metrics to quantify each component of the trade-off in terms of score-based impersonation attack for a fixed optimization solver, e.g., gradient estimation by NES.
\begin{itemize}[leftmargin=*]
    \item (Metric Projection) How close the solution of Eq.~\eqref{eq:LSB} is to the target voice, i.e., $d_{\mathcal{X}}(\mathfrak{v};G) := \min_{z \in \mathcal{Z}} d_{\mathcal{X}}(F(\mathfrak{v}), F(G(z)))$.

    \item (Optimization Trajectory) How many optimization steps are required to succeed in the attack, i.e., the smallest $t$ such that $d_{\mathcal{X}}(F(\mathfrak{v}), F(G(z_{t})) < \tau$ after $t$ score queries.    
\end{itemize}
The former examines whether the adversary can succeed in the attack with an unlimited query budget, whereas the latter measures the (expected) query complexity.

\begin{figure}[t!]
    \centering
    \includegraphics[width=\linewidth]{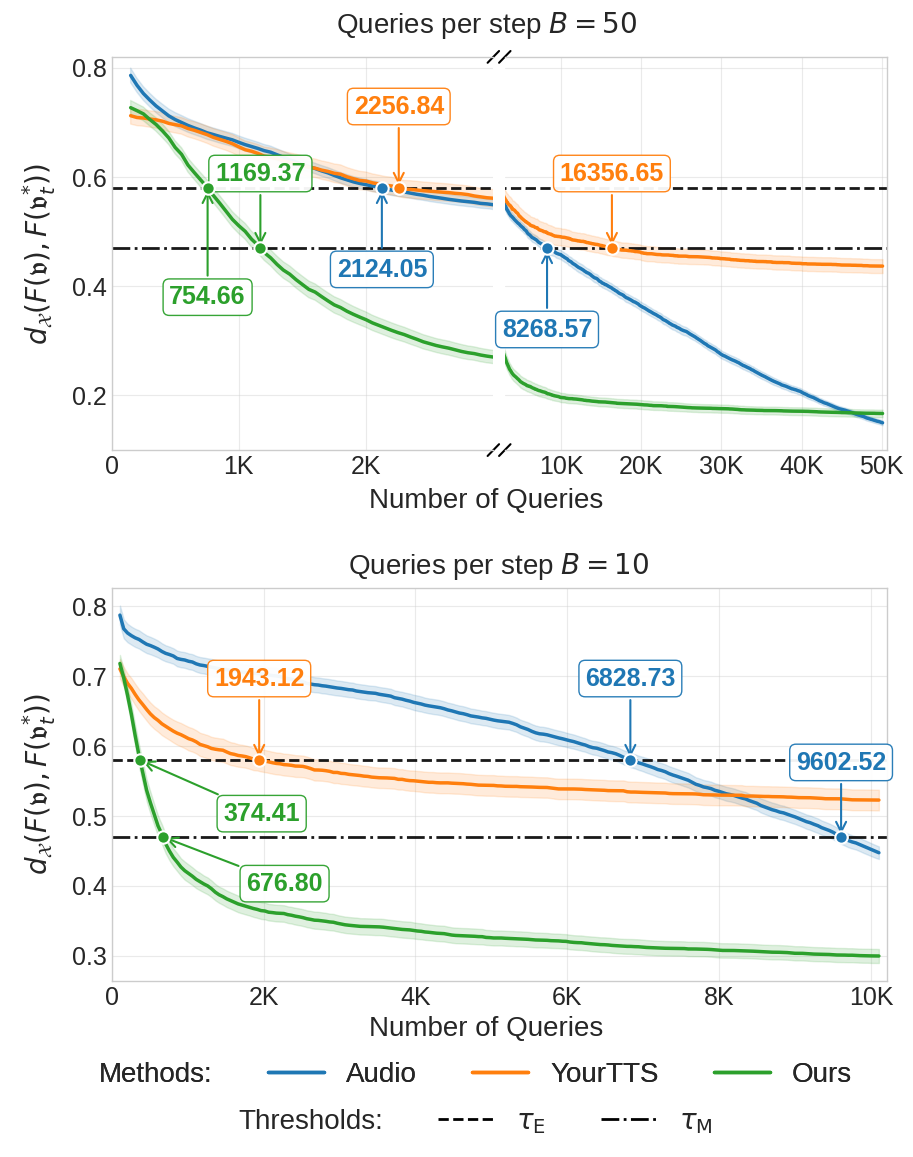}
    \vspace{-20pt}
    \caption{Optimization trajectories of score-based impersonation attacks on various search spaces. 
    Curves show the mean of the best cosine distance with 95\% CI. Best viewed in color.
}
    \label{fig:roundtrip_a1_a2}
\end{figure}

Figure~\ref{fig:roundtrip_a1_a2} compares the optimization trajectories of different inverse model candidates under using two different batch sizes (num. queries for calculating optimization gradients), \(B=50\) and \(B=10\), evaluated against the target model ($L$: feature extractor of YourTTS) in a black-box setting.
The x-axis denotes the cumulative number of score queries, and the y-axis reports the projection metric \(d_{\mathcal{X}}(F(\mathfrak{v}), F(\mathfrak{v}^{\ast}_t))\).

Audio-space optimization shows similar behavior across both budgets, requiring a large number of total queries to make noticeable progress. 
This reflects the high per-step query cost incurred by gradient estimation in the audio space, which limits the number of effective update steps.
Latent-space optimization substantially reduces this cost, but its effectiveness depends critically on how well the latent space aligns with the feature space.

This difference is evident when comparing YourTTS with our inverse model.
Although both methods operate in a low-dimensional latent space, YourTTS exhibits early saturation and slow improvement after an initial decrease in the projection metric.
This suggests that, while gradient estimation is feasible, the resulting gradients are weakly aligned with the impersonation objective.
In contrast, our inverse model maintains a consistently steep trajectory, indicating that its latent representation provides more informative gradients for speaker similarity optimization.

As a result, the gap between our method and YourTTS widens in the low-query regime.
While YourTTS benefits from dimensionality reduction, it fails to fully translate additional optimization steps into improved projection accuracy.
These results indicate that reducing dimensionality is necessary but not sufficient: the latent space must also be structured to reflect the geometry of the target feature space.

In the next sections, we present our method for building he inverse model and completing the impersonation attack.

\section{The Proposed Inverse Model}


Our approach constructs an inverse model by fine-tuning a pre-trained TTS model.
The training procedure consists of three key components: (1) a fixed-text strategy, (2) specialized loss functions, and (3) strategic parameter freezing.

\subsection{Framework: Fixed-Text Fine-Tuning}

\begin{figure}[h]
    \centering
    \includegraphics[width=.8\linewidth]{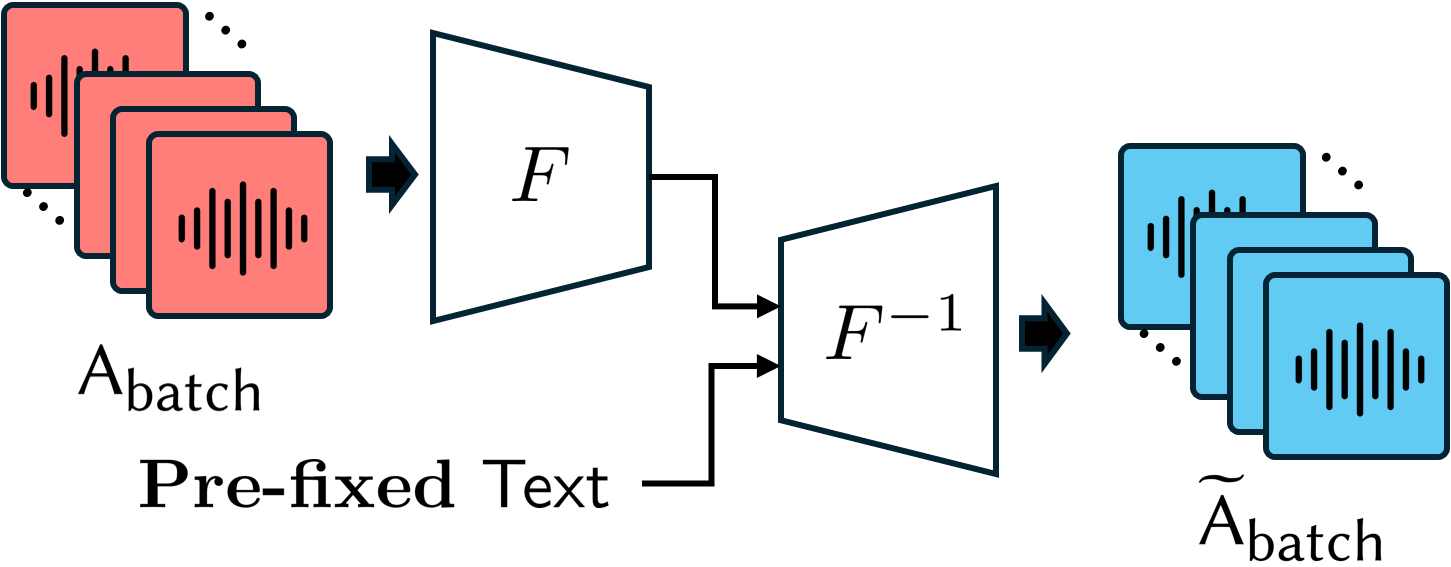}
    \caption{Fixed-text Strategy for Fine Tuning Inverse Model.}
    \label{fig:loss}
\end{figure}

Our core idea is to adapt a pre-trained TTS model to approximate the inverse of a given speaker recognition model $F$.
A fundamental obstacle is that speaker recognition datasets lack aligned text labels, rendering conventional TTS fine-tuning inapplicable.
To overcome this issue, we employ a \emph{fixed-text strategy}, which fixes a pre-defined text sequence as the TTS input during fine-tuning.

This strategy offers two key advantages.
First, it enables the use of large-scale, text-unlabeled speaker recognition datasets for training, allowing the inverse model to be optimized directly on the data distribution learned by the target model it aims to invert.
Second, by decoupling constant linguistic content from variable speaker identity, the model is forced to allocate its learning capacity exclusively to mapping a given speaker embedding to its corresponding vocal characteristics.
Fixing the linguistic content implicitly regularizes training by eliminating variability unrelated to speaker identity.

\subsection{Loss Function Design for the Inverse Model}
\label{subsec:loss_function_design}

To train the inverse model effectively, we introduce two complementary loss functions:
an \emph{ID-constraints loss} for enforcing sample-level fidelity and a \emph{structure-constraints loss} for preserving the geometry of the speaker embedding space.

\begin{figure}[h]
    \centering
    \includegraphics[width=.7\linewidth]{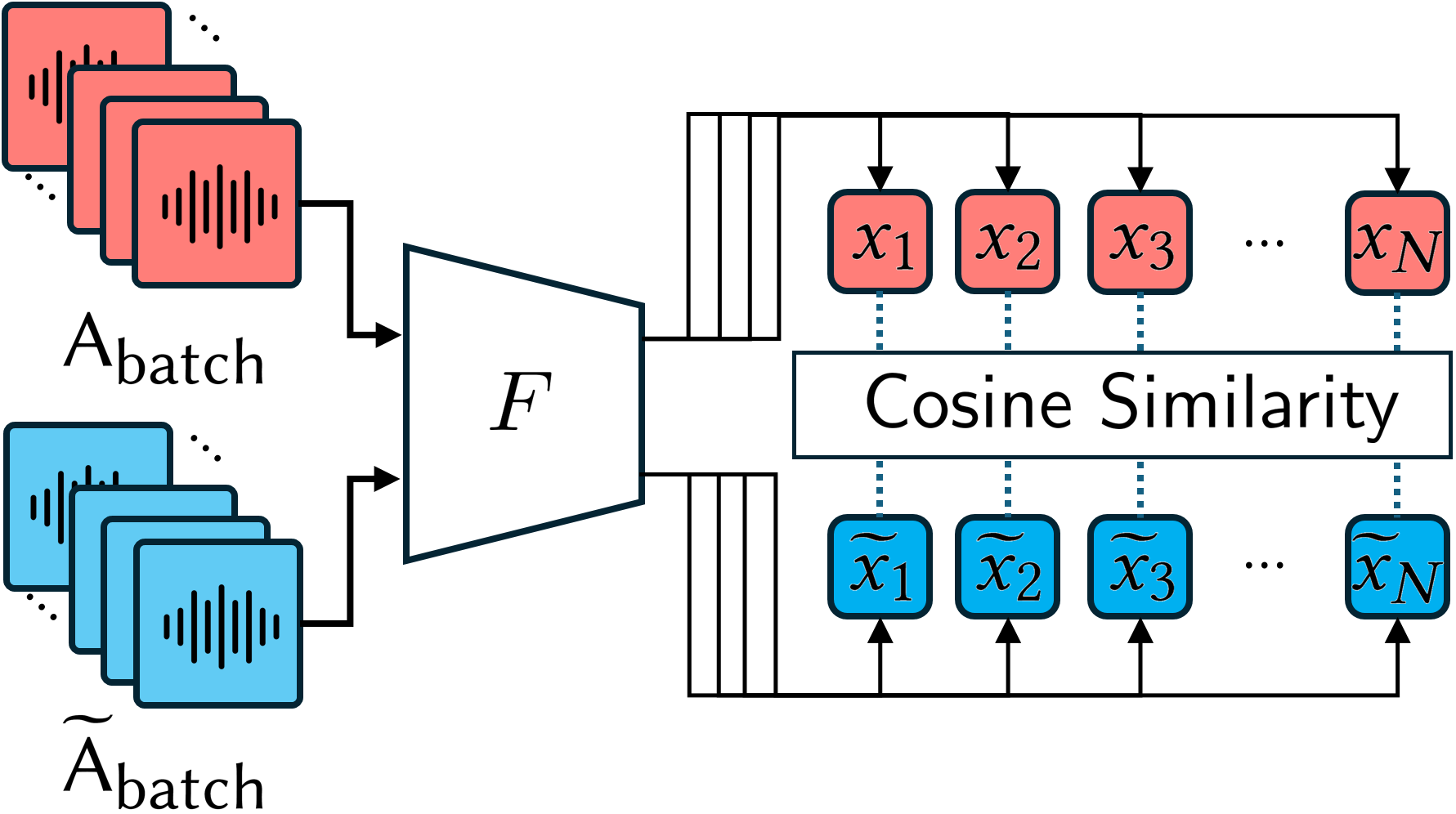}
    \caption{ID-Constraints Loss $L_{\mathsf{IC}}$. Given an audio batch $\mathsf{A}_\mathsf{batch}$, the cosine similarities for each sample pair are computed using the set of synthesized samples $\widetilde{\mathsf{A}}_\mathsf{batch}$ generated through the process illustrated in Figure~\ref{fig:loss}.}
    \label{fig:loss1}
\end{figure}

\subsubsection*{\bf ID-Constraints Loss ($L_{\mathsf{IC}}$).}
At the sample level, this loss enforces a one-to-one identity correspondence between an original audio sample $\mathsf{aud}_i$ and its synthesized counterpart $\widetilde{\mathsf{aud}}_i$ by maximizing the similarity of their speaker embeddings.
This loss directly promotes property \textbf{A1}.
\[
L_{\mathsf{IC}}=\frac{1}{N}\sum_{i=1}^{N} \left(1-\langle F(\mathsf{aud}_i), F(\widetilde{\mathsf{aud}}_i) \rangle\right),
\]
where $\widetilde{\mathsf{aud}}_i = F^{-1}(F(\mathsf{aud}_i), \mathsf{Text})$, $N$=size of batch.

\begin{figure}[h]
    \centering
    \includegraphics[width=.8\linewidth]{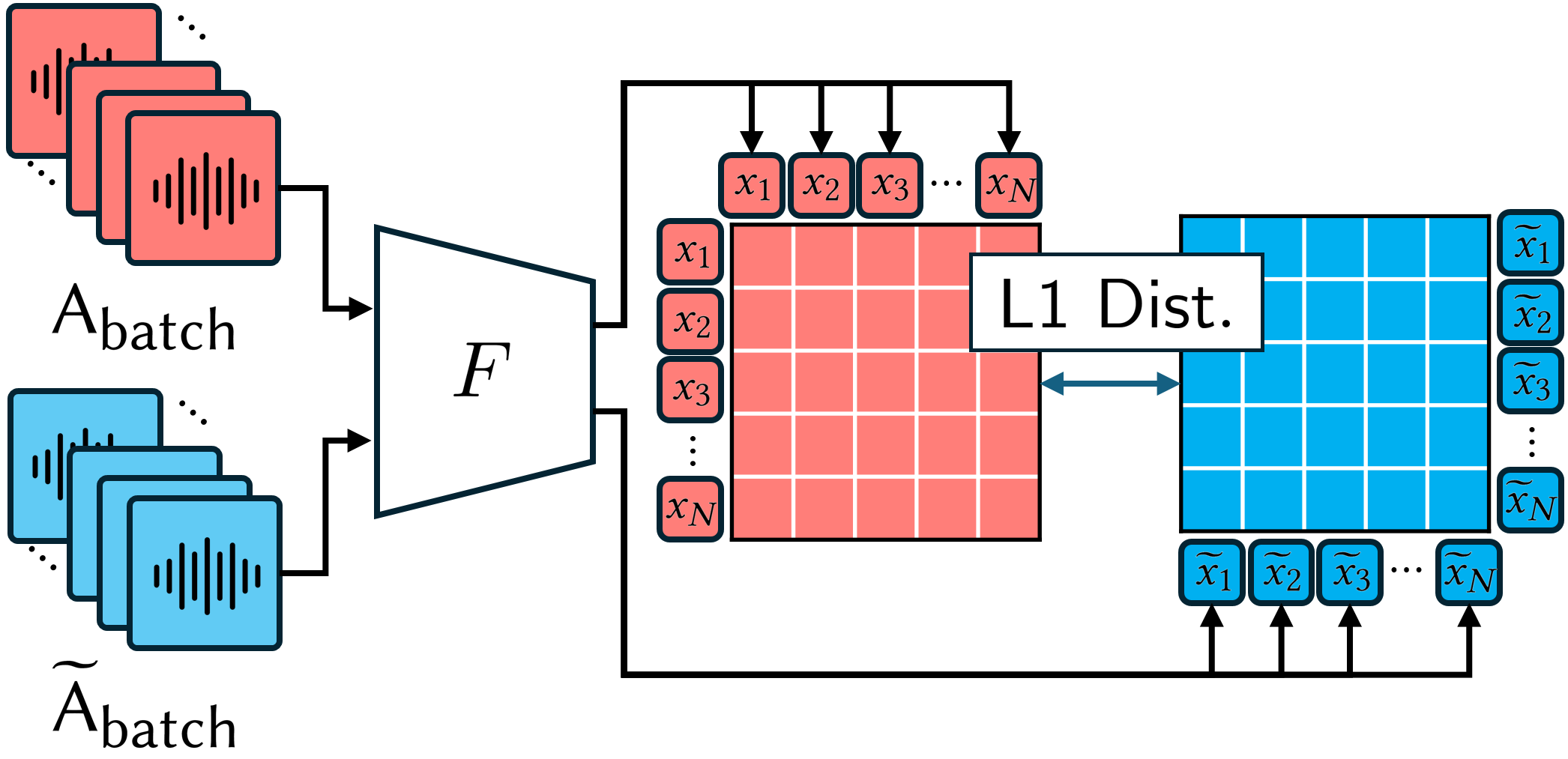}
    \caption{ID-Constraints Loss $L_{\mathsf{IC}}$. Given an audio batch $\mathsf{A}_\mathsf{batch}$, the cosine similarities for each sample pair are computed using the set of synthesized samples $\widetilde{\mathsf{A}}_\mathsf{batch}$ generated through the process illustrated in Figure~\ref{fig:loss}.}
    \label{fig:loss1}
\end{figure}

\subsubsection*{\bf Structure-Constraints Loss ($L_{\mathsf{SC}}$).}
At the batch level, this loss preserves the relational structure of the speaker embedding space.
Specifically, it enforces consistency between the pairwise similarity matrix of original samples, $S$, and that of synthesized samples, $\widetilde{S}$.
By preserving relative distances between speaker embeddings, this loss encourages the inverse model to respect the global geometry of the speaker space, which is critical for generalization and transferability (property \textbf{A2}).
\[
L_{\mathsf{SC}} = \frac{1}{N^2} \sum_{i=1}^{N} \sum_{j=1}^{N} \left| S_{i,j} - \widetilde{S}_{i,j} \right|,
\]
where $S_{i,j} = \langle F(\mathsf{aud}_i), F(\mathsf{aud}_j)\rangle$ and
$\widetilde{S}_{i,j} = \langle F(\widetilde{\mathsf{aud}}_i),F(\widetilde{\mathsf{aud}}_j) \rangle$.

\subsubsection*{\bf Final Loss.}
The final training objective is a weighted combination of the two losses:
\[
L_{\mathsf{Total}}= \lambda_{\mathsf{IC}}L_{\mathsf{IC}} + \lambda_{\mathsf{SC}}L_{\mathsf{SC}}.
\]
Together, these losses ensure that the inverse model learns a structured and transferable mapping aligned with the speaker embedding space, rather than merely memorizing individual identities.

\subsection{Setting of Trainable Parameters}
\label{subsec:setting_params}

Given the fixed-text strategy, we freeze all text-related components of the TTS model, including the text encoder and the duration predictor.
Since the linguistic content remains constant, their pre-trained representations are sufficient.
Training updates are therefore concentrated on the speaker encoding and audio synthesis components, focusing the model capacity on learning the mapping from a speaker embedding $F(\mathsf{aud}_i)$ to the corresponding synthesized audio $\widetilde{\mathsf{aud}}_i$.
This design yields a compact and stable optimization space, which is essential for query-efficient gradient estimation in subsequent black-box attacks.
Further details are provided in Appendix~\ref{subsubsec:app_trainable}.

\section{Experiments}\label{sec:exp}

In this section, we evaluate the proposed inverse model under score-based black-box speaker impersonation attacks.
We first introduce two attack instantiations that leverage the inverse model in complementary ways: one based on iterative score-based optimization, and the other based on a single-step inference exploitation of score-level structural information.
This is followed by a description of the experimental setup and evaluation protocol.

We present quantitative results for the two attack settings and, to better understand the factors underlying these results, we further analyze the impact of the inverse-model training objectives through controlled ablation experiments.
Finally, we examine how the proposed framework can be generalized across languages.

\subsection{Attack Instantiation Using the Proposed Inverse Model}
\label{subsec:using_inverse_model}

We evaluate the proposed inverse model under two score-based black-box speaker impersonation attacks, denoted as \textbf{Ours-NES} and \textbf{Ours-SP}.
In both settings, the attacker can only query a target speaker recognition system $T$ for similarity scores and has no access to its internal parameters.
The attacker additionally possesses a speaker recognition model $F$ and its inverse $F^{-1}$, constructed independently of $T$.
The two attacks differ in how the inverse model is utilized during the attack.

\subsubsection*{\bf Ours-NES}
Ours-NES performs adaptive score-based optimization in the latent space induced by the inverse model.
At each iteration, a latent variable $z \in \mathcal{Z}$ is updated using Natural Evolution Strategies (NES)~\cite{wierstra2014natural} based on score queries to $T$, and decoded into an audio sample via $b_t = F^{-1}(z_t)$.

Although Ours-NES operates under the same black-box threat model as FakeBob, the two attacks are fundamentally different in formulation.
FakeBob is proposed as an adversarial perturbation attack and enforces explicit noise-norm constraints around a source audio.
In contrast, Ours-NES directly optimizes for speaker impersonation without relying on a source audio or imposing any audio-domain perturbation constraints, treating the inverse model as a generative prior for producing valid attack samples.

\subsubsection*{\bf Ours-SP}
Ours-SP instantiates a non-adaptive impersonation attack~\cite{kim2024scores} that exploits the structural similarity between the target system and a locally available speaker recognition model.
Rather than performing iterative optimization, Ours-SP leverages the fact that similarity scores produced by models trained for the same task often preserve consistent geometric structure.

Concretely, the attacker queries the target system $T$ with a set of audio samples $\{\mathfrak{v}_i\}$ and observes the corresponding scores $s_i = T(\mathfrak{v}_i)$.
Assuming structural alignment between the target system and the local model $F$, these scores can be approximated as
$s_i \approx \langle F(\mathfrak{v}_i), F(e) \rangle$, where $e$ denotes the enrolled voice.
This relation allows the attacker to derive, in a purely mathematical manner, an estimate $\hat{x}$ of the victim’s speaker embedding under the local model by solving a linear system.
To ensure numerical stability, the query set is constructed to be approximately orthogonal in the embedding space, formalized as a $\delta$-orthogonal biometric set.

Crucially, recovering such a speaker embedding alone is insufficient to mount a practical impersonation attack.
Without a feasible inverse model, there is no direct way to convert the reconstructed embedding $\hat{x}$ into a valid audio sample.
The proposed inverse model fills this gap by mapping $\hat{x}$ back to the audio domain, yielding an impersonating audio sample $b^\star = F^{-1}(\hat{x})$.
In this sense, the inverse model serves as an essential tool that transforms score-level structural leakage into an executable speaker impersonation attack. (More details in Appendix~\ref{sec:analysisASR})

\subsubsection*{\bf Summary.}
Ours-NES and Ours-SP represent two complementary ways of exploiting the same inverse model for score-based black-box impersonation.
Ours-NES emphasizes adaptive optimization in a compact latent space, whereas Ours-SP relies on mathematically deriving a victim's speaker embedding in local model space from score queries and executing the attack via inversion.

\subsection{Experimental Setting}
To construct a comprehensive experimental environment, we selected five open-source SR models ($T_1$--$T_5$) as our targets. 
Our selection strategy was designed to establish diverse evaluation conditions based on model architecture, and the training data's relationship to our local model ($L$). 
Specifically, $T_1$ through $T_4$ were trained on a dataset completely disjoint from $L$, allowing evaluation in a context with no data domain overlap. 
In contrast, $T_5$ was trained on a substantially larger and more diverse collection of datasets that partially overlaps with our own, enabling evaluation against a more powerful and well-generalized target.

For each model, we set acceptance thresholds by calculating the standard Equal Error Rate (EER) and the minimum Detection Cost Function (minDCF). 
In Appendix~\ref{subsec:modelconfig}, we provide the specific details for each model, including their architectures, training datasets, performance, and the resulting acceptance thresholds ($\tau_\mathsf{E}$, $\tau_\mathsf{M}$).

\begin{table}[t]
\centering
\caption{Specifications of target SR models ($T_1$--$T_5$) and local model ($L$), including architectures, trainsets, and thresholds defined on EER, minDCF (resp. $\tau_\mathsf{E}$, $\tau_\mathsf{M}$). Detailed EER and minDCF performance metrics are in Table~\ref{tab:merged_sr_models}.}
\resizebox{\linewidth}{!}{
\begin{tabular}{c|c|c|c|c}
\toprule[1pt]
\textbf{Model} & \textbf{Architecture} & \textbf{Trainset} & $\tau_\mathsf{E}$ & $\tau_\mathsf{M}$ \\ \hline \hline
$T_1$ & Redim-S~\cite{yakovlev24_interspeech} & VoxBlink2 & 0.6605 & 0.5518 \\ \hline
$T_2$ & Redim-M~\cite{yakovlev24_interspeech} & VoxBlink2 & 0.6624 & 0.5437 \\ \hline
$T_3$ & SimAMResNet34~\cite{qin2022simple} & VoxBlink2 & 0.6256 & 0.5060 \\ \hline
$T_4$ & SimAMResNet100~\cite{qin2022simple} & VoxBlink2 & 0.6135 & 0.4932 \\ \hline
$T_5$ & Titanet-L~\cite{koluguri2022titanet} & VoxCeleb1,2+$\alpha$ & 0.6654 & 0.5215 \\ \hline \hline
$L$ & H/ASP~\cite{heo2020clova} & VoxCeleb1,2 & 0.5778 & 0.4667 \\ 

\bottomrule[1pt]
\end{tabular}
}
\label{tab:SRmodels}
\end{table}

We use the Attack Success Rate (ASR) as an evaluation metric for impersonation attacks, which is the proportion of generated attack audio samples ($\widehat{\mathsf{aud}}_i$) that the target SRS successfully verifies as belonging to the same identity as their corresponding target audio samples ($\mathsf{aud}_i$). The ASR can be formulated as follows:
$\text{ASR} =  \frac{\sum_{i=1}^{|\mathcal{D}_\text{test}|}{\mathbb{1} (D_{\mathcal{X}}(F_{T}(\mathsf{aud}_i), F_{T}(\widehat{\mathsf{aud}}_i)) < \tau_{\text{target}}})}{|\mathcal{D_\text{test}}|}$.
where $\mathcal{D}_\text{test}$ represents the set of test audio samples, $\mathsf{aud}_i \in \mathcal{D}_\text{test}$ refers to each target audio sample, $F_{T}$ is the target SRS's feature extractor, $\widehat{\mathsf{aud}}_i$ is the generated attack audio corresponding to $\mathsf{aud}_i$, $\tau_{\text{target}}$ is the target SRS's decision threshold for verification, and $\mathbb{1}(\cdot)$ is the indicator function (1 if true, 0 otherwise). 
An attack succeeds for $\mathsf{aud}_i$ if the similarity between $F_T(\mathsf{aud}_i)$ and $F_T(\widehat{\mathsf{aud}}_i)$ exceeds $\tau_{\text{target}}$. In our evaluation, $\mathcal{D}_\text{test}$ is instantiated using the VoxCeleb1 test set for English
and the CNCeleb test--enrollment split for Chinese.
For NES-based attacks, we randomly subsample 100 target audio samples from $\mathcal{D}_\text{test}$ due to the high query cost per sample, and compute ASR over this subset.

\subsection{Evalation of Ours-NES}

\label{subsec:eval_ours_nes}

\subsubsection*{\bf Baselines}

We compare \textbf{Ours-NES} against several score-based and generative impersonation baselines under the same black-box threat model.

\paragraph{Audio-NES}
Audio-NES estimates gradients directly in the waveform space by injecting random perturbations and observing score variations from the target system.
This approach is conceptually closest to prior score-based attacks such as FakeBob.
However, FakeBob is originally formulated as an adversarial perturbation attack and assumes the existence of a source utterance together with an explicit noise-norm constraint.

To ensure a fair comparison under our impersonation setting, where no source audio is not essential, we modify the setup as follows.
Audio-NES is initialized from a randomly sampled 3-second audio signal and performs unconstrained optimization without any perturbation budget.
This configuration removes advantages stemming from source-audio proximity and isolates the effect of optimizing directly in the audio space.

\paragraph{YourTTS-NES}
YourTTS-NES follows the same NES-based optimization as Ours-NES, but replaces the proposed inverse model with YourTTS.

\paragraph{Excluded candidates}
We do not include SV2TTS and Voxstructor as NES-based baselines.
Both methods require repeated inner-loop synthesis or reconstruction steps, making them prohibitively slow for iterative score-based optimization.
Moreover, our earlier analysis shows that their speaker identity control and local speaker-discriminative alignment are substantially weaker than those of YourTTS.
As a result, they are dominated both computationally and in attack effectiveness, and are therefore omitted from the NES-based comparison.

\paragraph{Implementation details.}
While all NES-based attacks follow the same score-based optimization paradigm, their concrete implementations differ depending on the underlying optimization space.
In particular, the parameterization, noise injection strategy, and decoding process vary across audio-space, and inverse-latent-space optimization.
We describe these differences and the corresponding algorithmic instantiations in detail in Appendix~\ref{subsec:implemetaiondetailss}.
For NES-based methods, we report results from the best-performing setting among the tested values of $B = 10, 50$ with 1K iteration steps and in case of comparable performance, we choose the case with fewer queries.

\subsubsection*{\bf Analysis of NES-based Impersonation Results}

\begin{figure}[t!]
    \centering
    \includegraphics[width=\linewidth]{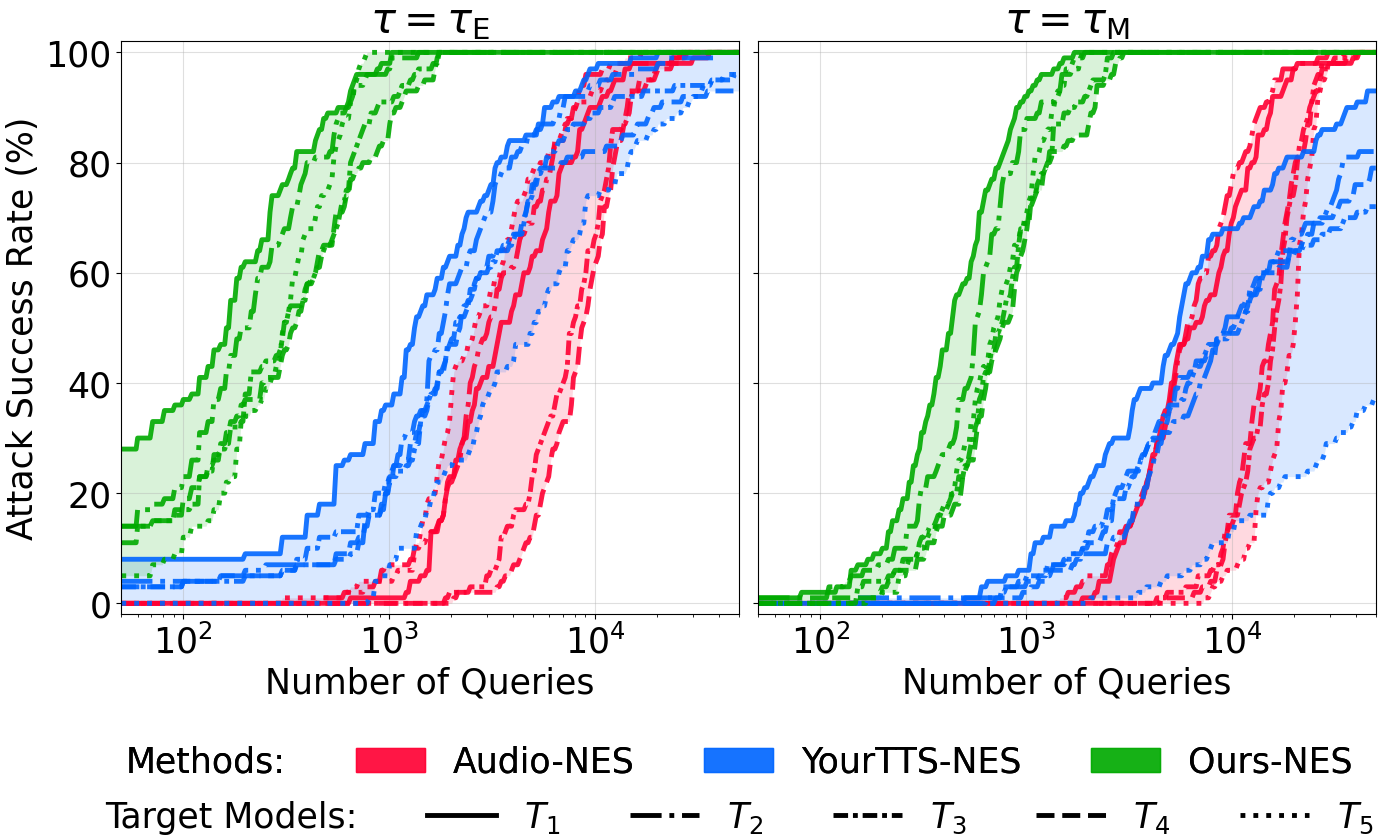}
    \caption{
Attack success rate as a function of the number of score queries for NES-based impersonation attacks.
Results are shown for two decision thresholds, $\tau_E$ (left) and $\tau_M$ (right), across five target speaker recognition models ($T_1$--$T_5$).
}
    \label{fig:eval_nes}
\end{figure}

Figure~\ref{fig:eval_nes} and Table~\ref{tab:main_results_styled} jointly summarize the effectiveness and query efficiency of NES-based speaker impersonation attacks.
Across all five target speaker recognition systems, \textbf{Ours-NES} consistently achieves superior performance in terms of both attack success rate (ASR) and query efficiency under the same black-box threat model.

\begin{table}[t]
\centering
\caption{Comparison of ASR and query efficiency under $\tau_{\text{target}} = \tau_\mathsf{E}$ and $\tau_\mathsf{M}$. The query budget is 50k. Each cell reports the \textbf{ASR} (top) and the \textbf{average number of queries calculated over successful attacks} (bottom).}
\resizebox{\linewidth}{!}{
\begin{tabular}{c|ccccc|c}
\toprule[1.5pt]
\textbf{Method} & \textbf{$T_1$} & \textbf{$T_2$} & \textbf{$T_3$} & \textbf{$T_4$} & \textbf{$T_5$} & \textbf{Avg.} \\ \hline \hline

\multicolumn{7}{c}{$\tau_{\text{target}} = \tau_{\mathsf{E}}$} \\ \hline

Audio-NES & 
\begin{tabular}[c]{@{}c@{}}\textbf{100.0}\% \\ (5.2k)\end{tabular} & 
\begin{tabular}[c]{@{}c@{}}\textbf{100.0}\% \\ (4.3k)\end{tabular} & 
\begin{tabular}[c]{@{}c@{}}\textbf{100.0}\% \\ (8.7k)\end{tabular} & 
\begin{tabular}[c]{@{}c@{}}\textbf{100.0}\% \\ (9.5k)\end{tabular} & 
\begin{tabular}[c]{@{}c@{}}\textbf{100.0}\% \\ (3.8k)\end{tabular} & 
\begin{tabular}[c]{@{}c@{}}\textbf{100.0}\% \\ (6.3k)\end{tabular} \\
\cline{1-1}\cline{2-7}

YourTTS-NES & 
\begin{tabular}[c]{@{}c@{}}\textbf{100.0}\% \\ (2.5k)\end{tabular} & 
\begin{tabular}[c]{@{}c@{}}\textbf{100.0}\% \\ (3.4k)\end{tabular} & 
\begin{tabular}[c]{@{}c@{}}96.0\% \\ (4.1k)\end{tabular} & 
\begin{tabular}[c]{@{}c@{}}93.0\% \\ (4.2k)\end{tabular} & 
\begin{tabular}[c]{@{}c@{}}96.0\% \\ (7.9k)\end{tabular} & 
\begin{tabular}[c]{@{}c@{}}97.0\% \\ (4.4k)\end{tabular} \\
\cline{1-1}\cline{2-7}

\rowcolor{gray!20}
\textbf{Ours-NES} & 
\begin{tabular}[c]{@{}c@{}}\textbf{100.0}\% \\ \textbf{(0.2k)}\end{tabular} & 
\begin{tabular}[c]{@{}c@{}}\textbf{100.0}\% \\ \textbf{(0.3k)}\end{tabular} & 
\begin{tabular}[c]{@{}c@{}}\textbf{100.0}\% \\ \textbf{(0.4k)}\end{tabular} & 
\begin{tabular}[c]{@{}c@{}}\textbf{100.0}\% \\ \textbf{(0.5k)}\end{tabular} & 
\begin{tabular}[c]{@{}c@{}}\textbf{100.0}\% \\ \textbf{(0.3k)}\end{tabular} & 
\begin{tabular}[c]{@{}c@{}}\textbf{100.0}\% \\ \textbf{(0.3k)}\end{tabular} \\
\hline \hline

\multicolumn{7}{c}{$\tau_{\text{target}} = \tau_{\mathsf{M}}$} \\ \hline

Audio-NES & 
\begin{tabular}[c]{@{}c@{}}\textbf{100.0}\% \\ (8.5k)\end{tabular} & 
\begin{tabular}[c]{@{}c@{}}\textbf{100.0}\% \\ (7.6k)\end{tabular} & 
\begin{tabular}[c]{@{}c@{}}\textbf{100.0}\% \\ (16.3k)\end{tabular} & 
\begin{tabular}[c]{@{}c@{}}\textbf{100.0}\% \\ (16.1k)\end{tabular} & 
\begin{tabular}[c]{@{}c@{}}\textbf{100.0}\% \\ (19.3k)\end{tabular} & 
\begin{tabular}[c]{@{}c@{}}\textbf{100.0}\% \\ (13.6k)\end{tabular} \\
\cline{1-1}\cline{2-7}

YourTTS-NES & 
\begin{tabular}[c]{@{}c@{}}93.0\% \\ (8.9k)\end{tabular} & 
\begin{tabular}[c]{@{}c@{}}82.0\% \\ (11.5k)\end{tabular} & 
\begin{tabular}[c]{@{}c@{}}72.0\% \\ (8.9k)\end{tabular} & 
\begin{tabular}[c]{@{}c@{}}79.0\% \\ (11.3k)\end{tabular} & 
\begin{tabular}[c]{@{}c@{}}37.0\% \\ (18.8k)\end{tabular} & 
\begin{tabular}[c]{@{}c@{}}72.6\% \\ (11.0k)\end{tabular} \\
\cline{1-1}\cline{2-7}

\rowcolor{gray!20}
\textbf{Ours-NES} & 
\begin{tabular}[c]{@{}c@{}}\textbf{100.0}\% \\ \textbf{(0.5k)}\end{tabular} & 
\begin{tabular}[c]{@{}c@{}}\textbf{100.0}\% \\ \textbf{(0.6k)}\end{tabular} & 
\begin{tabular}[c]{@{}c@{}}\textbf{100.0}\% \\ \textbf{(0.9k)}\end{tabular} & 
\begin{tabular}[c]{@{}c@{}}\textbf{100.0}\% \\ \textbf{(1.0k)}\end{tabular} & 
\begin{tabular}[c]{@{}c@{}}\textbf{100.0}\% \\ \textbf{(0.8k)}\end{tabular} & 
\begin{tabular}[c]{@{}c@{}}\textbf{100.0}\% \\ \textbf{(0.8k)}\end{tabular} \\

\bottomrule[1.5pt]
\end{tabular}
}
\label{tab:main_results_styled}
\end{table}

Under the EER threshold, all methods are eventually able to achieve high ASR given a sufficiently large query budget.
However, the required number of queries differs substantially across methods.
As shown in Table~\ref{tab:main_results_styled}, Audio-NES requires on average 6.3k queries to succeed, reflecting the difficulty of directly optimizing in the high-dimensional waveform space.
YourTTS-NES reduces this number to 4.4k queries on average by operating in a lower-dimensional generative latent space.
In contrast, Ours-NES achieves 100\% ASR across all target models with only 0.3k queries on average, corresponding to more than an order-of-magnitude reduction in query complexity.
This trend is consistent with the success curves in Figure~\ref{fig:eval_nes}, where Ours-NES exhibits rapid and stable convergence within a few hundred queries.

The performance gap becomes more pronounced under the stricter minDCF threshold.
While Audio-NES still attains 100\% ASR at the cost of a substantially increased query budget (13.6k queries on average), YourTTS-NES exhibits a notable degradation in attack success, achieving only 72.6\% ASR on average.
This degradation is most severe for $T_5$, where the ASR of YourTTS-NES drops to 37.0\% despite consuming up to 18.8k queries.
These results indicate that, beyond a certain point, increasing the query budget alone is insufficient to overcome structural misalignment between the optimization space and the target decision boundary.

The failure cases observed for YourTTS-NES can be interpreted as worst-case manifestations of the trade-off discussed in Section~\ref{sec:goodinverse}.
When the latent space of a generative model is insufficiently aligned with the speaker-discriminative embedding geometry of the target system, score-based optimization may converge to regions that produce natural-sounding speech but remain consistently below stringent acceptance thresholds.
In such poorly aligned scenarios, impersonation attacks may become largely ineffective rather than merely less query-efficient, as exemplified by the behavior on $T_5$ under the minDCF threshold.

\begin{table*}[t!]
    \centering
    \caption{ASRs comparing the proposed method against alternative baseline approaches attacking $T_1$--$T_5$. Target thresholds ($\tau_{\text{target}}$) are derived from EER ($\tau_\mathsf{E}$) and minDCF ($\tau_\mathsf{M}$) for each target model.}
    \resizebox{0.82\linewidth}{!}{
        \begin{tabular}{c|cc|cc|cc|cc|cc}
            \toprule[1.5pt]
            \multirow{2}{*}{\textbf{Method}} & 
            \multicolumn{2}{c|}{\textbf{$T_1$}} & 
            \multicolumn{2}{c|}{\textbf{$T_2$}} & 
            \multicolumn{2}{c|}{\textbf{$T_3$}} & 
            \multicolumn{2}{c|}{\textbf{$T_4$}} & 
            \multicolumn{2}{c}{\textbf{$T_5$}} \\ \cline{2-11}
             & $\tau_\mathsf{E}$ & $\tau_\mathsf{M}$ 
             & $\tau_\mathsf{E}$ & $\tau_\mathsf{M}$ 
             & $\tau_\mathsf{E}$ & $\tau_\mathsf{M}$ 
             & $\tau_\mathsf{E}$ & $\tau_\mathsf{M}$ 
             & $\tau_\mathsf{E}$ & $\tau_\mathsf{M}$ \\ \hline \hline

            SV2TTS 
            & 1.57\% & 0.04\% 
            & 0.87\% & 0.02\% 
            & 0.59\% & 0.00\% 
            & 0.42\% & 0.00\% 
            & 0.02\% & 0.00\% \\ \cline{1-1}\cline{2-11}

            Voxstructor 
            & 0.11\% & 0.00\% 
            & 0.02\% & 0.00\% 
            & 0.02\% & 0.00\% 
            & 0.01\% & 0.00\% 
            & 0.00\% & 0.00\% \\ \cline{1-1}\cline{2-11}

            Audio-GD 
            & 4.65\% & 0.15\% 
            & 6.37\% & 0.19\% 
            & 1.38\% & 0.02\% 
            & 0.30\% & 0.00\% 
            & 34.36\% & 4.89\% \\ \cline{1-1}\cline{2-11}

            YourTTS 
            & 15.99\% & 2.70\% 
            & 11.62\% & 1.57\% 
            & 10.58\% & 1.72\% 
            & 7.99\% & 1.57\% 
            & 0.28\% & 0.00\% \\ \hline \hline

            \rowcolor{gray!20}
            \textbf{Ours-SP} 
            & \textbf{91.65\%} & \textbf{59.83\%} 
            & \textbf{85.47\%} & \textbf{44.33\%} 
            & \textbf{76.23\%} & \textbf{34.54\%} 
            & \textbf{62.23\%} & \textbf{19.86\%} 
            & \textbf{68.46\%} & \textbf{17.03\%} \\ 
            
            \bottomrule[1.5pt]
        \end{tabular}
    }
    \label{tab:attack_comparison}
\end{table*}

In contrast, Ours-NES maintains 100\% ASR across all target systems and thresholds, including these worst-case settings, with an average of only 0.8k queries under the minDCF criterion.
This robustness stems from the inverse-model-induced latent space, which is explicitly trained to preserve speaker identity with respect to a speaker recognition model.
By aligning the optimization space with speaker-discriminative structure, Ours-NES enables stable gradient estimation and effective score exploitation even under stringent decision boundaries.

\subsubsection*{\bf Takeaway.}
Overall, these results highlight that the effectiveness of score-based speaker impersonation is fundamentally determined by the choice of the optimization space.
Simply adopting a generic generative prior is insufficient to ensure reliable attacks across target systems.
In contrast, an inverse-model-induced latent space that is explicitly aligned with speaker-discriminative feature geometry enables both query-efficient and robust black-box impersonation.
This alignment allows stable score exploitation even under stringent decision thresholds, making the inverse model a key component of practical score-based impersonation attacks.

\subsection{Evaluation of Ours-SP}
\label{subsec:eval_ours_sp}

\subsubsection*{\bf Baselines}
Ours-SP is a non-adaptive, one-step impersonation attack that reconstructs a victim's feature vector in local space from score queries and executes the attack via inversion. Since no iterative optimization is involved, all baselines can be evaluated under the same query budget(=50, detailed in Apppendix~\ref{ovs}). \cite{wierstra2014natural}

\paragraph{SV2TTS, Voxstructor and YourTTS}
For SV2TTS, Voxstructor, and YourTTS, we generate attack utterances by conditioning the models on speaker representations estimated via subspace projection in the local feature space.
The projected speaker vector is used as the conditioning input to synthesize a single utterance per attack.

\paragraph{Audio-GD}
Audio-GD (Gradient Descent) operates in the same search space as Audio-NES, namely the raw audio domain, but differs fundamentally in its objective.
Instead of optimizing the target system’s score directly, Audio-GD minimizes the distance between the embedding of the generated audio and a reconstructed \emph{local} speaker embedding under the attacker’s local model. (Detailed in Appendix~\ref{subsub:audiogd})
This setting isolates the effect of local embedding recovery while removing the need for an inverse mapping.

\subsubsection*{\bf Analysis of SP-based Impersonation Results.}

Table~\ref{tab:attack_comparison} summarizes the attack success rates (ASRs) of Ours-SP and baseline methods against five target speaker recognition systems under both the EER-derived threshold $\tau_\mathsf{E}$ and the stricter minDCF-derived threshold $\tau_\mathsf{M}$.
All methods are evaluated under the same query budget and generate a single attack utterance, allowing for a direct comparison of their practical effectiveness.

Across all target systems and both threshold settings, \textbf{Ours-SP} consistently achieves substantially higher ASRs than all baseline approaches.
For example, on $T_1$, Ours-SP attains an ASR of 91.65\% at $\tau_\mathsf{E}$ and 59.83\% at $\tau_\mathsf{M}$, whereas the strongest baseline (YourTTS) reaches only 15.99\% and 2.70\%, respectively.
Similar performance gaps are observed across $T_2$--$T_5$, indicating that the proposed method generalizes well across heterogeneous target systems even without iterative score optimization.

Baseline TTS-based approaches, including SV2TTS, Voxstructor, and YourTTS, rely purely on the generative capability of pretrained models conditioned on inferred speaker representations.
Since these methods do not exploit score-level structural information from the target system, their success rates remain low, particularly under the stricter threshold $\tau_\mathsf{M}$.
This limitation highlights that generative priors alone are insufficient to reliably produce utterances that cross target decision boundaries.

Audio-GD represents a stronger baseline that attempts to recover a speaker embedding by minimizing the distance between the generated audio and a reconstructed \emph{local} speaker embedding.
While this approach partially leverages speaker-discriminative information, it still exhibits low and unstable ASRs across most target systems.
This behavior suggests that local embedding recovery alone does not guarantee effective transfer to the target system, especially when the local and target embedding spaces are misaligned.
In such cases, the absence of an identity-preserving inverse mapping prevents the reconstructed representation from being translated into a successful impersonation.

An exception is observed for $T_5$, where Audio-GD achieves relatively higher ASRs.
This behavior can be attributed to improved alignment between the local and target embedding spaces, potentially due to partial overlap in training data or architectural similarity.
Even in this more favorable setting, however, Audio-GD remains substantially less effective than Ours-SP, underscoring the importance of effective inversion.

\subsubsection*{\bf Takeaway} Overall, these results show that reconstructing speaker-discriminative representations from score queries alone is insufficient for practical impersonation.
Without an effective inverse mapping, recovered embeddings often fail to translate into speech that crosses the target decision boundary.
The proposed inverse model enables identity-preserving translation from reconstructed representations to speech, making one-step SP-based impersonation feasible under realistic query budgets.

\subsection{Effect of Inverse-Model Training Objectives}

Table~\ref{tab:ablation_loss} analyzes how the proposed training objectives influence the behavior of the inverse model.
In the NES-based setting, incorporating the identity-constraint loss ($L_\mathsf{IC}$) substantially improves attack success and query efficiency.
However, applying $L_\mathsf{IC}$ in isolation may still encourage overfitting to local speaker-discriminative features.
This effect is most clearly manifested in the results on $T_5$, where the NES-based attack exhibits a severe drop in success rate despite reduced query counts.
Such behavior suggests that overfitting induced by $L_\mathsf{IC}$ can distort the structural similarity required for effective score-based optimization against different target systems.
By further incorporating the structure-constraint loss ($L_\mathsf{SC}$), the inverse model preserves the geometry of the speaker embedding space, leading to consistently high success rates across all targets and a significant reduction in the number of required queries.

A similar but more implicit trend is observed in the SP-based setting.
Although SP-based attacks do not involve iterative optimization, adding $L_\mathsf{IC}$ alone already enables effective one-shot impersonation by enforcing speaker identity.
Nevertheless, the additional gains achieved by incorporating $L_\mathsf{SC}$ indicate that structural regularization also improves cross-model transferability in the one-shot regime.
Taken together, these results confirm that $L_\mathsf{IC}$ and $L_\mathsf{SC}$ play complementary roles: $L_\mathsf{IC}$ ensures identity fidelity, while $L_\mathsf{SC}$ mitigates overfitting and preserves structural compatibility necessary for robust and transferable impersonation.

\begin{table}[t]
\centering
\caption{
Ablation study on inverse-model training objectives.
For NES-based, we report ASR (\%) at $\tau_\mathsf{M}$ together with the average number of score queries over successful attacks (in parentheses).
For SP-based, we report ASR (\%) at $\tau_\mathsf{E}$.
$L_\mathsf{IC}$: Identity-Constraint loss,
$L_\mathsf{SC}$: Structure-Constraint loss.
}
\resizebox{\linewidth}{!}{
\begin{tabular}{c|l|ccccc}
\toprule[1.2pt]
\multirow{2}{*}{\begin{tabular}[c]{@{}c@{}}\textbf{Attack} \\ \textbf{Method}\end{tabular}} &
\multirow{2}{*}{$\boldsymbol{F^{-1}}$} &
\multicolumn{5}{c}{\textbf{Target SR models}} \\ \cline{3-7}
 &  & $T_1$ & $T_2$ & $T_3$ & $T_4$ & $T_5$ \\ \hline \hline

\multirow{3}{*}{NES}
& YourTTS
& \begin{tabular}[c]{@{}c@{}}93.0\% \\ (8.9k)\end{tabular}
& \begin{tabular}[c]{@{}c@{}}82.0\% \\ (11.5k)\end{tabular}
& \begin{tabular}[c]{@{}c@{}}72.0\% \\ (8.9k)\end{tabular}
& \begin{tabular}[c]{@{}c@{}}79.0\% \\ (11.3k)\end{tabular}
& \begin{tabular}[c]{@{}c@{}}37.0\% \\ (18.8k)\end{tabular} \\

& $+ L_\mathsf{IC}$
& \begin{tabular}[c]{@{}c@{}}100\% \\ (1.7k)\end{tabular}
& \begin{tabular}[c]{@{}c@{}}100\% \\ (1.9k)\end{tabular}
& \begin{tabular}[c]{@{}c@{}}100\% \\ (2.5k)\end{tabular}
& \begin{tabular}[c]{@{}c@{}}99.0\% \\ (2.8k)\end{tabular}
& \begin{tabular}[c]{@{}c@{}}6.0\% \\ (1.3k)\end{tabular} \\

& \cellcolor{gray!20}$+ L_\mathsf{IC} + L_\mathsf{SC}$ 
& \cellcolor{gray!20}\begin{tabular}[c]{@{}c@{}}\textbf{100\%} \\ \textbf{(0.5k)}\end{tabular}
& \cellcolor{gray!20}\begin{tabular}[c]{@{}c@{}}\textbf{100\%} \\ \textbf{(0.6k)}\end{tabular}
& \cellcolor{gray!20}\begin{tabular}[c]{@{}c@{}}\textbf{100\%} \\ \textbf{(0.9k)}\end{tabular}
& \cellcolor{gray!20}\begin{tabular}[c]{@{}c@{}}\textbf{100\%} \\ \textbf{(1.0k)}\end{tabular}
& \cellcolor{gray!20}\begin{tabular}[c]{@{}c@{}}\textbf{100\%} \\ \textbf{(0.8k)}\end{tabular} \\
\hline \hline
\multirow{3}{*}{SP}
& YourTTS
& 15.99\% & 11.62\% & 10.58\% & 7.99\% & 0.34\% \\

& $+ L_\mathsf{IC}$
& 71.37\% & 56.29\% & 45.58\% & 29.21\% & 64.80\% \\

& \cellcolor{gray!20}$+ L_\mathsf{IC} + L_\mathsf{SC}$
& \cellcolor{gray!20}\textbf{91.65\%}
& \cellcolor{gray!20}\textbf{85.47\%}
& \cellcolor{gray!20}\textbf{76.23\%}
& \cellcolor{gray!20}\textbf{62.23\%}
& \cellcolor{gray!20}\textbf{68.46\%} \\ 

\bottomrule[1.2pt]
\end{tabular}
}
\label{tab:ablation_loss}
\end{table}

\subsection{Lingual Generalization: Chinese Speaker Recognition}
\label{subsec:chinese_sr}

Our framework is designed to operate in the speaker embedding space and does not rely on linguistic content.
As such, its applicability is not inherently tied to any specific language.
To further support this claim, we provide an additional evaluation on Chinese speaker recognition systems for providing additional empirical evidence for the language-agnostic nature of the proposed framework.

Specifically, we train an inverse model $L_{\mathsf{CN}}^{-1}$ of Chinese local speaker recognition model $L_{\mathsf{CN}}$ using CN-Celeb1,2.
We then evaluate the proposed attacks against two Chinese target systems ($C_1$, $C_2$) under the same black-box threat model, where the attacker can only observe similarity scores from the target system.

\begin{table}[t]
\centering
\caption{Specifications of CN target SR models ($C_1$-$C_2$) and local model ($L_\mathsf{CN}$), including architectures, trainsets, and thresholds defined on EER, minDCF (resp. $\tau_\mathsf{E}$, $\tau_\mathsf{M}$).}
\resizebox{.95\linewidth}{!}{
\begin{tabular}{c|c|c|c|c}
\toprule[1pt]
\textbf{Model} & \textbf{Architecture} & \textbf{Trainset} & $\tau_\mathsf{E}$ & $\tau_\mathsf{M}$ \\ \hline \hline
$C_1$ & CAM++~\cite{cam++} & N/A & 0.7064 & 0.4817 \\ \hline
$C_2$ & ERes2Net~\cite{eres2net} & N/A & 0.6350 & 0.4617 \\ \hline
$L_{\mathsf{CN}}$ & ResNet34~\cite{chung2020defence} & CN-Celeb 1,2 & 0.7349 & 0.5043 \\ 
\bottomrule[1pt]
\end{tabular}
}
\label{tab:CN_SRmodels}
\end{table}

\begin{table}[t]
\centering
\caption{
Attack success rates (ASR, \%) on Chinese speaker recognition systems.
For Ours-NES, the average number of score queries computed over successful attacks is reported in parentheses.
Target thresholds are derived from EER ($\tau_\mathsf{E}$) and minDCF ($\tau_\mathsf{M}$).
}
\resizebox{0.75\linewidth}{!}{
\begin{tabular}{c|cc|cc}
\toprule[1pt]
\multirow{2}{*}{\textbf{Method}} &
\multicolumn{2}{c|}{\textbf{$C_1$}} &
\multicolumn{2}{c}{\textbf{$C_2$}} \\ \cline{2-5}
 & $\tau_\mathsf{E}$ & $\tau_\mathsf{M}$ & $\tau_\mathsf{E}$ & $\tau_\mathsf{M}$ \\ \hline \hline

Ours-NES
& \begin{tabular}[c]{@{}c@{}}100\% \\ (0.2k)\end{tabular}
& \begin{tabular}[c]{@{}c@{}}100\% \\ (1.4k)\end{tabular}
& \begin{tabular}[c]{@{}c@{}}100\% \\ (0.2k)\end{tabular}
& \begin{tabular}[c]{@{}c@{}}100\% \\ (0.6k)\end{tabular} \\ \hline
Ours-SP & 71.43\% & 4.08\% & 87.76\% & 16.84\% \\
\bottomrule[1pt]
\end{tabular}
}
\label{tab:cn_results}
\end{table}

\subsubsection*{\bf Evaluation Settings.}
We consider two Chinese target speaker recognition systems ($C_1$, $C_2$) and one Chinese local model ($L_{\mathsf{CN}}$).
The target systems are adopted from the \textit{3D-Speaker}~\cite{chen20243d} toolkit and are trained on large-scale Chinese speech corpora.
While the exact training datasets are not publicly disclosed, the model descriptions report training on the order of $200\mathrm{k}$ speaker identities.
Although the exact training datasets are not publicly disclosed, these models are developed independently from our local training setup.

The detailed architectures, training data descriptions, verification performance, as well as experimental settings and implementation details for the Chinese experiments are provided in Appendix~\ref{subsec:implemetaiondetailss}.

\subsubsection*{\bf Evaluation Results.}
`Table~\ref{tab:cn_results} summarizes the attack results on the Chinese target systems.
Despite the language shift, Ours-NES achieves a 100\% ASR across both target systems under both EER- and minDCF-derived thresholds.
Moreover, successful attacks require only a small number of score queries on average, remaining within the same order of magnitude as observed in the English setting.
This indicates that the inverse-model-induced latent space remains well aligned with speaker-discriminative geometry even when trained on non-English voice data.

Ours-SP also remains effective in the Chinese setting, achieving high ASRs under the EER-based threshold and non-trivial success rates under the stricter minDCF-based threshold.
While performance naturally degrades as the decision threshold becomes more stringent, the attack remains practically viable.
Taken together, these results show that the effectiveness and query efficiency of inverse-model-based score exploitation are preserved across languages, reinforcing the generality of the proposed framework.

\section{Discussion}

We discuss representative defenses, an exploratory cross-lingual setting, and practical limitations regarding commercial API evaluation to contextualize our results. These analyses clarify the scope of our threat model and highlight directions for future work.

\subsection{Defense Methods}
\subsubsection*{\bf Deepfake Detection}
Since our attack produces synthetic speech via an inverse model rather than replaying genuine recordings, we briefly explored whether a deepfake detector~\cite{hamza2022deepfake} would flag such samples.
This experiment is intended as a lightweight, prototype-level probe rather than a systematic evaluation; detailed settings and results are provided in Appendix~\ref{sec:deepfake}.

Under this single tested configuration, we observed that audio generated by our inverse model was less frequently classified as spoofed than audio synthesized by the baseline YourTTS.
This observation should not be interpreted as a general limitation of deepfake detection, which remains one of the strongest candidate defenses against synthetic speech.
Rather, the result may reflect a distribution mismatch between the detector’s training data and the characteristics of audio produced by our inverse model.

In particular, our inverse model is optimized to satisfy feature-level identity constraints, which differs from the objectives of standard speech generation models.
As a result, the resulting audio may exhibit artifacts that are atypical from the perspective of existing detectors. 
From a defensive viewpoint, such samples could potentially be used as out-of-distribution or hard examples to improve detector robustness.
We leave a systematic investigation of this possibility to future work.

\subsubsection*{\bf Liveness Detection}

Another important line of defense is liveness detection, which aims to distinguish live human speech from replayed or synthesized inputs~\cite{7310809}, often via interactive or auxiliary signals beyond static audio.
We emphasize that our threat model does not incorporate such mechanisms: the proposed attack targets score-based speaker verification in non-interactive settings and does not attempt to bypass liveness-based defenses.
Accordingly, our should not be interpreted as evidence of effectiveness against liveness detection.
Nevertheless, liveness detection remains a strong and practically relevant defense, and understanding its interaction with feature-level attacks is an important direction for future work.

\subsection{Cross-Lingual Attack Evaluation}

We include a brief exploratory analysis to highlight an interesting cross-lingual scenario. In this setting, the local speaker recognition model, the inverse model, and the attack inputs are all based on English data, while only the target speaker verification system is trained on a different language (Chinese).
From the target system’s perspective, the resulting attack queries therefore constitute out-of-distribution inputs from the perspective of language.  (Detailed in Appendix~\ref{subsec:cross-lingual})

Despite this mismatch, we observe that impersonation attacks can still succeed under certain operating points.
Importantly, this observation reflects the cross-lingual properties of speaker representations learned by the target SRS, which our attack exploits, and should not be confused with the language-agnostic design of our overall framework, demonstrated in Section~\ref{subsec:chinese_sr}.
We emphasize that this experiment is intended to spark interest rather than provide a comprehensive evaluation, and we leave a systematic study of cross-lingual robustness to future work.

\subsection{Limitation on Commercial API Evaluation}

We acknowledge the importance of evaluating speaker impersonation attacks against commercial SRS APIs,
which are widely deployed in practice.
However, by early 2025, major providers had enforced strict Responsible AI policies that restrict
or prohibit academic testing of hosted speaker verification endpoints.
To comply with these policies and research ethics, we refrained from bypassing such restrictions
and instead pursued official authorization for academic evaluation, which was ultimately not granted.
Further details regarding our attempts and the corresponding policy constraints are provided in Appendix~\ref{subsec:commercial}.

Despite the lack of direct access to commercial APIs, we expect our findings to generalize to such systems.
Modern commercial SRSs are typically built upon deep learning--based architectures similar to those evaluated in this work, and our attack relies solely on score-based feedback without requiring gradient access.
Accordingly, even normalized confidence scores commonly returned by commercial APIs can be exploited by score-based optimization methods such as NES.
To approximate commercial-grade robustness, we therefore evaluate our attack on strong, publicly available models, including those released by Alibaba’s 3D-Speaker group ($C_1, C_2$) and NVIDIA ($T_5$).

\section{Conclusion}
In this work, we investigated score-based speaker impersonation attacks under a realistic black-box setting without access to victim voice samples. We show that the primary bottleneck of prior attacks lies not only in the high dimensionality of the audio space, but more fundamentally in the misalignment between the optimization space and speaker-discriminative embedding geometry.

Motivated by this insight, we reformulate score-based impersonation as a black-box optimization problem and identify feature-to-speech inversion as a critical missing component. We propose a feature-aligned inverse model trained via fixed-text fine-tuning with identity and structure constraints, explicitly aligning the latent space with speaker embeddings while preserving identity and cross-model compatibility.

Leveraging this inverse model, we enable two complementary attack paradigms: (i) a query-efficient NES-based attack in the inverse-induced latent space, and (ii) a non-adaptive subspace-projection attack that exploits score-level structural leakage and was previously infeasible. Extensive experiments show that both attacks achieve high success rates with orders-of-magnitude fewer queries than prior approaches and generalize across heterogeneous models and languages.

Overall, our results establish feature-aligned inversion as a key enabler for practical score-based speaker impersonation attacks and demonstrate that exposing similarity scores poses significant security risks, calling for careful system design and stronger defenses in deployed speaker recognition systems.

\newpage

\section*{Ethical Statement}

We attest that we have read the ethics-related discussions in the conference call for papers and the submission guidelines. The authors have considered the ethical implications of this work and believe that the research was conducted responsibly.
This work studies speaker impersonation attacks solely for research purposes, with the goal of evaluating the security of speaker recognition systems. We do not conduct attacks against real-world systems.
All experiments are performed using publicly available datasets and open-source models. No private or sensitive data is collected, and thus institutional review board (IRB) approval was not required. Details on dataset collection, processing, and access are provided.
To mitigate potential misuse, we release only a minimal implementation sufficient to reproduce our experimental results. The released code is intentionally limited and is not designed for direct use against real-world speaker recognition or authentication systems.

\section*{Open Sciences}

We provide all the implementation to reproduce our experimental results through the anonymized GitHub link: \url{https://anonymous.4open.science/r/Scores-Know-Bobs-Voice-5B14}. We release a minimal attack pipeline that is sufficient to validate the correctness of our experimental results. The provided implementation is intentionally scoped for research evaluation and is not designed for direct use against real-world speaker recognition or authentication systems. Our anonymized repository documents how to obtain the publicly available datasets and open-source speaker recognition, text-to-speech models used in our experiments. No proprietary data, trained attack models, or end-to-end attack systems are distributed.
The full source code will be made publicly available upon paper acceptance.


\bibliographystyle{ACM-Reference-Format}
\bibliography{bib}

@article{BAI202165,
title = {Speaker recognition based on deep learning: An overview},
journal = {Neural Networks},
volume = {140},
pages = {65-99},
year = {2021},
author = {Zhongxin Bai and Xiao-Lei Zhang},
}

@inproceedings{liu2024disentangling,
  title={Disentangling voice and content with self-supervision for speaker recognition},
  author={Liu, Tianchi and Lee, Kong Aik and Wang, Qiongqiong and Li, Haizhou},
  booktitle={Thirty-seventh Conference on Neural Information Processing Systems (NeurIPS)},
  volume={36},
  pages = {50221--50236},
  year={2023}
}

@inproceedings{yakovlev24_interspeech,
  title     = {Reshape Dimensions Network for Speaker Recognition},
  author    = {Ivan Yakovlev and Rostislav Makarov and Andrei Balykin and Pavel Malov and Anton Okhotnikov and Nikita Torgashov},
  year      = {2024},
  booktitle = {Interspeech 2024},
  pages     = {3235--3239},
  doi       = {10.21437/Interspeech.2024-2116},
  issn      = {2958-1796},
}

@inproceedings{kinnunen2019can,
  title={Can we use speaker recognition technology to attack itself? enhancing mimicry attacks using automatic target speaker selection},
  author={Kinnunen, Tomi and Hautam{\"a}ki, Rosa Gonz{\'a}lez and Vestman, Ville and Sahidullah, Md},
  booktitle={ICASSP 2019-2019 IEEE International Conference on Acoustics, Speech and Signal Processing (ICASSP)},
  pages={6146--6150},
  year={2019},
  organization={IEEE}
}

@article{yoon2020new,
  title={A new replay attack against automatic speaker verification systems},
  author={Yoon, Sung-Hyun and Koh, Min-Sung and Park, Jae-Han and Yu, Ha-Jin},
  journal={IEEE Access},
  volume={8},
  pages={36080--36088},
  year={2020},
  publisher={IEEE}
}

@article{jia2018transfer,
  title={Transfer learning from speaker verification to multispeaker text-to-speech synthesis},
  author={Jia, Ye and Zhang, Yu and Weiss, Ron and Wang, Quan and Shen, Jonathan and Ren, Fei and Nguyen, Patrick and Pang, Ruoming and Lopez Moreno, Ignacio and Wu, Yonghui and others},
  journal={Advances in neural information processing systems},
  volume={31},
  year={2018}
}

@inproceedings{pani2023voice,
  title={Voice Morphing: Two Identities in One Voice},
  author={Pani, Sushanta K and Chowdhury, Anurag and Sandler, Morgan and Ross, Arun},
  booktitle={2023 International Conference of the Biometrics Special Interest Group (BIOSIG)},
  pages={1--6},
  year={2023},
  organization={IEEE}
}

@inproceedings{zuo2024advtts,
  title={AdvTTS: Adversarial Text-to-Speech Synthesis Attack on Speaker Identification Systems},
  author={Zuo, Chu-Xiao and Jia, Zhi-Jun and Li, Wu-Jun},
  booktitle={ICASSP 2024-2024 IEEE International Conference on Acoustics, Speech and Signal Processing (ICASSP)},
  pages={4840--4844},
  year={2024},
  organization={IEEE}
}

@inproceedings{chen2021real,
  title={Who is real bob? adversarial attacks on speaker recognition systems},
  author={Chen, Guangke and Chenb, Sen and Fan, Lingling and Du, Xiaoning and Zhao, Zhe and Song, Fu and Liu, Yang},
  booktitle={2021 IEEE Symposium on Security and Privacy (SP)},
  pages={694--711},
  year={2021},
  organization={IEEE}
}

@inproceedings{du2020sirenattack,
  title={Sirenattack: Generating adversarial audio for end-to-end acoustic systems},
  author={Du, Tianyu and Ji, Shouling and Li, Jinfeng and Gu, Qinchen and Wang, Ting and Beyah, Raheem},
  booktitle={Proceedings of the 15th ACM Asia conference on computer and communications security},
  pages={357--369},
  year={2020}
}

@inproceedings{chen2023qfa2sr,
  title={$\{$QFA2SR$\}$:$\{$Query-Free$\}$ Adversarial Transfer Attacks to Speaker Recognition Systems},
  author={Chen, Guangke and Zhang, Yedi and Zhao, Zhe and Song, Fu},
  booktitle={32nd USENIX Security Symposium (USENIX Security 23)},
  pages={2437--2454},
  year={2023}
}

@inproceedings{zheng2021black,
  title={Black-box adversarial attacks on commercial speech platforms with minimal information},
  author={Zheng, Baolin and Jiang, Peipei and Wang, Qian and Li, Qi and Shen, Chao and Wang, Cong and Ge, Yunjie and Teng, Qingyang and Zhang, Shenyi},
  booktitle={Proceedings of the 2021 ACM SIGSAC conference on computer and communications security},
  pages={86--107},
  year={2021}
}

@article{kinnunen2017asvspoof,
  title={The ASVspoof 2017 challenge: Assessing the limits of replay spoofing attack detection},
  author={Kinnunen, Tomi and Sahidullah, Md and Delgado, H{\'e}ctor and Todisco, Massimiliano and Evans, Nicholas and Yamagishi, Junichi and Lee, Kong Aik},
  year={2017},
  publisher={ISCA (the International Speech Communication Association)}
}

@article{todisco2019asvspoof,
  title={ASVspoof 2019: Future horizons in spoofed and fake audio detection},
  author={Todisco, Massimiliano and Wang, Xin and Vestman, Ville and Sahidullah, Md and Delgado, H{\'e}ctor and Nautsch, Andreas and Yamagishi, Junichi and Evans, Nicholas and Kinnunen, Tomi and Lee, Kong Aik},
  journal={arXiv preprint arXiv:1904.05441},
  year={2019}
}

@article{yamagishi2021asvspoof,
  title={ASVspoof 2021: accelerating progress in spoofed and deepfake speech detection},
  author={Yamagishi, Junichi and Wang, Xin and Todisco, Massimiliano and Sahidullah, Md and Patino, Jose and Nautsch, Andreas and Liu, Xuechen and Lee, Kong Aik and Kinnunen, Tomi and Evans, Nicholas and others},
  journal={arXiv preprint arXiv:2109.00537},
  year={2021}
}

@article{wang2024asvspoof,
  title={ASVspoof 5: Crowdsourced speech data, deepfakes, and adversarial attacks at scale},
  author={Wang, Xin and Delgado, H{\'e}ctor and Tak, Hemlata and Jung, Jee-weon and Shim, Hye-jin and Todisco, Massimiliano and Kukanov, Ivan and Liu, Xuechen and Sahidullah, Md and Kinnunen, Tomi and others},
  journal={arXiv preprint arXiv:2408.08739},
  year={2024}
}

@inproceedings{mills2022replay,
  title={Replay attack detection based on voice and non-voice sections for speaker verification},
  author={Mills, Ananda Garin and Kaewcharuay, Patthranit and Sathirasattayanon, Pannathorn and Duangpummet, Suradej and Galajit, Kasorn and Karnjana, Jessada and Aimmanee, Pakinee},
  booktitle={2022 Asia-Pacific Signal and Information Processing Association Annual Summit and Conference (APSIPA ASC)},
  pages={221--226},
  year={2022},
  organization={IEEE}
}

@article{zhou2022voice,
  title={Voice spoofing countermeasure for voice replay attacks using deep learning},
  author={Zhou, Jincheng and Hai, Tao and Jawawi, Dayang NA and Wang, Dan and Ibeke, Ebuka and Biamba, Cresantus},
  journal={Journal of Cloud Computing},
  volume={11},
  number={1},
  pages={51},
  year={2022},
  publisher={Springer}
}

@inproceedings{pianese2022deepfake,
  title={Deepfake audio detection by speaker verification},
  author={Pianese, Alessandro and Cozzolino, Davide and Poggi, Giovanni and Verdoliva, Luisa},
  booktitle={2022 IEEE International Workshop on Information Forensics and Security (WIFS)},
  pages={1--6},
  year={2022},
  organization={IEEE}
}

@article{tak2022automatic,
  title={Automatic speaker verification spoofing and deepfake detection using wav2vec 2.0 and data augmentation},
  author={Tak, Hemlata and Todisco, Massimiliano and Wang, Xin and Jung, Jee-weon and Yamagishi, Junichi and Evans, Nicholas},
  journal={arXiv preprint arXiv:2202.12233},
  year={2022}
}

@article{hamza2022deepfake,
  title={Deepfake audio detection via MFCC features using machine learning},
  author={Hamza, Ameer and Javed, Abdul Rehman Rehman and Iqbal, Farkhund and Kryvinska, Natalia and Almadhor, Ahmad S and Jalil, Zunera and Borghol, Rouba},
  journal={IEEE Access},
  volume={10},
  pages={134018--134028},
  year={2022},
  publisher={IEEE}
}

@inproceedings{wu2022adversarial,
  title={Adversarial sample detection for speaker verification by neural vocoders},
  author={Wu, Haibin and Hsu, Po-Chun and Gao, Ji and Zhang, Shanshan and Huang, Shen and Kang, Jian and Wu, Zhiyong and Meng, Helen and Lee, Hung-Yi},
  booktitle={ICASSP 2022-2022 IEEE international conference on acoustics, speech and signal processing (ICASSP)},
  pages={236--240},
  year={2022},
  organization={IEEE}
}

@article{chen2023detection,
  title={On the detection of adaptive adversarial attacks in speaker verification systems},
  author={Chen, Zesheng},
  journal={IEEE Internet of Things Journal},
  volume={10},
  number={18},
  pages={16271--16283},
  year={2023},
  publisher={IEEE}
}

@article{joshi2022advest,
  title={Advest: Adversarial perturbation estimation to classify and detect adversarial attacks against speaker identification},
  author={Joshi, Sonal and Kataria, Saurabh and Villalba, Jes{\'u}s and Dehak, Najim},
  journal={arXiv preprint arXiv:2204.03848},
  year={2022}
}

@inproceedings{casanova2022yourtts,
  title={Yourtts: Towards zero-shot multi-speaker tts and zero-shot voice conversion for everyone},
  author={Casanova, Edresson and Weber, Julian and Shulby, Christopher D and Junior, Arnaldo Candido and G{\"o}lge, Eren and Ponti, Moacir A},
  booktitle={International conference on machine learning},
  pages={2709--2720},
  year={2022},
  organization={PMLR}
}

@inproceedings{lu2021voxstructor,
  title={Voxstructor: Voice reconstruction from voiceprint},
  author={Lu, Panpan and Li, Qi and Zhu, Hui and Sovernigo, Giuliano and Lin, Xiaodong},
  booktitle={Information Security: 24th International Conference, ISC 2021, Virtual Event, November 10--12, 2021, Proceedings 24},
  pages={374--397},
  year={2021},
  organization={Springer}
}

@inproceedings{kim2024scores,
  title={Scores Tell Everything about Bob: Non-adaptive Face Reconstruction on Face Recognition Systems},
  author={Kim, Sunpill and Tan, Yong Kiam and Jeong, Bora and Mondal, Soumik and Aung, Khin Mi Mi and Seo, Jae Hong},
  booktitle={2024 IEEE Symposium on Security and Privacy (SP)},
  pages={1684--1702},
  year={2024},
  organization={IEEE}
}

@INPROCEEDINGS{8461375,
  author={Snyder, David and Garcia-Romero, Daniel and Sell, Gregory and Povey, Daniel and Khudanpur, Sanjeev},
  booktitle={2018 IEEE International Conference on Acoustics, Speech and Signal Processing (ICASSP)}, 
  title={X-Vectors: Robust DNN Embeddings for Speaker Recognition}, 
  year={2018},
  volume={},
  number={},
  pages={5329-5333},
  doi={10.1109/ICASSP.2018.8461375}}

@inproceedings{desplanques20_interspeech,
  title     = {ECAPA-TDNN: Emphasized Channel Attention, Propagation and Aggregation in TDNN Based Speaker Verification},
  author    = {Brecht Desplanques and Jenthe Thienpondt and Kris Demuynck},
  year      = {2020},
  booktitle = {Interspeech 2020},
  pages     = {3830--3834},
  doi       = {10.21437/Interspeech.2020-2650},
  issn      = {2958-1796},
}

@article{mai2018reconstruction,
  title={On the reconstruction of face images from deep face templates},
  author={Mai, Guangcan and Cao, Kai and Yuen, Pong C and Jain, Anil K},
  journal={IEEE transactions on pattern analysis and machine intelligence},
  volume={41},
  number={5},
  pages={1188--1202},
  year={2018},
  publisher={IEEE}
}

@inproceedings{duong2020vec2face,
  title={Vec2face: Unveil human faces from their blackbox features in face recognition},
  author={Duong, Chi Nhan and Truong, Thanh-Dat and Luu, Khoa and Quach, Kha Gia and Bui, Hung and Roy, Kaushik},
  booktitle={Proceedings of the IEEE/CVF Conference on Computer Vision and Pattern Recognition},
  pages={6132--6141},
  year={2020}
}

@incollection{NEURIPS2019_9015,
title = {PyTorch: An Imperative Style, High-Performance Deep Learning Library},
author = {Paszke, Adam and Gross, Sam and Massa, Francisco and Lerer, Adam and Bradbury, James and Chanan, Gregory and Killeen, Trevor and Lin, Zeming and Gimelshein, Natalia and Antiga, Luca and Desmaison, Alban and Kopf, Andreas and Yang, Edward and DeVito, Zachary and Raison, Martin and Tejani, Alykhan and Chilamkurthy, Sasank and Steiner, Benoit and Fang, Lu and Bai, Junjie and Chintala, Soumith},
booktitle = {Advances in Neural Information Processing Systems 32},
pages = {8024--8035},
year = {2019},
publisher = {Curran Associates, Inc.},
url = {http://papers.neurips.cc/paper/9015-pytorch-an-imperative-style-high-performance-deep-learning-library.pdf}
}

@misc{ravanelli2021speechbraingeneralpurposespeechtoolkit,
      title={SpeechBrain: A General-Purpose Speech Toolkit}, 
      author={Mirco Ravanelli and Titouan Parcollet and Peter Plantinga and Aku Rouhe and Samuele Cornell and Loren Lugosch and Cem Subakan and Nauman Dawalatabad and Abdelwahab Heba and Jianyuan Zhong and Ju-Chieh Chou and Sung-Lin Yeh and Szu-Wei Fu and Chien-Feng Liao and Elena Rastorgueva and François Grondin and William Aris and Hwidong Na and Yan Gao and Renato De Mori and Yoshua Bengio},
      year={2021},
      eprint={2106.04624},
      archivePrefix={arXiv},
      primaryClass={eess.AS},
      url={https://arxiv.org/abs/2106.04624}, 
}

@software{CoquiTTS,
  title = {Coqui TTS},
  author = {Gölge Eren and The Coqui TTS Team},
  year = {2021},
  month = {01},
  version = {1.4},
  doi = {10.5281/zenodo.6334862},
  license = {MPL-2.0},
  url = {https://www.coqui.ai},
  repository = {https://github.com/coqui-ai/TTS}
}

@article{casanova2024xtts,
  title={Xtts: a massively multilingual zero-shot text-to-speech model},
  author={Casanova, Edresson and Davis, Kelly and G{\"o}lge, Eren and G{\"o}knar, G{\"o}rkem and Gulea, Iulian and Hart, Logan and Aljafari, Aya and Meyer, Joshua and Morais, Reuben and Olayemi, Samuel and others},
  journal={arXiv preprint arXiv:2406.04904},
  year={2024}
}

@article{ren2020fastspeech,
  title={Fastspeech 2: Fast and high-quality end-to-end text to speech},
  author={Ren, Yi and Hu, Chenxu and Tan, Xu and Qin, Tao and Zhao, Sheng and Zhao, Zhou and Liu, Tie-Yan},
  journal={arXiv preprint arXiv:2006.04558},
  year={2020}
}

@inproceedings{osman2022emo,
  title={Emo-tts: Parallel transformer-based text-to-speech model with emotional awareness},
  author={Osman, Mohamed},
  booktitle={2022 5th International Conference on Computing and Informatics (ICCI)},
  pages={169--174},
  year={2022},
  organization={IEEE}
}

@article{ju2024naturalspeech,
  title={Naturalspeech 3: Zero-shot speech synthesis with factorized codec and diffusion models},
  author={Ju, Zeqian and Wang, Yuancheng and Shen, Kai and Tan, Xu and Xin, Detai and Yang, Dongchao and Liu, Yanqing and Leng, Yichong and Song, Kaitao and Tang, Siliang and others},
  journal={arXiv preprint arXiv:2403.03100},
  year={2024}
}

@inproceedings{wan2018generalized,
  title={Generalized end-to-end loss for speaker verification},
  author={Wan, Li and Wang, Quan and Papir, Alan and Moreno, Ignacio Lopez},
  booktitle={2018 IEEE International Conference on Acoustics, Speech and Signal Processing (ICASSP)},
  pages={4879--4883},
  year={2018},
  organization={IEEE}
}

@article{jung2022pushing,
  title={Pushing the limits of raw waveform speaker recognition},
  author={Jung, Jee-weon and Kim, You Jin and Heo, Hee-Soo and Lee, Bong-Jin and Kwon, Youngki and Chung, Joon Son},
  journal={arXiv preprint arXiv:2203.08488},
  year={2022}
}

@article{heo2020clova,
  title={Clova baseline system for the voxceleb speaker recognition challenge 2020},
  author={Heo, Hee Soo and Lee, Bong-Jin and Huh, Jaesung and Chung, Joon Son},
  journal={arXiv preprint arXiv:2009.14153},
  year={2020}
}

@article{li2023styletts,
  title={Styletts 2: Towards human-level text-to-speech through style diffusion and adversarial training with large speech language models},
  author={Li, Yinghao Aaron and Han, Cong and Raghavan, Vinay and Mischler, Gavin and Mesgarani, Nima},
  journal={Advances in Neural Information Processing Systems},
  volume={36},
  pages={19594--19621},
  year={2023}
}

@article{ping2017deep,
  title={Deep voice 3: Scaling text-to-speech with convolutional sequence learning},
  author={Ping, Wei and Peng, Kainan and Gibiansky, Andrew and Arik, Sercan O and Kannan, Ajay and Narang, Sharan and Raiman, Jonathan and Miller, John},
  journal={arXiv preprint arXiv:1710.07654},
  year={2017}
}

@inproceedings{shen2018natural,
  title={Natural tts synthesis by conditioning wavenet on mel spectrogram predictions},
  author={Shen, Jonathan and Pang, Ruoming and Weiss, Ron J and Schuster, Mike and Jaitly, Navdeep and Yang, Zongheng and Chen, Zhifeng and Zhang, Yu and Wang, Yuxuan and Skerrv-Ryan, Rj and others},
  booktitle={2018 IEEE international conference on acoustics, speech and signal processing (ICASSP)},
  pages={4779--4783},
  year={2018},
  organization={IEEE}
}

@ARTICLE{10760244,
  author={Wang, Shuai and Chen, Zhengyang and Lee, Kong Aik and Qian, Yanmin and Li, Haizhou},
  journal={IEEE/ACM Transactions on Audio, Speech, and Language Processing}, 
  title={Overview of Speaker Modeling and Its Applications: From the Lens of Deep Speaker Representation Learning}, 
  year={2024},
  volume={32},
  number={},
  pages={4971-4998},
  doi={10.1109/TASLP.2024.3492793}}

@article{chung2018voxceleb2,
  title={Voxceleb2: Deep speaker recognition},
  author={Chung, Joon Son and Nagrani, Arsha and Zisserman, Andrew},
  journal={arXiv preprint arXiv:1806.05622},
  year={2018}
}

@misc{lin2024voxblink2100kspeakerrecognition,
      title={VoxBlink2: A 100K+ Speaker Recognition Corpus and the Open-Set Speaker-Identification Benchmark}, 
      author={Yuke Lin and Ming Cheng and Fulin Zhang and Yingying Gao and Shilei Zhang and Ming Li},
      year={2024},
      eprint={2407.11510},
      archivePrefix={arXiv},
      primaryClass={eess.AS},
      url={https://arxiv.org/abs/2407.11510}, 
}

@inproceedings{fan2020cn,
  title={Cn-celeb: a challenging chinese speaker recognition dataset},
  author={Fan, Yue and Kang, JW and Li, LT and Li, KC and Chen, HL and Cheng, ST and Zhang, PY and Zhou, ZY and Cai, YQ and Wang, Dong},
  booktitle={ICASSP 2020-2020 IEEE International Conference on Acoustics, Speech and Signal Processing (ICASSP)},
  pages={7604--7608},
  year={2020},
  organization={IEEE}
}

@inproceedings{kim2021conditional,
  title={Conditional variational autoencoder with adversarial learning for end-to-end text-to-speech},
  author={Kim, Jaehyeon and Kong, Jungil and Son, Juhee},
  booktitle={International Conference on Machine Learning},
  pages={5530--5540},
  year={2021},
  organization={PMLR}
}

@InProceedings{Nagrani17,
	author       = "Nagrani, A. and Chung, J.~S. and Zisserman, A.",
	title        = "VoxCeleb: a large-scale speaker identification dataset",
	booktitle    = "INTERSPEECH",
	year         = "2017",
}

@article{wijewardena2022fingerprint,
  title={Fingerprint template invertibility: Minutiae vs. deep templates},
  author={Wijewardena, Kanishka P and Grosz, Steven A and Cao, Kai and Jain, Anil K},
  journal={IEEE Transactions on Information Forensics and Security},
  volume={18},
  pages={744--757},
  year={2022},
  publisher={IEEE}
}

@inproceedings{kauba2020inverse,
  title={Inverse biometrics: Reconstructing grayscale finger vein images from binary features},
  author={Kauba, Christof and Kirchgasser, Simon and Mirjalili, Vahid and Uhl, Andreas and Ross, Arun},
  booktitle={2020 IEEE International Joint Conference on Biometrics (IJCB)},
  pages={1--10},
  year={2020},
  organization={IEEE}
}

@inproceedings{ahmad2020resist,
  title={Resist: Reconstruction of irises from templates},
  author={Ahmad, Sohaib and Fuller, Benjamin},
  booktitle={2020 IEEE International Joint Conference on Biometrics (IJCB)},
  pages={1--10},
  year={2020},
  organization={IEEE}
}

@article{yan2024toward,
  title={Toward comprehensive and effective palmprint reconstruction attack},
  author={Yan, Licheng and Wang, Fei and Leng, Lu and Teoh, Andrew Beng Jin},
  journal={Pattern Recognition},
  volume={155},
  pages={110655},
  year={2024},
  publisher={Elsevier}
}

@article{SpeakerGuard,
  author    = {Guangke Chen and
               Zhe Zhao and
               Fu Song and
               Sen Chen and
               Lingling Fan and
               Feng Wang and 
               Jiashui Wang},
  title     = {Towards Understanding and Mitigating Audio Adversarial Examples for Speaker Recognition},
  journal   = {IEEE Transactions on Dependable and Secure Computing},
  year      = {2022}
}

@article{oord2016wavenet,
  title={Wavenet: A generative model for raw audio},
  author={Oord, Aaron van den and Dieleman, Sander and Zen, Heiga and Simonyan, Karen and Vinyals, Oriol and Graves, Alex and Kalchbrenner, Nal and Senior, Andrew and Kavukcuoglu, Koray},
  journal={arXiv preprint arXiv:1609.03499},
  year={2016}
}

@inproceedings{liu2024towards,
  title={Towards quantifying and reducing language mismatch effects in cross-lingual speech anti-spoofing},
  author={Liu, Tianchi and Kukanov, Ivan and Pan, Zihan and Wang, Qiongqiong and Sailor, Hardik B and Lee, Kong Aik},
  booktitle={2024 IEEE Spoken Language Technology Workshop (SLT)},
  pages={1185--1192},
  year={2024},
  organization={IEEE}
}

@article{kong2020hifi,
  title={Hifi-gan: Generative adversarial networks for efficient and high fidelity speech synthesis},
  author={Kong, Jungil and Kim, Jaehyeon and Bae, Jaekyoung},
  journal={Advances in neural information processing systems},
  volume={33},
  pages={17022--17033},
  year={2020}
}

@inproceedings{wang2023wespeaker,
  title={Wespeaker: A research and production oriented speaker embedding learning toolkit},
  author={Wang, Hongji and Liang, Chengdong and Wang, Shuai and Chen, Zhengyang and Zhang, Binbin and Xiang, Xu and Deng, Yanlei and Qian, Yanmin},
  booktitle={ICASSP 2023-2023 IEEE International Conference on Acoustics, Speech and Signal Processing (ICASSP)},
  pages={1--5},
  year={2023},
  organization={IEEE}
}

@inproceedings{qin2022simple,
  title={Simple attention module based speaker verification with iterative noisy label detection},
  author={Qin, Xiaoyi and Li, Na and Weng, Chao and Su, Dan and Li, Ming},
  booktitle={ICASSP 2022-2022 IEEE International Conference on Acoustics, Speech and Signal Processing (ICASSP)},
  pages={6722--6726},
  year={2022},
  organization={IEEE}
}

@article{liu2025nes2net,
  title={Nes2Net: A Lightweight Nested Architecture for Foundation Model Driven Speech Anti-spoofing},
  author={Liu, Tianchi and Truong, Duc-Tuan and Das, Rohan Kumar and Lee, Kong Aik and Li, Haizhou},
  journal={arXiv preprint arXiv:2504.05657},
  year={2025}
}

@inproceedings{wang2024fraudwhistler,
  title={$\{$FraudWhistler$\}$: A Resilient, Robust and Plug-and-play Adversarial Example Detection Method for Speaker Recognition},
  author={Wang, Kun and Xu, Xiangyu and Lu, Li and Ba, Zhongjie and Lin, Feng and Ren, Kui},
  booktitle={33rd USENIX Security Symposium (USENIX Security 24)},
  pages={7303--7320},
  year={2024}
}

@inproceedings{cieri-etal-2004-fisher,
    title = "The Fisher Corpus: a Resource for the Next Generations of Speech-to-Text",
    author = "Cieri, Christopher  and
      Miller, David  and
      Walker, Kevin",
    editor = "Lino, Maria Teresa  and
      Xavier, Maria Francisca  and
      Ferreira, F{\'a}tima  and
      Costa, Rute  and
      Silva, Raquel",
    booktitle = "Proceedings of the Fourth International Conference on Language Resources and Evaluation ({LREC}{'}04)",
    month = may,
    year = "2004",
    address = "Lisbon, Portugal",
    publisher = "European Language Resources Association (ELRA)",
    url = "https://aclanthology.org/L04-1500/"
}

@INPROCEEDINGS{225858,
  author={Godfrey, J.J. and Holliman, E.C. and McDaniel, J.},
  booktitle={[Proceedings] ICASSP-92: 1992 IEEE International Conference on Acoustics, Speech, and Signal Processing}, 
  title={SWITCHBOARD: telephone speech corpus for research and development}, 
  year={1992},
  volume={1},
  number={},
  pages={517-520 vol.1},
  doi={10.1109/ICASSP.1992.225858}}

@misc{nvidia2023zero,
author = {NVIDIA},
title = {Overview of Zero-Shot Multi-Speaker TTS Systems: Top Q\&As},
howpublished = {Speech AI Summit},
year = {2023},
note = {https://developer.nvidia.com/blog/overview-of-zero-shot-multi-speaker-tts-systems-top-qas/}
}

@techreport{nist_sre08,
  title        = {The NIST Year 2008 Speaker Recognition Evaluation Plan},
  author       = {{National Institute of Standards and Technology}},
  year         = {2008},
  institution  = {National Institute of Standards and Technology},
  address      = {Gaithersburg, MD, USA},
  note         = {Available: \url{https://www.nist.gov/document/sre08evalplanrelease4pdf}}
}

@inproceedings{koluguri2022titanet,
  title={Titanet: Neural model for speaker representation with 1d depth-wise separable convolutions and global context},
  author={Koluguri, Nithin Rao and Park, Taejin and Ginsburg, Boris},
  booktitle={ICASSP 2022-2022 IEEE international conference on acoustics, speech and signal processing (ICASSP)},
  pages={8102--8106},
  year={2022},
  organization={IEEE}
}

@online{azure_speaker,
  author = {{Microsoft}},
  title = {Speaker Recognition API - Cognitive Services},
  year = {2024},
  url = {https://learn.microsoft.com/en-us/azure/ai-services/speech-service/speaker-recognition-overview},
  note = {Explicitly returns confidence scores to support adjustable security thresholds.}
}

@online{aws_transcribe,
  author = {{Amazon Web Services}},
  title = {Amazon Transcribe - Speaker Identification},
  year = {2024},
  url = {https://docs.aws.amazon.com/transcribe/latest/dg/speaker-identification.html},
  note = {Provides confidence scores for identified speaker segments.}
}

@techreport{owasp_api,
  author = {{OWASP Foundation}},
  title = {OWASP API Security Top 10: API3:2019 Excessive Data Exposure},
  year = {2019},
  institution = {OWASP},
  url = {https://owasp.org/www-project-api-security/},
  note = { highlights that API endpoints often expose full data objects relying on client-side filtering.}
}

@book{owasp_mstg,
  author = {{OWASP Mobile Security Project}},
  title = {OWASP Mobile Security Testing Guide (MSTG)},
  year = {2023},
  publisher = {OWASP Foundation},
  note = {Describes testing procedures for local authentication using function hooking and dynamic binary instrumentation.}
}

@article{voigt2017eu,
  title={The eu general data protection regulation (gdpr)},
  author={Voigt, Paul and Von dem Bussche, Axel},
  journal={A practical guide, 1st ed., Cham: Springer International Publishing},
  volume={10},
  number={3152676},
  pages={10--5555},
  year={2017},
  publisher={Springer}
}

@misc{iso2022iso24745,
  title={{ISO/IEC} 24745:2022. Information security, cybersecurity and privacy protection--Biometric information protection},
  author={ISO},
  year={2022},
  publisher={International Organization for Standardization Geneva, Switzerland}
}

@inproceedings{chen2017zoo,
  title={Zoo: Zeroth order optimization based black-box attacks to deep neural networks without training substitute models},
  author={Chen, Pin-Yu and Zhang, Huan and Sharma, Yash and Yi, Jinfeng and Hsieh, Cho-Jui},
  booktitle={Proceedings of the 10th ACM workshop on artificial intelligence and security},
  pages={15--26},
  year={2017}
}

@article{alzantot2018did,
  title={Did you hear that? adversarial examples against automatic speech recognition},
  author={Alzantot, Moustafa and Balaji, Bharathan and Srivastava, Mani},
  journal={arXiv preprint arXiv:1801.00554},
  year={2018}
}

@inproceedings{guo2019simple,
  title={Simple black-box adversarial attacks},
  author={Guo, Chuan and Gardner, Jacob and You, Yurong and Wilson, Andrew Gordon and Weinberger, Kilian},
  booktitle={International conference on machine learning},
  pages={2484--2493},
  year={2019},
  organization={PMLR}
}

@article{cam++,
  title={CAM++: A Fast and Efficient Network for Speaker Verification Using Context-Aware Masking},
  author={Hui Wang and Siqi Zheng and Yafeng Chen and Luyao Cheng and Qian Chen},
  booktitle={Interspeech 2023},
  year={2023},
  organization={IEEE}
}

@article{eres2net,
  title={An Enhanced Res2Net with Local and Global Feature Fusion for Speaker Verification},
  author={Yafeng Chen and Siqi Zheng and Hui Wang and Luyao Cheng and Qian Chen and Jiajun Qi},
  booktitle={Interspeech 2023},
  year={2023},
  organization={IEEE}
}

@inproceedings{razzhigaev2020black,
  title={Black-box face recovery from identity features},
  author={Razzhigaev, Anton and Kireev, Klim and Kaziakhmedov, Edgar and Tursynbek, Nurislam and Petiushko, Aleksandr},
  booktitle={European Conference on Computer Vision},
  pages={462--475},
  year={2020},
  organization={Springer}
}

@inproceedings{vendrow2021realistic,
  title={Realistic face reconstruction from deep embeddings},
  author={Vendrow, Edward and Vendrow, Joshua},
  booktitle={NeurIPS 2021 Workshop Privacy in Machine Learning},
  year={2021}
}

@INPROCEEDINGS{8698539,
  author={Bontrager, Philip and Roy, Aditi and Togelius, Julian and Memon, Nasir and Ross, Arun},
  booktitle={2018 IEEE 9th International Conference on Biometrics Theory, Applications and Systems (BTAS)}, 
  title={DeepMasterPrints: Generating MasterPrints for Dictionary Attacks via Latent Variable Evolution}, 
  year={2018},
  volume={},
  number={},
  pages={1-9},
  doi={10.1109/BTAS.2018.8698539}}

@inproceedings{nguyen2023analysis,
  title={Analysis of master vein attacks on finger vein recognition systems},
  author={Nguyen, Huy H and Le, Trung-Nghia and Yamagishi, Junichi and Echizen, Isao},
  booktitle={Proceedings of the IEEE/CVF Winter Conference on Applications of Computer Vision},
  pages={1900--1908},
  year={2023}
}

@article{chen20243d,
  title={3D-Speaker-Toolkit: An Open Source Toolkit for Multi-modal Speaker Verification and Diarization},
  author={Chen, Yafeng and Zheng, Siqi and Wang, Hui and Cheng, Luyao and others},
  booktitle={ICASSP},
  year={2025}
}

@inproceedings{ilyas2018black,
  title={Black-box adversarial attacks with limited queries and information},
  author={Ilyas, Andrew and Engstrom, Logan and Athalye, Anish and Lin, Jessy},
  booktitle={International conference on machine learning},
  pages={2137--2146},
  year={2018},
  organization={PMLR}
}

@article{wierstra2014natural,
  title={Natural evolution strategies},
  author={Wierstra, Daan and Schaul, Tom and Glasmachers, Tobias and Sun, Yi and Peters, Jan and Schmidhuber, J{\"u}rgen},
  journal={The Journal of Machine Learning Research},
  volume={15},
  number={1},
  pages={949--980},
  year={2014},
  publisher={JMLR. org}
}

@inproceedings{chung2020defence,
  title={In defence of metric learning for speaker recognition},
  author={Chung, Joon Son and Huh, Jaesung and Mun, Seongkyu and Lee, Minjae and Heo, Hee Soo and Choe, Soyeon and Ham, Chiheon and Jung, Sunghwan and Lee, Bong-Jin and Han, Icksang},
  booktitle={Proc. Interspeech},
  year={2020},
  pages={2977--2981}
}

@ARTICLE{7310809,
  author={Akhtar, Zahid and Micheloni, Christian and Foresti, Gian Luca},
  journal={IEEE Security \& Privacy}, 
  title={Biometric Liveness Detection: Challenges and Research Opportunities}, 
  year={2015},
  volume={13},
  number={5},
  pages={63-72},
  doi={10.1109/MSP.2015.116}}
\clearpage
\appendix
\section{Additional Related Works}
\label{sec:additional_related}
Various threats and attack scenarios have been proposed for speaker recognition systems (SRSs) ranging from compromising their training data to physical attacks against deployed systems.
Given the broad range of threats, we will primarily focus our related work discussion on the \emph{impersonation attack} setting, where an attacker presents an attack sample in order to impersonate an identity enrolled in the targeted SRS.
A summary of prior research on speaker impersonation is shown in Table~\ref{tab:rlwork_full}; as can be seen from the table, there are numerous impersonation vulnerabilities in SRSs through attacks with diverse assumptions and methodologies.

\paragraph{Known-Victim and Physical Attacks.}
One prominent line of research operates under the assumption that the attacker has direct access to the victim's voice samples or related information. This category includes attacks involving human voice mimicry~\cite{kinnunen2019can}, replay attacks using recorded speech~\cite{yoon2020new}, or voice synthesis methods that generate speech recognized as identical to two identities simultaneously~\cite{pani2023voice}. While these methods share the functionality of attempting impersonation based on prior information, they differ in their specific objectives and execution, ranging from direct human performance and simple playback to the computational generation of ambiguous, multi-identity audio.

\paragraph{Generative and Transfer Attacks.}
Other studies have attempted to generate victim voices by leveraging knowledge about the targeted system or through more complex interactions. For instance, the voice synthesis approach of~\citet{jia2018transfer} operates in a white-box setting, utilizing internal model information to generate speech recognized as a specific identity.
In contrast, Voxstructor~\cite{lu2021voxstructor} targets black-box models but requires performing an extensive number ($\geq100K$) of non-adaptive queries to reconstruct the victim's feature vector, which is then used to train a separate model for generating the target voice.
Notably, Voxstructor requires \emph{feature vector} queries from the black-box SRS, which is challenging to apply against deployed systems because such vectors are usually carefully stored as private user biometric data.
Transfer-based attack strategies have also been explored, which utilize victim information during preparation but do not directly query the target model during the attack phase. AdvTTS~\cite{zuo2024advtts}, for example, trained a model to generate adversarial noise that causes a source audio sample to be recognized as a target audio sample.
QFA2SR~\cite{chen2023qfa2sr} enhanced their transferability to black-box environments using ensemble techniques involving multiple surrogate models.

\begin{table*}[t]
\caption{Comparison of prior attack methods enabling speaker impersonation against SRSs, summarizing their methodologies, requirements (victim voice, target model access), and query strategies. The details of FakeBob are given for its \emph{speaker verification} setting, which we denote as FakeBob$^\dagger$ in this table.} 
\resizebox{.9\linewidth}{!}{
\begin{tabular}{c|c|c|c|c|cc}
\hline
\multirow{2}{*}{Method} & \multirow{2}{*}{Impersonation Type} & \multirow{2}{*}{Target Model} & \multirow{2}{*}{\begin{tabular}[c]{@{}c@{}}Victim's Voice\\Requirement\end{tabular}} & \multirow{2}{*}{Strategy} & \multicolumn{2}{c}{Query to Target Model} \\ \cline{6-7}
 &  &  &  &  & \multicolumn{1}{c|}{Output Format} & Number \\ \hline \hline
 \cite{kinnunen2019can}& Voice Mimicry & Black-box & $\bigcirc$ & Human Mimicry & \multicolumn{1}{c|}{-} & 0 \\ \hline
 \cite{yoon2020new}& Replay Attack & Black-box & $\bigcirc$ & Playback & \multicolumn{1}{c|}{-} & 0 \\ \hline
 \cite{pani2023voice}& Voice Morphing & Black-box & $\bigcirc$ & Identity Morphing & \multicolumn{1}{c|}{-} & 0 \\ \hline
SV2TTS~\cite{jia2018transfer} & Speech Synthesis & White-box & $\bigcirc$ & Training-based & \multicolumn{1}{c|}{-} & 0 \\ \cline{1-7}
Voxstructor~\cite{lu2021voxstructor} & Speech Synthesis & Black-box & $\bigcirc$ & Training-based & \multicolumn{1}{c|}{Feature vector} & $\geq100K$\\ \cline{1-7}
AdvTTS~\cite{zuo2024advtts} & Adversarial Example & Black-box & $\bigcirc$ & Transfer-based & \multicolumn{1}{c|}{-} & 0 \\ \cline{1-7}
QFA2SR~\cite{chen2023qfa2sr} & Adversarial Example & Black-box & $\bigcirc$ & Transfer-based & \multicolumn{1}{c|}{-} & 0 \\ \cline{1-7}
SirenAttack~\cite{du2020sirenattack} & Adversarial Example & Black-box & $\bigcirc$ & Adaptive Query & \multicolumn{1}{c|}{Score} & $\geq7500$\\ \cline{1-7}
Occam~\cite{zheng2021black} & Adversarial Example & Black-box & $\bigcirc$ & Adaptive Query & \multicolumn{1}{c|}{Decision} & $\geq10K$\\ \hline \hline
FakeBob$^\dagger$~\cite{chen2021real} & Adversarial Example & Black-box & $\bigtimes$ & Adaptive Query & \multicolumn{1}{c|}{Score} & $\geq10K$ \\ \hline
\textbf{Ours-NES} & \textbf{Speech Synthesis} & \textbf{Black-box} & \textbf{$\bigtimes$} & \textbf{Adaptive Query} & \multicolumn{1}{c|}{\textbf{Score}} & $\approx$ \textbf{500} \\ \hline
\textbf{Ours-SP} & \textbf{Speech Synthesis} & \textbf{Black-box} & \textbf{$\bigtimes$} & \textbf{Non-Adaptive Query} & \multicolumn{1}{c|}{\textbf{Score}} & \textbf{50} \\ \hline
\end{tabular}
}
\label{tab:rlwork_full}
\end{table*}

\paragraph{Black-Box Attacks.}
Black-box attacks involve making queries to the target model SRS, but often require some pre-existing victim voice information and predominantly rely on adaptive query strategies. Representative examples include SirenAttack~\cite{du2020sirenattack}, which generates adversarial samples based on score feedback through approximately 7,500 adaptive queries; and Occam~\cite{zheng2021black}, which optimizes attacks based on decision feedback using around $10K$ adaptive queries.
These methods employ gradual attack refinement based on query feedback and necessitate a substantial number of queries.
The closest work to our paper's setting is FakeBob~\cite{chen2021real}, which performs black-box impersonation attacks via score queries without prior victim voice information.
Their work demonstrated the feasibility of score-based attacks with no \emph{a priori} information of the victim.
However, it relies on an adaptive query strategy and requires a relatively large number of queries ($\geq 10K$) for successful impersonation.
We provide a comprehensive comparison using FakeBob against state-of-the-art SRS models in Section~\ref{sec:exp}.

\paragraph{Countermeasures.}
Alongside the emerging impersonation attacks, several countermeasures have been proposed to mitigate them, including the ASVspoof challenges~\cite{kinnunen2017asvspoof,todisco2019asvspoof,yamagishi2021asvspoof,wang2024asvspoof}.
A widely-used defense strategy against impersonation attacks is to deploy attack detectors, e.g., replay attack detection \cite{zhou2022voice,mills2022replay}, deepfake detection \cite{hamza2022deepfake,tak2022automatic,pianese2022deepfake,liu2025nes2net}, and adversarial attack detection \cite{joshi2022advest,chen2023detection,wu2022adversarial, wang2024fraudwhistler}. In particular, \cite{chen2023detection} is noteworthy as a detection technique specialized in adaptive query-based attacks, such as Fakebob~\cite{chen2021real}.

\paragraph{Non-Adaptive Attack on Face Recognition}
A non-adaptive attack that can independently and simultaneously generate all attack queries without any prior information was recently proposed in the domain of face recognition by~\citet{kim2024scores}.
A key observation of their work is that the similarity scores which are output by two well-trained face recognition models $A$ and $B$ should be similar.
Thus, if one can obtain queried scores from a target model $B$, then reconstruction can be carried out by transfering those scores to a local model $A$ and using a so-called inverse model $A^{-1}$ that reverses the image-to-feature mapping of $A$.
They demonstrate successful attacks against commercial and open-source targets using only 100 similarity scores obtained via non-adaptive queries. 
Our work investigates, for the first time, how ideas from this prior attack can be deployed against speaker recognition systems.
Notably, the adaptation is not straightforward---unlike face recognition, there is no effective technique to obtain a neural network which acts as the inverse function of an arbitrary speaker recognition model.
We remark that if an attacker can make an unlimited number of queries, there have been several approaches~\cite{mai2018reconstruction,duong2020vec2face} to train an inverse neural network by obtaining pseudo-labels, e.g., feature vectors, for all the facial image data that the attacker has. However, these approaches require an enormous number ($\geq 1M$) of feature vector queries, which we consider to be impractical.

\section{Additional Description of Inverse Models on Speaker Recognition}
\label{appsec:existingtts}
In this section, we provide (1) the candidate selection for inverse models used for comparison in the main paper, (2) implementation details of the selected candidates, (3) technical challenges of building custom inverse models, and (4) alternatives for inverse models.

\subsection{Use of Existing TTS Models as Inverse Model Candidates}\
\label{subsec:use_of_exist_TTS}
In designing our attack methodology, the selection of an effective inverse model (responsible for generating audio from a speaker embedding and text) is crucial.
While our main experiments in Table~\ref{tab:attack_comparison} compared the proposed method against selected zero-shot TTS (ZS-TTS) baselines, our initial investigation considered a wider array of contemporary Text-to-Speech (TTS) models as potential candidates for this role.
A key requirement for a TTS model to function as a suitable inverse model within our specific attack framework in Figure~\ref{alg:Pipeline} is its ability to generate audio conditioned solely on a target speaker embedding and a predefined text, without requiring additional inputs like a reference audio sample.

We evaluated several prominent TTS architectures based on this criterion. However, many state-of-the-art TTS systems, while powerful for synthesis, do not meet this specific requirement, rendering them unsuitable for direct use as the inverse model in our attack pipeline. Table~\ref{tab:otherTTS} provides details on some of these considered but ultimately unselected TTS candidates, summarizing their input requirements and inherent limitations for our purpose.

\begin{table*}[]
\caption{Other considered TTS models for Inverse Model Candidates.}
\resizebox{.9\linewidth}{!}{
\begin{tabular}{c|c|c|c|c}
\hline
\textbf{TTS} & \textbf{Input} & \textbf{(Speaker) Encoder} & \begin{tabular}[c]{@{}c@{}}\textbf{Auxiliary Input}\end{tabular} &\textbf{Output} \\ \hline \hline
Tacotron2~\cite{shen2018natural} & Text & None & None & \begin{tabular}[c]{@{}c@{}}\ \makecell{Mel-spectrogram \\ Waveform}\end{tabular}
\\ \hline
FastSpeech2~\cite{ren2020fastspeech} & Text & None & None & \begin{tabular}[c]{@{}c@{}}\makecell{Mel-spectrogram \\ Waveform}\end{tabular} \\ \hline
XTTS~\cite{casanova2024xtts} & \begin{tabular}[c]{@{}c@{}}Text, Voice Sample\end{tabular} & \begin{tabular}[c]{@{}c@{}}H/ASP~\cite{heo2020clova},\\ Perceiver Conditioner\end{tabular}   & \begin{tabular}[c]{@{}c@{}}Speaker Feature Vector\\ Identity, Style, Prosody, etc.\end{tabular} & Waveform  \\ \hline
EmoTTS~\cite{osman2022emo}  & \begin{tabular}[c]{@{}c@{}}Text, Voice Sample\end{tabular} & \begin{tabular}[c]{@{}c@{}}SER model \\ (based on Deep Voice3~\cite{ping2017deep})\end{tabular} & Emotion, Speaker Embedding & Mel-spectrogram        \\ \hline
NaturalSpeech3~\cite{ju2024naturalspeech} & \begin{tabular}[c]{@{}c@{}}Text, Voice Sample\end{tabular} & Convolutional Blocks & \begin{tabular}[c]{@{}c@{}} Content, Prosody, \\ Timbre, Acoustic Details\end{tabular} & Waveform      \\ \hline
StyleTTS2~\cite{li2023styletts} & \begin{tabular}[c]{@{}c@{}}Text, Voice Sample\end{tabular} &  \begin{tabular}[c]{@{}c@{}}Acoustic and Prosodic,\\ Style Encoders\end{tabular} & Style Embedding & Waveform                 \\ \hline
\end{tabular}}
\label{tab:otherTTS}
\end{table*}

Models like Tacotron2~\cite{shen2018natural} and FastSpeech2~\cite{ren2020fastspeech} are not designed to accept speaker embeddings or feature vectors as conditioning input for speaker identity control. Another group of models, including advanced ZS-TTS systems like XTTS~\cite{casanova2024xtts}, EmoTTS~\cite{osman2022emo}, NaturalSpeech3~\cite{ju2024naturalspeech}, and StyleTTS2~\cite{li2023styletts}, necessitate a reference voice sample during inference. This sample is used by their respective encoders to extract auxiliary acoustic information (such as style, prosody, timbre details, etc.) beyond just the core speaker identity represented by a standard embedding. Since our attack scenario presupposes access only to the target speaker embedding (not a full voice sample), these models cannot generate the required audio output based solely on the embedding and text, thus disqualifying them as inverse model candidates for our study.
The above consideration justifies our focus on the models evaluated in our main paper's comparison and the development of our proposed inversion approach.

\subsection{Additional Details of Selected ZS-TTS Inverse Model Candidates}
\label{subsec:detail_zs-tts-imp}
 This section details the specific implementations and configurations used for the chosen inverse model candidates evaluated in our main comparisons: SV2TTS, Voxstructor, and YourTTS. These models were configured to accept a speaker embedding and text as input to generate audio, simulating their role as potential inversion mechanisms.

\paragraph{SV2TTS~\cite{jia2018transfer}.} This model adapts the Tacotron2~\cite{shen2018natural} architecture for zero-shot voice conversion. The primary modification relevant to its use as an inverse model is that its encoder takes the concatenation of text embeddings and a speaker embedding as input. It generates a mel-spectrogram, which is then converted to a waveform using a pre-trained WaveNet vocoder \cite{oord2016wavenet}. For our experiments, we utilized the official implementation provided by the authors.

\paragraph{Voxstructor~\cite{lu2021voxstructor}.} This model was proposed as a method to adapt SV2TTS for use with arbitrary black-box speaker encoders, through the inroduction of a mapping layer which transforms the embedding from a target speaker encoder into the GE2E~\cite{wan2018generalized} embedding space used internally by SV2TTS.
Since no official implementation was readily available that matched our specific target encoder choice, we implemented and trained the Voxstructor mapping component ourselves. Specifically, we designated RawNet3~\cite{jung2022pushing} as the representative black-box target encoder. The mapping layer was constructed using three fully connected layers, each followed by Batch Normalization and a ReLU activation function. We trained this mapping layer for 80 epochs on the VoxCeleb1 trainset, using the Smooth L1 loss between the mapped RawNet3 embeddings and the corresponding ground-truth GE2E embeddings. The Adam optimizer was used with a learning rate of 0.001. The rest of the SV2TTS architecture (decoder and vocoder) remained the same as the original SV2TTS implementation.

\paragraph{YourTTS~\cite{casanova2022yourtts}.} This model is based on the VITS~\cite{kim2021conditional} architecture, incorporating modifications for improved zero-shot TTS. It employs a Transformer-based text encoder, takes a speaker embedding as input for conditioning, utilizes a stochastic duration predictor for alignment, and generates the final waveform using a flow-based decoder followed by a HiFi-GAN vocoder~\cite{kong2020hifi}. For our evaluation, we employed the implementation available in Coqui.

\paragraph{Statistics for ZS-TTS ID-Constraint Evaluation} 
\label{app:zs-tts-id-detail}
This section provides the detailed statistical data (mean $\pm$ standard deviation) corresponding to the score histograms presented in Figure~\ref{fig:id-consL}, which evaluate the suitability of standard Zero-Shot Text-to-Speech (ZS-TTS) models as direct inverse models. Table~\ref{tab:id-cons-stat} lists these statistics for Positive, Negative, and ID-constraints scores across representative ZS-TTS models (SV2TTS, Voxstructor, YourTTS) on the VoxCeleb1 testset.

The key ID-constraints scores for the ZS-TTS models quantitatively support the visual observation from Fig.~\ref{fig:id-consL}. These values highlight the significant gap compared to ideal identity preservation (represented by Positive scores) and confirm the limitations of using these off-the-shelf ZS-TTS models directly for high-fidelity speaker embedding inversion as required in our attack context.

\begin{table}
\caption{ID-constraint test statistics for selected ZS-TTS on VoxCeleb1 test dataset.}
\resizebox{\linewidth}{!}{

\begin{tabular}{c||c|c|c}
 \hline
Scores & SV2TTS & Voxstructor & YourTTS \\ \hline \hline
Pos. & $0.7795\pm0.0771$ & $0.6470\pm0.1212$ & $0.6549\pm0.1007$ \\
Neg. & $0.5766\pm0.0707$ & $0.0267\pm0.1119$ & $0.1383\pm0.1095$ \\
ID-const. & $0.6944\pm0.0761$ & $0.0472\pm0.0879$ & $0.4015\pm0.0923$\\ \hline
\end{tabular}
}
\label{tab:id-cons-stat}
\end{table}

\section{Additional Experimental Details on ID-Constraints and Transferability}
\label{app:id_constraints}

This appendix provides implementation details and complete experimental results for the \emph{ID-Constraints Test} and its \emph{transferability variant}, which are only briefly described in the main paper.
All quantitative results omitted from the main text are fully reported in Figure~\ref{fig:entire_experiment}.

\subsection{ID-Constraints Test}
\label{app:id_constraints_basic}

The ID-Constraints Test evaluates whether the inverse model $F^{-1}$ preserves speaker identity when performing a round-trip mapping from audio to speaker embedding and back to audio.
Given a speaker recognition (SR) feature extractor
$F: \mathcal{A} \rightarrow \mathbb{S}^{d-1}$ and its corresponding inverse model
$F^{-1}: \mathbb{S}^{d-1} \times \mathcal{T} \rightarrow \mathcal{A}$.

Concretely, for an input utterance $\mathsf{aud}$, we first extract its speaker embedding
$x = F(\mathsf{aud})$.
We then synthesize a new audio sample using the inverse model conditioned on a predefined text prompt $\mathsf{Text}$, and re-extract its embedding via the same SR model.
Identity preservation is quantified by the cosine similarity
$s = \langle x, x' \rangle$ between the original and reconstructed embeddings.

\paragraph{Experimental setup.}
The test is conducted on the VoxCeleb1 test set, consisting of 4,708 utterances from 40 distinct identities.
For all zero-shot TTS (ZS-TTS) models, the text prompt is fixed to:
\textit{``Hello Google, Hi Bixby, Hey Siri''}.
This ensures that variations in linguistic content do not confound identity evaluation.

\paragraph{Reference distributions.}
To contextualize the ID-Constraints scores, we additionally compute:
(i) \emph{positive scores}, obtained from pairs of utterances belonging to the same identity, and
(ii) \emph{negative scores}, obtained from utterances of different identities.
These distributions serve as upper and lower reference bounds, respectively.

\subsection{Transfer ID-Constraints Test}
\label{app:id_constraints_transfer}

While the basic ID-Constraints Test evaluates identity preservation under the same SR model used to extract the conditioning embedding, it does not capture cross-model generalization.

In this setting, the inverse model $F^{-1}$ is instantiated with a local SR model $F$, while identity preservation is evaluated using a distinct test-time SR extractor $F_{\text{test}}: \mathcal{A} \rightarrow \mathbb{S}^{d-1}$.
Importantly, $F_{\text{test}}$ may differ architecturally and parametrically from $F$, reflecting a more realistic scenario in which the attacker does not know the target system.

Given an utterance $\mathsf{aud}$, we compute $x = F_{\text{test}}(\mathsf{aud})$ and $\widetilde{x} = F_{\text{test}}(F^{-1}(F(\mathsf{aud}), \mathsf{Text}))$, and report their cosine similarity.
A high similarity score indicates that the identity information injected by $F^{-1}$ generalizes beyond the local SR model used during inversion.

\subsection{Complete Results}
\label{app:id_constraints_results}

\begin{figure*}[t]
    \centering
    \begin{subfigure}[b]{.9\textwidth}
        \centering
        \includegraphics[width=\linewidth]{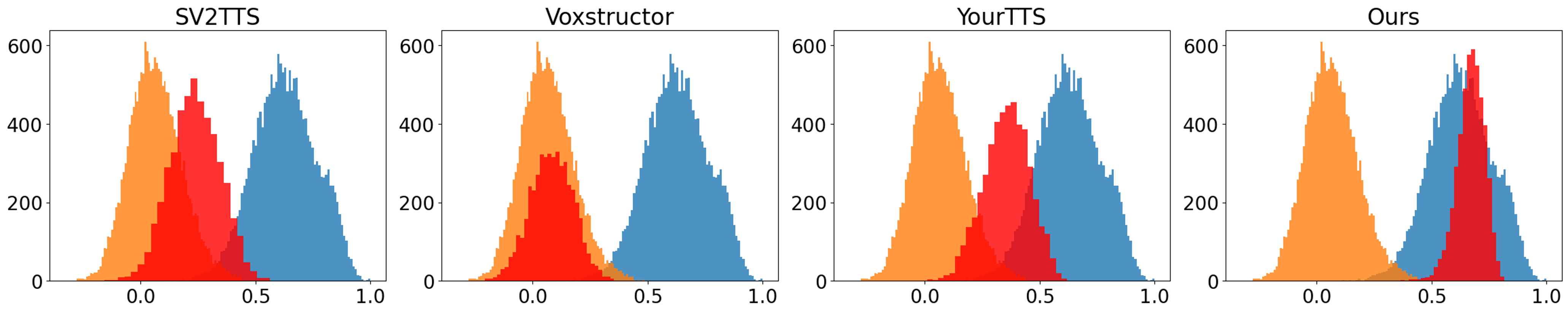}
        \caption{$F_T = T_1$} 
        \label{fig:row1}
    \end{subfigure}   
    \par\bigskip 
    \begin{subfigure}[b]{.9\textwidth}
        \centering
        \includegraphics[width=\linewidth]{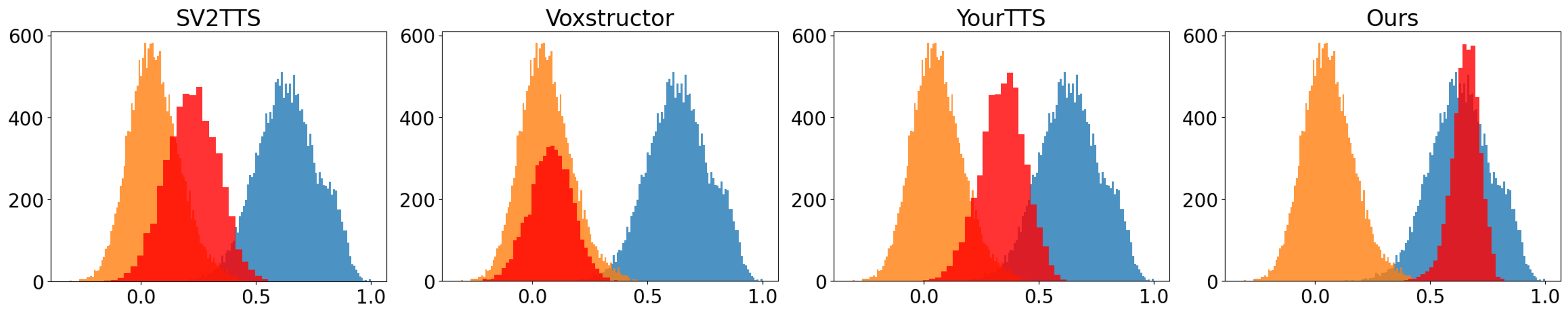}
        \caption{$F_T = T_2$} 
        \label{fig:row2}
    \end{subfigure}
    \par\bigskip 
    \begin{subfigure}[b]{.9\textwidth}
        \centering
        \includegraphics[width=\linewidth]{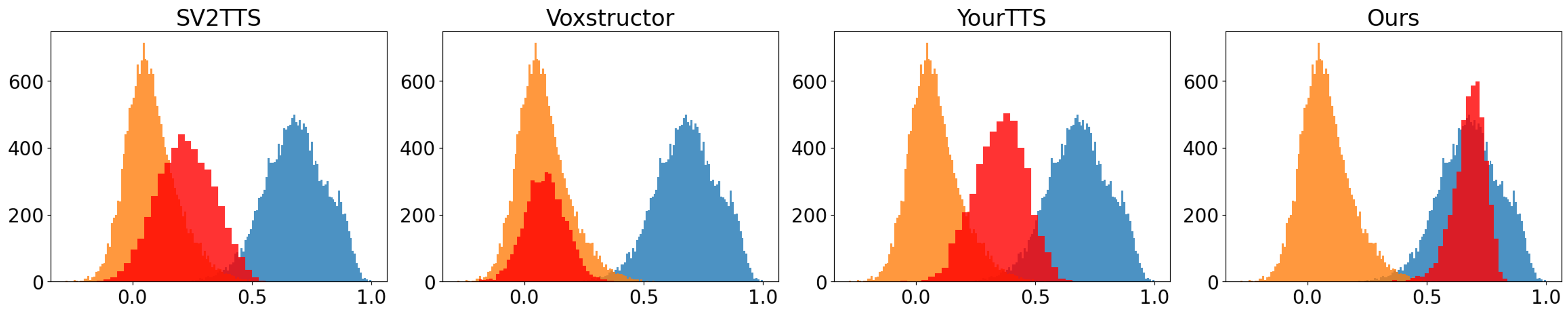}
        \caption{$F_T = T_3$} 
        \label{fig:row3}
    \end{subfigure}
        \begin{subfigure}[b]{.9\textwidth}
        \centering
        \includegraphics[width=\linewidth]{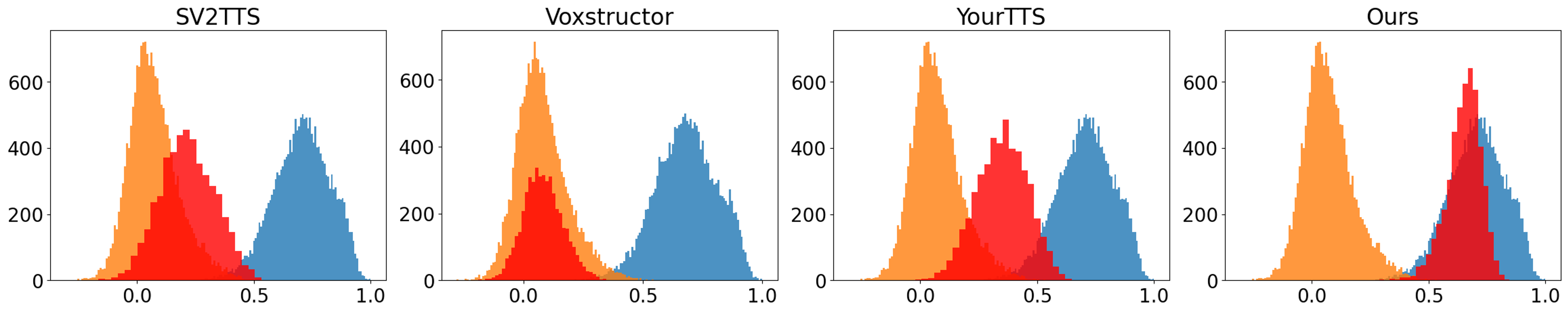}
        \caption{$F_T = T_4$} 
        \label{fig:row1}
    \end{subfigure}   
    \par\bigskip 
    \begin{subfigure}[b]{.9\textwidth}
        \centering
        \includegraphics[width=\linewidth]{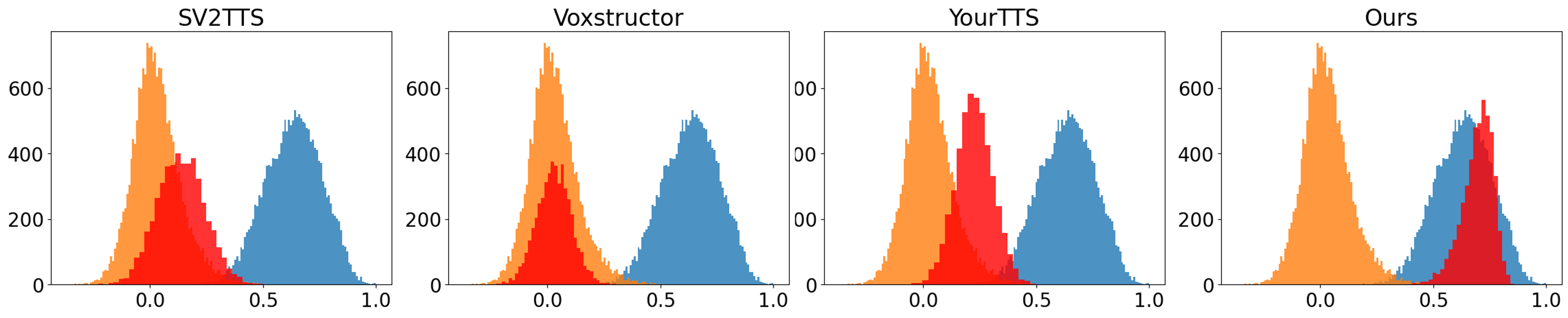}
        \caption{$F_T = T_5$} 
        \label{fig:row2}
    \end{subfigure}
    \caption{Distributions of round-trip similarity scores $s_T(\mathfrak{v})$ (red) evaluated on the each $F_T$.
    Positive (blue) and negative (orange) reference distributions are shown for comparison.}
    \label{fig:entire_experiment}
\end{figure*}

Figure~\ref{fig:entire_experiment} presents the full distributions of similarity scores for all evaluated systems and test-time extractors $\{F_T\}_{T=1}^{5}$.
For each $F_T$, we report: (i) positive reference distributions (blue), (ii) negative reference distributions (orange), and
(iii) round-trip ID-Constraints scores under transfer conditions (red).

All quantitative findings related to identity preservation and transferability are contained in this figure.
The main paper focuses on high-level trends and implications, while detailed distributional behaviors and cross-model comparisons are deferred to this appendix for completeness.

\section{Additional Experiment Setup and Implementation Details}
In this section, we provide a detailed description of the experimental setup and implementation details used throughout this study. Specifically, we elaborate on the following components: (1) the evaluation metrics for speaker recognition performance, (2) the fine-tuning procedure employed for training the inverse model, including the setup of trainable parameters for fine-tuning and construction of $\delta$-OVS, (3) the configuration of both the target models and the local surrogate models utilized for executing the proposed attack, and (4) the implementation details of FakeBob.

\subsection{Evaluation Metrics} 
\label{subsec:evaluation}
For evaluation of speaker recogniton, common metrics include the Equal Error Rate (EER) and the minimum Detection Cost Function (minDCF); EER is the error rate at which the False Acceptance Rate (FAR) and the False Rejection Rate (FRR) are equal; minDCF is a cost-based metric widely used in practical applications, particularly in the annual speaker recognition challenge held by NIST~\cite{nist_sre08}. According to the official evaluation plan, one commonly used parameter setting in the audio track is \texttt{id:1}, which corresponds to $P_{\mathsf{target}}=0.01, C_{\mathsf{FalseAlarm}}=C_{\mathsf{Miss}}=1$. Both metrics can be expressed in terms of FAR and FRR, where $\mathsf{FRR}(\tau) = 1 - \mathsf{TAR}(\tau)$, as follows:
\begin{gather}
    \mathsf{EER} = \frac{\mathsf{FRR}(\hat{\tau}) + \mathsf{FAR}(\hat{\tau})}{2}, \nonumber \\ 
    \ \text{where} \ \hat{\tau} = \arg \min_{\tau} \left| \mathsf{FAR}(\tau) - \mathsf{FRR}(\tau) \right|. \nonumber 
\end{gather}
\begin{gather}  
\mathsf{minDCF} = \min_{\tau} \Big( 
C_{\mathsf{miss}} \cdot \mathsf{FRR}(\tau) \cdot (1 - P_{\text{target}}) \nonumber \\
+ C_{\mathsf{FalseAlarm}} \cdot \mathsf{FAR}(\tau) \cdot P_{\mathsf{target}} 
\Big). \nonumber  
\end{gather}
As these definitions suggest, lower EER and minDCF indicate better discrimination ability, meaning the model performs more effectively in distinguishing speakers.

\subsection{Experiment and Implementation Details} 
\label{subsec:implemetaiondetailss}
All experiments were performed on a single NVIDIA A100 GPU using PyTorch~\cite{NEURIPS2019_9015} and SpeechBrain~\cite{ravanelli2021speechbraingeneralpurposespeechtoolkit}. Full utterances were used for all inferences. In addition, when using open source speaker recognition systems (SRS), the official inference codes provided by each model were utilized.

\subsubsection{Fine-Tuning an Inverse Model.(EN)} We used YourTTS as the foundation for our local inverse model (implementation provided by Coqui~\cite{CoquiTTS}). To apply our fine-tuning technique, we used the training set of VoxCeleb1~\cite{Nagrani17} with a batch size of 64, and the gradient accumulation step set at 4. Also, $\lambda_\mathsf{IC}=5$ and $\lambda_\mathsf{SC}=1$. We employed AdamW as the optimizer, and the learning rate was gradually lowered from 1e-3 to 1e-5 via CosineAnnealing scheduling.

\subsubsection{Fine-Tuning an Inverse Model (CN)}
We used YourTTS as the foundation for our local inverse model (implementation provided by Coqui~\cite{CoquiTTS}). To apply our fine-tuning technique, we used the training set of CNCeleb1,2~\cite{fan2020cn}. To bridge the dimensionality gap between the output of $L_\mathsf{CN}$ ($256$) and the YourTTS speaker encoder ($512$), we integrated a residual \textit{LinkBlock} consisting of linear projections, LayerNorm, and GELU activation. This block is trained end-to-end, with the SC loss applied between its input and output to preserve the topological structure of the feature space. The training settings included a batch size of 64, a gradient accumulation step of 4, $\lambda_\mathsf{IC}=5$, and $\lambda_\mathsf{SC}=1$. We employed AdamW as the optimizer, with the learning rate gradually lowered from 1e-3 to 1e-5 via CosineAnnealing scheduling.


\subsubsection{Parameter Configuration for Inverse Model Fine-Tuning.}
\label{subsubsec:app_trainable}
This paragraph provides further details on the specific parameter configuration used when fine-tuning the pre-trained YourTTS model~\cite{casanova2022yourtts} to act as our inverse model $F^{-1}$, as mentioned in Sec.~\ref{subsec:setting_params}. The goal is to leverage the fixed-text strategy and focus the learning process on adapting the model to generate audio $\widetilde{\mathsf{aud}}_i$ that reflects the input speaker embedding $F(\mathsf{aud}_i)$, while preserving the linguistic quality derived from the fixed input text $\mathsf{Text}$.

YourTTS builds upon the VITS architecture~\cite{kim2021conditional} and comprises several interconnected modules.
Our fine-tuning strategy focuses learning by selectively freezing specific components related to the fixed input text. Specifically, the Text Encoder, responsible for converting the input text $\mathsf{Text}$ into linguistic representations (including any associated Language Embeddings if applicable for the specific pre-trained model), is kept entirely frozen. Since the text input and its corresponding language are constant in our setup, preserving the pre-trained text and language representations is crucial for maintaining linguistic consistency. Similarly, the Duration Predictor, which predicts the duration for each linguistic unit based on the text content, is also kept frozen to maintain the prosodic structure associated with the fixed input text. The Posterior Encoder, a VITS component primarily used during training, is less critical for adaptation in this context as our focus is on the prior pathway driven by the speaker embedding.

Conversely, the core generative pathway, encompassing the prior encoder, decoder, and normalizing flow, is made trainable. This pathway is responsible for synthesizing the output Mel-spectrogram conditioned on the linguistic representations, predicted durations, and, critically, the input speaker embedding $F(\mathsf{aud}_i)$. Making these components trainable allows the model to learn the crucial mapping from the speaker embedding $F(\mathsf{aud}_i)$ to the corresponding vocal characteristics needed for generating $\widetilde{\mathsf{aud}}_i$. The gradient updates originating from our specialized loss functions in Section~\ref{subsec:loss_function_design} primarily influence these parameters, adjusting how the model synthesizes the Mel-spectrogram according to the target speaker identity.

Furthermore, the parameters of the vocoder (e.g., HiFi-GAN), which transforms the generated Mel-spectrogram into the final waveform $\widetilde{\mathsf{aud}}_i$, are also made trainable in our approach. Fine-tuning the vocoder allows the waveform generation process itself to adapt to the target speaker identity and the fine-tuning objective. This potentially enables the model to capture finer acoustic details directly at the waveform synthesis stage, further enhancing the identity preservation driven by our loss functions.

In summary, by freezing the text-related components (text Encoder, including language aspects, and duration predictor), we concentrate the fine-tuning updates on the core acoustic model responsible for Mel-spectrogram generation conditioned on $F(\mathsf{aud}_i)$, as well as the vocoder responsible for the final waveform synthesis. This targeted approach aims to maximize the model's adaptation towards learning the inverse mapping for speaker identity across the entire generation pipeline.

\subsubsection{Details of Fixed Text}

For training the inverse model, we employ fixed text sets in both English and Chinese.
The English text set consists of the following utterances: \emph{``Can you hear my voice clearly now''}, \emph{``Hello Google, Hi Siri, Hey Bixby''}, and \emph{``Up down left right start now again''}. For training Chinese inverse model, text set consists of the following utterances:
\emph{``Hello Google, Hi Siri, Hey Bixby''} and \emph{``Ni hao ya, Gu ge, bang wo yi xia''}.

Among these, we use \emph{``Hello Google, Hi Siri, Hey Bixby''} as the primary fixed text for reporting the inverse model training results, as it is shared across both languages.
We observe that restricting the inverse model to a fixed text does not introduce a noticeable change in the overall attack behavior or performance trends.
This indicates that the proposed attack is largely insensitive to the specific choice of fixed text.

\subsubsection{Details of Audio-NES}

Algorithm~\ref{alg:audio_fakebob_scoreonly} describes the Audio-NES attack, which performs score-based optimization
directly in the waveform space.
The attacker interacts with the target speaker recognition system $T$ exclusively through a score-only black-box oracle
$\textsc{Query}(\cdot)$, which returns a similarity score between a query audio and the victim utterance.
No access to internal representations, model parameters, or gradients of $T$ is assumed.

The optimization variable is a waveform $\mathsf{aud}\in\mathbb{R}^{48000}$, corresponding to 3 seconds of audio sampled at 16~kHz.
At each iteration, Audio-NES estimates a gradient direction using Natural Evolution Strategies (NES) with $B$ random perturbations.
The gradient estimate is then used to update the waveform via sign-based descent.

The total query budget $Q$ is determined by the product of the number of perturbation samples per iteration $B$
and the maximum number of iterations: $Q = B \times texttt{max\_iter}$.
Unless stated otherwise, all reported Audio-NES results correspond to the configuration that achieves the best attack performance
among the tested values of $B$ (cf.~Table~\ref{tab:audio_hparams}).
If multiple configurations yield comparable performance, we report the result with the smaller total query count.

\begin{algorithm}[t]
\caption{Audio-NES}
\label{alg:audio_fakebob_scoreonly}
\begin{algorithmic}[1]
\Require target oracle $\textsc{Query}(T,\cdot,\cdot)$,
victim utterance $\mathsf{aud}_{\text{vic}}\in\mathcal{A}$,
query budget $Q$, threshold $\tau$,
step size $\alpha$, noise scale $\sigma$, samples per draw $B$
\Ensure adversarial audio $\mathsf{aud}^\star$ (or best found)
\State initialize $\mathsf{aud}_0 \sim \mathcal{N}(0,I_{48000})$ and clip to $[-1,1]$
\For{$t=1,2,\dots$ while cumulative queries $\le Q$}
    \State Sample $\epsilon_1,\dots,\epsilon_N \sim \mathcal{N}(0,I)$ \Comment{optionally antithetic}
    \For{$i=1$ to $B$}
        \State $\tilde{\mathsf{aud}}_i \leftarrow \mathrm{clip}(\mathsf{aud}_{t-1}+\sigma\epsilon_i)$
        \State $s_i \leftarrow \textsc{Query}(T, \tilde{\mathsf{aud}}_i, \mathsf{aud}_{\text{vic}})$
        \State $\ell_i \leftarrow 1 - s_i$
    \EndFor
    \State $\hat g \leftarrow \frac{1}{B\sigma}\sum_{i=1}^B \ell_i\,\epsilon_i$
    \State $\mathsf{aud}_t \leftarrow \mathrm{clip}\!\left(\mathsf{aud}_{t-1}-\alpha\cdot \mathrm{sign}(\hat g)\right)$
    \State $s \leftarrow \textsc{Query}(T, \mathsf{aud}_t, \mathsf{aud}_{\text{vic}})$; \;\; $\ell \leftarrow 1-s$
    \If{$\ell \le \tau$} \State \textbf{return} $\mathsf{aud}_t$ \EndIf
\EndFor
\State \Return best found $\mathsf{aud}^\star$
\end{algorithmic}
\end{algorithm}

\begin{table}[t]
\centering
\caption{Hyperparameters for Audio-NES}
\label{tab:audio_hparams}
\begin{tabular}{lc}
\toprule
\textbf{Hyperparameter} & \textbf{Value} \\
\midrule
Search variable                    & waveform $\mathsf{aud}\in\mathbb{R}^{48000}$ \\
Initialization                     & $\mathcal{N}(0,I_{48000})$ \\
Candidate pool size $C$            & 100 \\
Selected candidates $S$            & 1 \\
Maximum iterations                 & 1000 \\
Samples per draw $B$               & 10, 50 \\
Noise scale $\sigma$               & $1\times 10^{-3}$ \\
Initial learning rate $\alpha_0$   & $1\times 10^{-1}$ \\
Minimum learning rate              & $1\times 10^{-4}$ \\
Momentum coefficient               & 0.9 \\
Update rule                        & sign-based descent \\
\bottomrule
\end{tabular}
\end{table}

\subsubsection{Details of Latent-NES}
\label{subsubsec:latent_nes_details}

Algorithm~\ref{alg:latent_fakebob_scoreonly} presents the Latent-NES attack, which operates in a low-dimensional latent space
and synthesizes attack audio via an inverse decoder.
This algorithm is used for both \emph{YourTTS-NES} and \emph{Ours-NES}, differing only in the choice of the inverse model $G$.
The optimization procedure itself remains identical across these two instantiations.

Latent-NES optimizes a latent vector $z$ constrained to the unit hypersphere $\mathbb{S}^D$.
At each iteration, spherical NES perturbations are applied to $z$, and each perturbed latent vector is decoded into audio
using $G(z,\mathsf{Text})$, where the text input $\mathsf{Text}$ is fixed throughout the attack.
The resulting audio samples are evaluated via the same score-only black-box oracle $\textsc{Query}(\cdot)$ as in Audio-NES.

As in the audio-space case, the total query budget is expressed as
$Q = B \times \texttt{max\_iter}$ where $B$ denotes the number of NES samples per iteration.
All Latent-NES results reported in the main paper are obtained using the configuration that yields the best attack performance
among different values of $B$ (see Table~\ref{tab:latent_hparams}).
When multiple configurations achieve similar success, the one with fewer queries is selected.

We emphasize that, despite operating in a latent space, Latent-NES strictly adheres to the same score-based black-box threat model as Audio-NES.
The attacker never observes speaker embeddings or internal features of the target system and relies solely on similarity scores
returned by the oracle.

\begin{algorithm}[t]
\caption{Latent-NES (for YourTTS, Ours-NES)}
\label{alg:latent_fakebob_scoreonly}
\begin{algorithmic}[1]
\Require target oracle $\textsc{Query}(T,\cdot,\cdot)$,
inverse decoder $G:\mathbb{R}^{D}\times\mathcal{T}\to\mathcal{A}$,
victim utterance $\mathsf{aud}_{\text{vic}}$,
fixed text $\mathsf{Text}\in\mathcal{T}$,
query budget $Q$, threshold $\tau$,
step size $\alpha$, noise scale $\sigma$, samples per draw $B$
\Ensure latent $z^\star$ and audio $\mathsf{aud}^\star=G(z^\star,\mathsf{Text})$ (or best found)
\State initialize $z_0 \sim \mathcal{N}(0,I_D)$ and normalize $z_0 \leftarrow z_0/\|z_0\|_2$
\For{$t=1,2,\dots$ while cumulative queries $\le Q$}
    \State Sample $\epsilon_1,\dots,\epsilon_B \sim \mathcal{B}(0,I_D)$
    \For{$i=1$ to $B$}
        \State $\tilde z_i \leftarrow \frac{z_{t-1}+\sigma\epsilon_i}{\|z_{t-1}+\sigma\epsilon_i\|_2}$
        \State $\tilde{\mathsf{aud}}_i \leftarrow G(\tilde z_i,\mathsf{Text})$
        \State $s_i \leftarrow \textsc{Query}(T, \tilde{\mathsf{aud}}_i, \mathsf{aud}_{\text{vic}})$
        \State $\ell_i \leftarrow 1 - s_i$
    \EndFor
    \State $\hat g \leftarrow \frac{1}{B\sigma}\sum_{i=1}^B \ell_i\,\epsilon_i$
    \State $z_t \leftarrow \frac{z_{t-1}-\alpha\cdot \hat g}{\|z_{t-1}-\alpha\cdot \hat g\|_2}$
    \State $\mathsf{aud}_t \leftarrow G(z_t,\mathsf{Text})$
    \State $s \leftarrow \textsc{Query}(T, \mathsf{aud}_t, \mathsf{aud}_{\text{vic}})$; \;\; $\ell \leftarrow 1-s$
    \If{$\ell \le \tau$} \State \textbf{return} $z_t,\,\mathsf{aud}_t$ \EndIf
\EndFor
\State \Return best found $z^\star$ and $\mathsf{aud}^\star=G(z^\star,\mathsf{Text})$
\end{algorithmic}
\end{algorithm}

\begin{table}[t]
\centering
\caption{Hyperparameters for latent(spherical) NES. (for Ours-NES, YourTTS-NES)}
\label{tab:latent_hparams}
\begin{tabular}{lc}
\toprule
\textbf{Hyperparameter} & \textbf{Value} \\
\midrule
Search variable                    & latent vector $z\in\mathbb{S}^D$ \\
Initialization                     & $\mathcal{N}(0,I_D)$ + $\ell_2$ normalization \\
Candidate pool size $C$            & 100 \\
Selected candidates $S$            & 1 \\
Maximum iterations                 & 1000 \\
Samples per draw $B$               & 10, 50 \\
Noise scale $\sigma$               & $5.0$ \\
Initial learning rate $\alpha_0$   & $5\times 10^{-2}$ \\
Minimum learning rate              & $1\times 10^{-6}$ \\
Momentum coefficient               & 0.9 \\
Update rule                        & spherical gradient descent \\ 
\bottomrule
\end{tabular}
\end{table}

\subsubsection{Construction of $\delta$-Orthogonal Voice Set.(Ours-SP)}
\label{ovs}
To construct the $\delta$-OVS, we used VoxCeleb1~\cite{Nagrani17} and trainset as datasets. First, we performed pre-processing to extract feature vectors for all utterances in the dataset using the speaker recognition model used as the speaker encoder of YourTTS. After that, we constructed a set satisfying the $\delta$-OVS constraint in Definition~\ref{def:ovs}, where $\delta \approx 0.2$ and the OVS size is 50.

\subsubsection{Details of Audio-GD}
\label{subsub:audiogd}
Audio-GD is a white-box baseline that generates impersonation audio by directly optimizing the waveform in the audio domain. Given a target speaker embedding $t$, the attack iteratively updates an audio signal \(\tilde{\mathfrak{v}}\) via gradient descent to minimize the cosine distance between the extracted embedding \(F(\mathfrak{v})\) and \(t\). Gradients are computed through the speaker recognition model and applied to the audio samples until convergence or a predefined iteration limit. This approach serves as a tool for inversion via optimization in the high-dimensional audio space, without relying on generative priors or inverse models.

\begin{algorithm}[t]
    \caption{Attack Audio Generation via Gradient Descent (Audio-GD)}
    \begin{algorithmic}[1]
        \Require SR feature extractor $F: \mathcal{A} \rightarrow \mathbb{S}^{d-1}$, original audio $\mathsf{aud}_{\text{orig}} \in \mathcal{A}$, target embedding $x \in \mathbb{S}^{d-1}$, learning rate $\eta$, max iterations $K_{\max}$, optional convergence threshold $\epsilon$

        \Ensure Optimized attack audio $\tilde{a} \in \mathcal{A}$

        \State Initialize audio: $\tilde{a} \leftarrow \tilde{a}_0$
        \State Initialize step counter: $K \leftarrow 0$
        \State Initialize previous loss: $\mathcal{L}_{\text{prev}} \leftarrow \infty$

        \Repeat
            \State Compute current embedding: $\tilde{x} \leftarrow F(\tilde{a})$
            \State Compute loss: $\mathcal{L}_{\text{now}} \leftarrow 1 - \frac{\langle \tilde{x},x \rangle} {\lVert\tilde{x}\rVert \lVert x \rVert}$
            \State Compute gradient of loss w.r.t. audio input: $\nabla_{\tilde{a}} \mathcal{L}_{\text{now}}$
            \State Update audio signal: $\tilde{a} \leftarrow \tilde{a} - \eta \cdot \nabla_{\tilde{a}} \mathcal{L}_{\text{now}}$
            \State $K \leftarrow K + 1$
            \State Check for convergence $\Delta \mathcal{L} = |\mathcal{L}_{\text{prev}} - \mathcal{L}_{\text{now}}|$
            \State $\mathcal{L}_{\text{prev}} \leftarrow \mathcal{L}_{\text{now}}$
        \Until{$K \ge K_{\max}$ or $\Delta \mathcal{L} \le \epsilon$}
        \State \Return Optimized audio $\tilde{a}$
    \end{algorithmic}
    \label{alg:gd_inverse}
\end{algorithm}

\subsection{Model Configurations}
\label{subsec:modelconfig}

\begin{table*}[t]
\centering
\caption{Specifications and performance of the speaker recognition models used in this paper, including the target models ($T_1$--$T_5$) and the attacker's local models ($L_\mathsf{SV2TTS}$, $L_\mathsf{Vox}$, and $L_\mathsf{Ours}$). Note that `$L_\mathsf{Ours}$' is referred to as `$L$' in the main paper. Performance is evaluated using EER, minDCF, and the corresponding thresholds $\tau_\mathsf{E}$ and $\tau_\mathsf{M}$ on the VoxCeleb1 test set. For the Chinese speaker recognition models ($C_1$, $C_2$) and the local model ($L_\mathsf{CN}$), performance is evaluated on the CN-Celeb test set.
}
\resizebox{.92\linewidth}{!}{
\begin{tabular}{c|c|c|c|c|c|c}
\hline
Index & Architecture & Trainset & EER(\%) & minDCF & $\tau_\mathsf{E}$ & $\tau_\mathsf{M}$ \\ \hline \hline
$T_1$ & Redim-S & VoxBlink2 & 1.54 & 0.1415 & 0.6605 & 0.5518 \\ \hline
$T_2$ & Redim-M & VoxBlink2& 1.31 & 0.1003 & 0.6624 & 0.5437 \\ \hline
$T_3$ & SimAMResNet34 & VoxBlink2 & 1.16 & 0.1039 & 0.6258 & 0.5060 \\ \hline
$T_4$ & SimAMResNet100 & VoxBlink2 & 0.76 & 0.0767 & 0.6137 & 0.4932 \\ \hline
$T_5$ & Titanet-L & VoxCeleb1,2, Fisher, Switchboard, LibriSpeech & 0.82 & 0.1353 & 0.6654 & 0.5215 \\ \hline \hline
$C_1$ & CAM++ & N/A & 4.70 & 0.3120 & 0.7064 & 0.4817 \\ \hline
$C_2$ & ERE2Net & N/A & 2.79 & 0.1779 & 0.6350 & 0.4617 \\ \hline \hline
$L_\mathsf{SV2TTS}$ & GE2E & VoxCeleb1,2 & 8.28 & 0.6810 & 0.3232 & 0.2100 \\ \hline
$L_\mathsf{Vox}$ & Rawnet3 & VoxCeleb1,2 & 0.83 & 0.1111 & 0.6619 & 0.5236 \\ \hline
$L_\mathsf{CN}$ & ResNet34 & CNCeleb & 7.54 & 0.4575 & 0.7349 & 0.5043  \\ \hline
$L_\mathsf{Ours}$ & H/ASP & VoxCeleb1,2 & 1.18 & 0.1339 & 0.5778 & 0.4667 \\ \hline
\end{tabular}
}
\label{tab:merged_sr_models}
\end{table*}

Table~\ref{tab:merged_sr_models} summarizes the speaker recognition models used in our study, categorized into two groups: target models ($T_1$--$T_5$) and attacker-side local models ($L_\mathsf{SV2TTS}$, $L_\mathsf{Vox}$, and $L_\mathsf{Ours}$). The target models employ a diverse range of architectures and size—including RedimNet~\cite{yakovlev24_interspeech}, SimAMResNet~\cite{qin2022simple}, and Titanet~\cite{koluguri2022titanet}—trained on either VoxBlink2~\cite{lin2024voxblink2100kspeakerrecognition} or large-scale aggregated corpora such as VoxCeleb1~\cite{Nagrani17}, VoxCeleb2~\cite{chung2018voxceleb2}, Fisher~\cite{cieri-etal-2004-fisher}, Switchboard~\cite{225858}, LibriSpeech~\cite{ravanelli2021speechbraingeneralpurposespeechtoolkit}. This variation ensures coverage of a wide spectrum of real-world systems in our evaluation.

On the attacker side, the local models are configured to reflect realistic black-box constraints, with both architectural and training data separation from the target models. Specifically, $L_\mathsf{SV2TTS}$ utilizes a GE2E~\cite{wan2018generalized}-based speaker encoder, $L_\mathsf{Vox}$ adopts the RawNet3~\cite{jung2022pushing} architecture optimized for end-to-end verification, and $L_\mathsf{Ours}$ employs a hybrid structure integrating heuristic attention and ASP~\cite{heo2020clova} modules. For simplicity, $L_\mathsf{Ours}$ is referred to as $L$ throughout the main text.

In addition to model architecture and training data, the table also includes key performance metrics: Equal error rate (EER), minimum detection cost function (minDCF), and their respective decision thresholds ($\tau_\mathsf{E}$ for EER and $\tau_\mathsf{M}$ for minDCF); all evaluated on the VoxCeleb1 test set.

\subsection{Additional Experiments: Original FakeBob}
\label{app:fakebob}

For the baseline comparison presented in Tab.~\ref{tab:rlwork}, we reproduced the FakeBob~\cite{chen2021real} attack following the methodology provided in SpeakerGuard's~\cite{SpeakerGuard} official codebase. The attack was configured to perform targeted impersonation against our targeted systems acting as speaker verification systems ($T_1$--$T_5$).
Consistent with FakeBob's requirements, each attack trial commenced with an initial seed utterance and aimed to iteratively refine it by querying the target system's scores until it surpassed the pre-determined EER decision threshold. This iterative optimization process was capped at a maximum of 1000 iterations. Within each iteration, the gradient direction was estimated by sampling 50 perturbation samples, generated using Gaussian noise with a standard deviation of 0.001, and querying their scores from the target system; these queries were processed efficiently in a single batch. Updates were performed using a momentum of 0.9 and an adaptive learning rate strategy. The learning rate started at 0.001 and was halved if no score improvement was observed for 5 consecutive iterations, down to a minimum of 1e-6. Furthermore, early stopping was enabled, terminating the process if the best achieved score did not change for 100 iterations. Other configuration parameters included using an $\epsilon$ value of 0.002 and processing the samples individually. Crucially, the query counts reported in Table~\ref{tab:fakebob} encompass only those consumed during this iterative attack generation phase, excluding any potentially required for initial threshold estimation, thereby facilitating a direct comparison with the query efficiency of our proposed method.

\begin{table}[t]
\caption{FakeBob attack performance against target models $T_1$--$T_5$, showing attack successes (out of 50 trials) and the number of queries required solely for the attack generation phase (excluding threshold estimation).}
\resizebox{\linewidth}{!}{
\begin{tabular}{c|c|c|c|c|c}
\hline
Methods & $T_1$ & $T_2$ & $T_3$ & $T_4$ & $T_5$\\ \hline \hline
Ours & 46/50 & 44/50 & 37/50 & 36/50 & 36/50\\ \hline \hline
FakeBob & 43/50 & 43/50 & 32/50 & 29/50 & 34/50\\ \hline
\# of Queries & $\approx14233$ & $\approx14067$ & $\approx14073$ & $\approx13181$ & $\approx11552$ \\ \hline
\end{tabular}
}
\label{tab:fakebob}
\end{table}

\section{In-depth Analysis of Ours-SP}
\label{sec:analysisASR}
In this section, we provide a detailed analysis of our attack. We begin by outlining the overall attack pipeline and discussing specific errors or gaps encountered at each stage of the process. Then we analyze how these issues influenced the overall success rate of the attack. Furthermore, we present an evaluation of black-box transfer settings by assessing their success rates against target models that were not explicitly used during the attack generation phase.

We present a score-based non-adaptive impersonation attack within the speaker recognition domain by generalizing the framework originally established by \cite{kim2024scores} for attacking face recognition systems.
Let $T:\mathcal{B} \rightarrow [-1,1]$ be a target black-box biometric recognition system with an enrolled biometric datum $e$. 
The attacker has a local feature extractor $F_L:\mathcal{B} \rightarrow \mathbb{S}^{d-1}$ and a corresponding inverse model $F_L^{-1}:\mathbb{S}^{d-1} \rightarrow \mathcal{B}$, both constructed independently of $T$. The impersonation attack consists of the following steps:

\begin{enumerate}
    \item \textbf{Feature Space Reduction via Target Queries.}
    The attacker queries $T$ with a local set of samples $(b_1, \dots, b_n)$ and collects corresponding scores $s_i = T(b_i)$. Assuming that models trained for the same task behave similarly, we expect the local extractor to align well with the target system in similarity scores. Based on this assumption, we approximate $s_i \approx \langle F_L(b_i), F_L(e) \rangle$.

    \item \textbf{Solving Linear System for the Feature Vector.}
    Treating $F_L(b_i)$ as known row vectors, setting $A$ such that it's $i$-th row is $s_i = F_L(b_i)$,
    the system can be expressed as $Ax = y$, where $y = (s_1, \dots, s_n)^\top$, and solved via least squares to approximate the (local) feature vector of the enrolled identity $ \hat{x} \approx F_L(e)$.
    To ensure numerical stability, a well-conditioned query set is crucial. We define:

    \begin{Definition}[$\delta$-Orthogonal Biometric Set ($\delta$-OBS)]
        Given $F:\mathcal{B} \rightarrow \mathbb{S}^{d-1}$, a set $\mathcal{O} = \{ o_1, \dots, o_m \}$ is 
        a $\delta$-\text{OBS} \\ if $\left| \langle F(o_i), F(o_j) \rangle \right| \leq \delta$ for all $i \neq j$.
        \label{def:ovs}
    \end{Definition}


    \item \textbf{Pre-image Reconstruction.}
    Using the solved feature vector $\hat{v}$, the attacker generates an input sample $b'$ such that $F_L(b') \approx \hat{x}$ 
    , in order to satisfy $T(b') > \tau$, where $\tau$ is the acceptance threshold. In the SR domain, this step requires a feasible $F_L^{-1}$ model inverting $\hat{x}$ into audio.
\end{enumerate}

\subsection{Error Analysis of the Attack Pipeline}

\begin{figure}[h]
\fbox{
\parbox{.95\linewidth}{
\begin{center}
    \textbf{Non-Adaptive Black-Box Impersonation Attack}\\
    \textbf{(Ours-SP)}
\end{center}
\textbf{Inputs}:
\begin{itemize}[leftmargin=10pt]
    \item The black-box access target SRS $T: \mathcal{A} \rightarrow [-1,1]$ with enrolled voice $\mathsf{aud}$.
    \item The local SR feature extractor $F: \mathcal{A} \rightarrow \mathbb{S}^{d-1}$ and its inverse model $F^{-1}: \mathbb{S}^{d-1} \rightarrow \mathcal{A}$.
    \item The local voice dataset $\mathcal{D}$ where $|\mathcal{D}|=n$ and $n  \in \mathbb{N}$ .
    \item The pre-defined parameter $\delta \in [-1, 1]$ and $m \in \mathbb{N}$.
\end{itemize}
\textbf{Output}: $\widehat{\mathsf{aud}}$ \\
\rule{\linewidth}{0.1mm}

\textbf{Pre-processing ($\delta$-OVS generation):}
\begin{algorithmic}[1]
        \State Set $f_i \leftarrow F(\mathcal{D}_i)$ for $\forall i \in [n]$
        \State Find a set of indices $I=\{ I_{1},\cdots,I_{m} \} \subseteq [n]$ of size $m$ such that $\underset{j,k \in I}{\max}{|\langle f_j, f_k \rangle|} \leq \delta$
        \State $\mathcal{O} \leftarrow \{o_i = \mathcal{D}_{I_i} | i \in [m] \}$
\end{algorithmic}
\rule{\linewidth}{0.1mm}
\textbf{Algorithm:}
\begin{algorithmic}[1]
        \State Query and set $s_{i} \gets T(o_{i}) $ for $\forall i\in[m]$.\
        \State Set $s \in [-1,1]^{m}$, whose $i$'th element is $s_{i}$ for $\forall i\in[m]$.
        \State Extract and set $A_i \gets F(o_{i})$ for $\forall i\in[m]$.
        \State Set $A \gets [A_1, \cdots , A_m]^T$.
        \State $r \leftarrow A^{\dagger}s$
        \State $\widehat{\mathsf{aud}} \leftarrow {F}^{-1}(r/ \left\lVert r \right\rVert )$
        \State \textbf{Return} $\widehat{\mathsf{aud}}$.
\end{algorithmic}
}
}
\caption{Algorithm for Non-Adaptive Black-box Impersonation Attack on an SRS.}
\label{alg:Pipeline}
\end{figure}

\begin{figure*}[t]
    \centering
        \includegraphics[width=\linewidth]{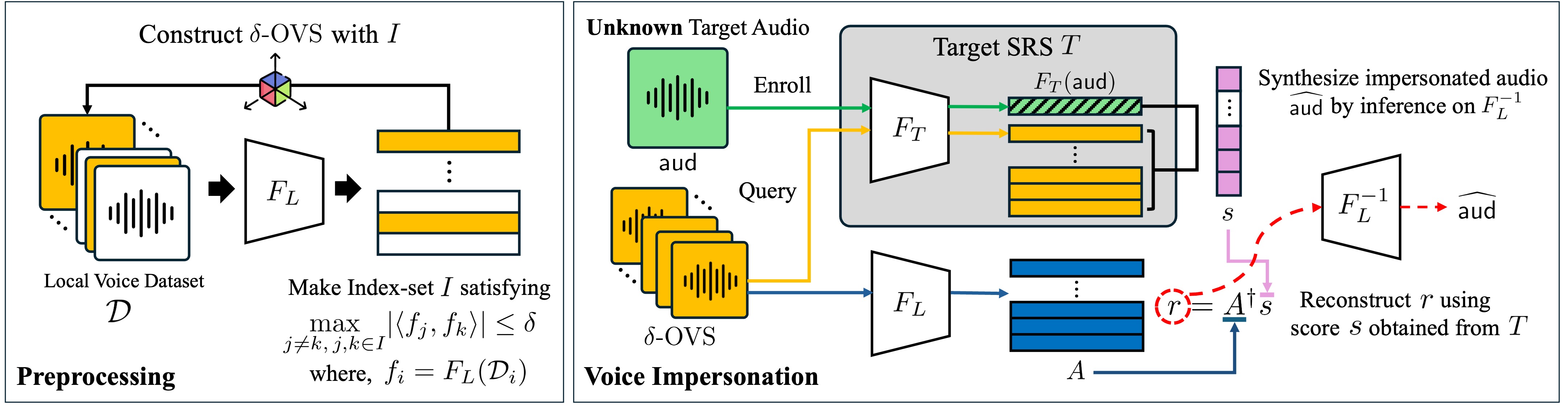}
    \caption{Flow diagram of the proposed Non-Adaptive Speaker Impersonation Attack applied to SRSs (Ours-SP). Key stages include pre-processing for $\delta$-OVS generation, querying the target SRS, and synthesizing the impersonation audio. Note that the attacker cannot directly access the black-box target $T$ (and $F_T$). The local models ($F_L$ and $F_L^{-1}$) are used as surrogate for the attack.}
    \label{fig:pipeline}
\end{figure*}

The flow of our attack is illustrated in Figure~\ref{fig:pipeline}, and the detailed algorithm is described in Figure~\ref{alg:Pipeline}. The attack consists of three main steps: (1) reducing the problem from the target model $T$ to a local feature space, (2) solving for the feature representation in this local space, and (3) synthesizing a biometric sample $b'$ that can successfully impersonate the target $T$. Specifically, lines 1--4 in the algorithm correspond to (Step 1), line 5 implements (Step 2), and line 6 executes (Step 3).

The overall attack pipeline comprises three key stages whose effectiveness depends on managing and mitigating the cumulative impact of errors introduced throughout the process. While these errors do not propagate in a purely additive manner, we decompose them step-by-step to better understand which components most significantly influence attack success and where improvements could yield the greatest benefit. This breakdown offers a more comprehensive view into the attack dynamics, highlighting how different sources of error interplay to shape the final outcome.

In (Step 1), discrepancies arise between the local and target models due to imperfect alignment of their feature space (more accurately, similarity scores), which can distort the recovered objective. (Step 2) introduces numerical inaccuracies during the pseudo-inversion process used to estimate the enrolled feature vector. (Step 3) involves synthesizing an audio sample from the recovered feature such that it both conforms to the input requirements of the target model and preserves alignment with the intended feature representation.

\begin{table}[t]
\caption{Statistical summary of score discrepancies between local and target models ($T_1$--$T_5$) on the VoxCeleb1 test set. Each entry reports the mean and std. of the absolute differences in cosine similarity scores across all pairs.}
\resizebox{.9\linewidth}{!}{
\begin{tabular}{c|c|c|c}
\hline
$T$ & $L_{\mathsf{SV2TTS}}$ & $L_{\mathsf{Vox}}$ & $L_{\mathsf{Ours}}$ \\ \hline \hline
$T_1$ & $0.3304\pm0.2017$ & $0.0617\pm0.0479$ & $0.0741\pm0.0582$ \\ \hline
$T_2$ & $0.3305\pm0.2076$ & $0.0588\pm0.0461$ & $0.0757\pm0.0594$ \\ \hline
$T_3$ & $0.2999\pm0.2143$ & $0.0675\pm0.0517$ & $0.0691\pm0.0523$ \\ \hline
$T_4$ & $0.2935\pm0.2248$ & $0.0722\pm0.0528$ & $0.0760\pm0.0548$ \\ \hline
$T_5$ & $0.3461\pm0.2191$ & $0.0484\pm0.0387$ & $0.0804\pm0.0664$ \\ \hline
\end{tabular}
}
\label{tab:score_discrepancy}
\end{table}

\begin{figure*}[t]
    \centering
    \begin{subfigure}[b]{0.3\linewidth}
        \centering
        \includegraphics[width=\linewidth]{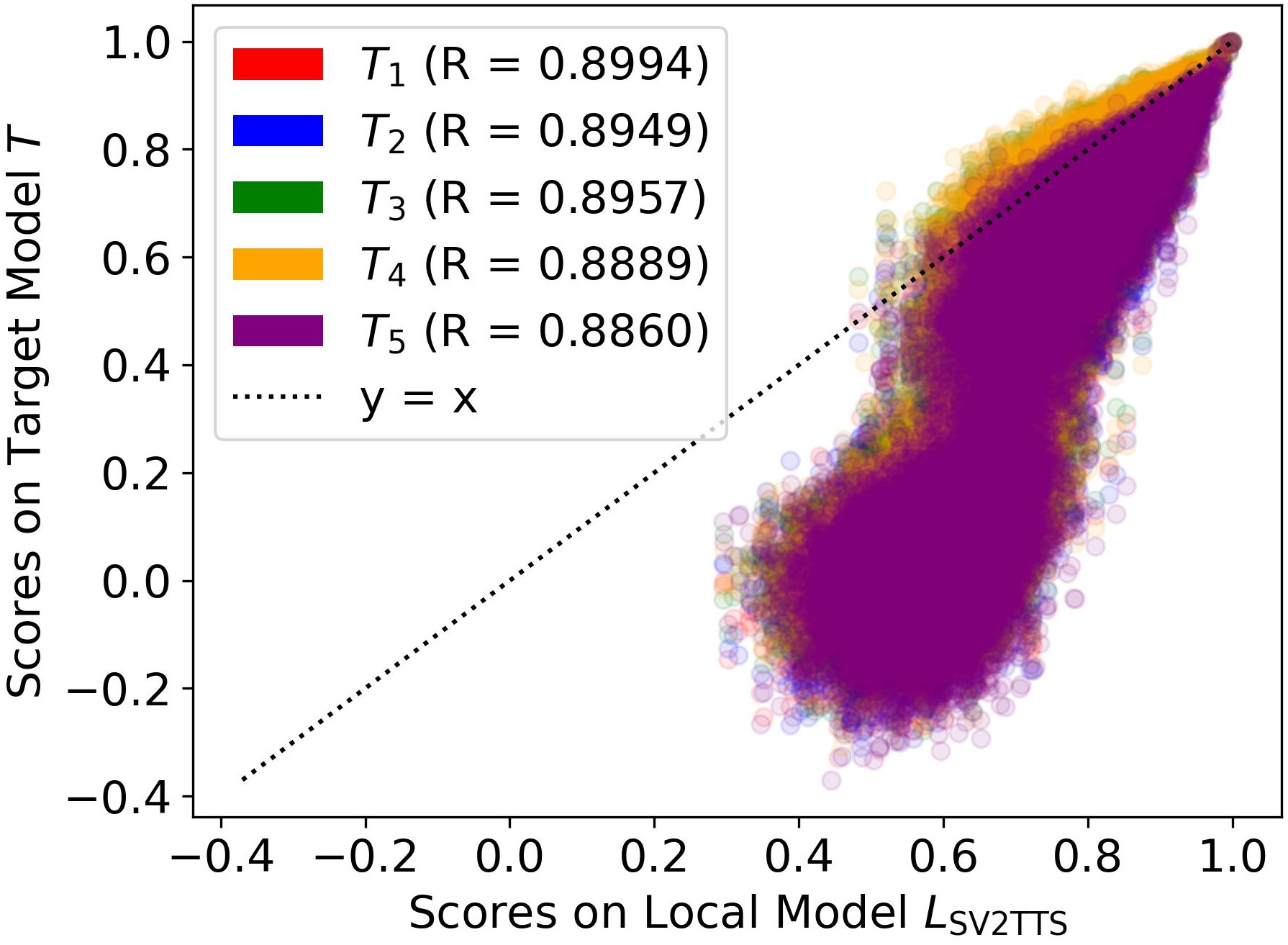}
        \caption{$L_\mathsf{SV2TTS}$ : Local model of SV2TTS}
        \label{fig:img1}
    \end{subfigure}
    \hspace{0.03\textwidth}
    \begin{subfigure}[b]{0.3\linewidth}
        \centering
        \includegraphics[width=\linewidth]{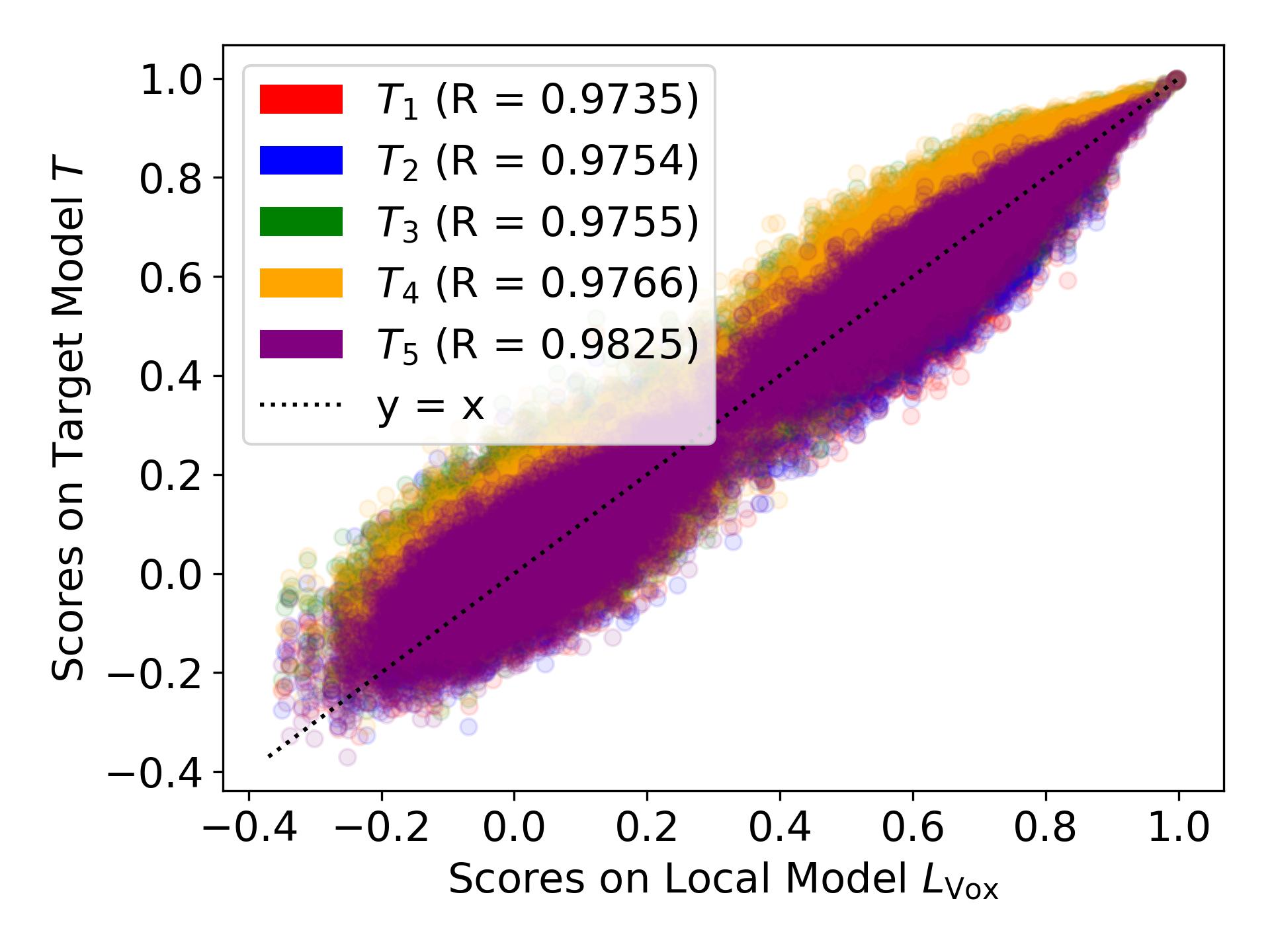}
        \caption{$L_\mathsf{Vox}$ : Local model of Voxstructor}
        \label{fig:img2}
    \end{subfigure}
    \hspace{0.03\textwidth}
    \begin{subfigure}[b]{0.3\linewidth}
        \centering
        \includegraphics[width=\linewidth]{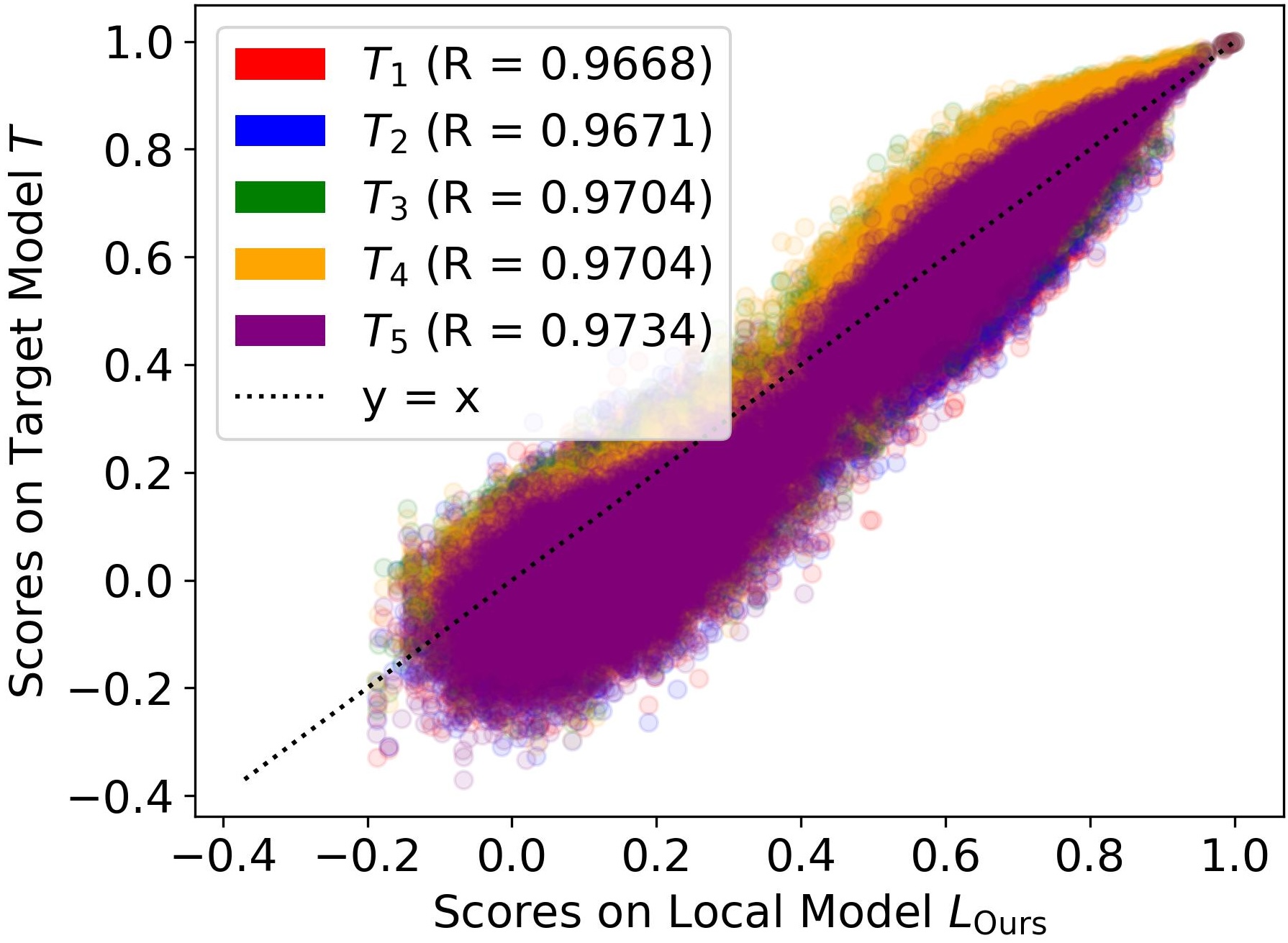}
        \caption{$L_\mathsf{Our}$ : Local model of Ours}
        \label{fig:img3}
    \end{subfigure}
    \caption{Scatter plots of the cosine similarity values output by local models for all pairs from the Voxceleb1 testset; R is the correlation coefficient between scores output by $L$ different target models ($T_1$-$T_5$).}
    \label{fig:score_scatter}
\end{figure*}

\subsection{Model Discrepancy}

Table~\ref{tab:score_discrepancy} presents the mean and standard deviation of the absolute differences in cosine similarity scores between each local model and the five target models ($T_1$–$T_5$), evaluated over all verification pairs in the VoxCeleb1 test set. Among the three local models, $L_{\mathsf{SV2TTS}}$ exhibits significantly larger score discrepancies, with mean deviations exceeding $0.29$ in all cases. In contrast, $L_{\mathsf{Vox}}$ and $L_{\mathsf{Ours}}$ demonstrate much smaller deviations than typically $0.08$, indicating closer alignment with the scoring behavior of the target models.

This observation is further supported by the scatter plots shown in Figure~\ref{fig:score_scatter}, which visualize the relationship between similarity scores produced by local models and their corresponding target models for all test pairs. Ideally, if a local model accurately approximates the target model scoring behavior, the points should be along the identity line ($y = x$), and the correlation between the two scores should be high. However, the scatter plots for $L_{\mathsf{SV2TTS}}$ reveal a large dispersion from the identity line, indicating low agreement in absolute score values. Furthermore, Pearson's correlation coefficients $R$ which quantify the linear agreement between the local and target model scores are consistently lower for $L_{\mathsf{SV2TTS}}$ compared to the other two models. This suggests that the absolute scores are not only misaligned, but also that the relative ranking or trend of scores is poorly preserved.

In contrast, both $L_{\mathsf{Vox}}$ and $L_{\mathsf{Ours}}$ achieve higher $R$ values in all target models, indicating that they better preserve the similarity of test pairs. This is consistent with their lower score discrepancies and visually tighter clustering around the diagonal in the scatter plots. Together, these results suggest that $L_{\mathsf{Vox}}$ and $L_{\mathsf{Ours}}$ offer a more faithful approximation of the behavior of the target model in terms of the alignment of the absolute scores. This is critical for the effectiveness of (Step 1) in the attack pipeline.

\subsection{Numerical Error in Pseudo-Inverse}

\begin{table}[t]
\caption{Cosine similarity between the recovered feature $x'$ and original feature $x$ using a fixed pseudo-inversion matrix $A$ constructed from $L_\mathsf{Ours}$. The condition number of $A$ is $\mathrm{cond}(A) = 3.8692 $ and remains constant across all rows.}
\centering
\begin{tabular}{c|c}
\hline
$F$ &  $\langle x', x \rangle$ \\
\hline
$T_1$ & 0.3134 \\ \hline
$T_2$ & 0.3112 \\ \hline
$T_3$ & 0.3114 \\ \hline
$T_4$ & 0.3004 \\ \hline
$T_5$ & 0.3151 \\ \hline
\end{tabular}
\label{tab:cosine_recovery}
\end{table}

We now analyze the numerical stability of the pseudo-inversion step in our pipeline, where a linear system of the form $Ax = y$ is solved to recover the enrolled feature vector $x$ from a score vector $y$ and a matrix $A$ constructed from local features. In practice, the score vector may contain small perturbations, i.e., $y' = y + \epsilon$, resulting in an approximate solution $x'$ upon inversion. The propagation of such perturbations through the pseudo-inversion is governed by the condition number of $A$, defined as
\[
\mathrm{cond}(A) = \|A\| \cdot \|A^{\dagger}\|,
\]
where $A^{\dagger}$ is the Moore–Penrose pseudo-inverse. This yields the classical upper bound on the relative error:
\[
\frac{\|x' - x\|}{\|x\|} \leq \mathrm{cond}(A) \cdot \frac{\|y' - y\|}{\|y\|}.
\]
A large condition number implies that even small perturbations in the input can lead to significant deviations in the recovered feature vector, indicating poor numerical stability of the inversion.

In our setting, the matrix $A$ is constructed once from features generated by the local model $L_\mathsf{Ours}$ and remains fixed for all evaluations. Consequently, the condition number depends solely on the local model and is invariant across different score sources. For the local model used in our attack, we empirically find $\mathrm{cond}(A) = 3.8692$.

We restrict this analysis to $L_\mathsf{Ours}$, as preliminary experiments have shown that other baselines—such as SV2TTS and Voxstructor—fail to preserve identity consistency across embeddings or high model discrepancy. These models exhibit semantically inconsistent behavior, rendering pseudo-inversion unreliable regardless of its numerical stability.

To assess how the mismatch between score source and the local feature space affects the recovered vector, we compute the cosine similarity between the recovered feature $x'$ and the ground truth feature $x$ across different target models. Each rows reflect black-box cases where $y$ is computed using each of the target models $T_1$–$T_5$.

\subsection{Implicit Noise Purification Effect}
In practice, the recovered feature vectors from the pseudo-inversion step inevitably contain noise, and directly using these noisy features can result in unsuccessful attacks in some cases. However, we observe an implicit noise purification effect when the recovered features are passed through the inverse model. Table~\ref{tab:noise} demonstrates this effect under both synthetic noise (added with increasing angular perturbations) and real attack noise.

Although we do not offer a complete theoretical account of this phenomenon, it can be intuitively understood. The inverse models, as well as the foundation model (YourTTS), is trained to produce outputs corresponding to clean feature vectors extracted from actual speech data. In contrast, the recovered vectors are corrupted and therefore unlikely to lie on the same distribution as genuine features. As a result, the trained inverse model tends to map these noisy inputs toward outputs that are more consistent with the training distribution—that is, it effectively "denoises" the input by projecting it onto the learned feature manifold.

This behavior is in line with the general observation that deep learning models often demonstrate robustness to distribution shifts, especially when the test samples are mapped to high-density regions of the training distribution. In our context, this manifests as a denoising effect that increases identity-consistency and improves the effectiveness of the attack, even under considerable input perturbations.

We believe that further investigation into this implicit purification effect—both in theoretical and empirical terms—would be a promising direction for future research, particularly in understanding how neural models implicitly regularize noisy or out-of-distribution representations.
\begin{table}[t]
\caption{Analysis of ID-constraint robustness under noise with random noise ($10^\circ$ to $40^\circ$).}
\resizebox{\linewidth}{!}
{
\begin{tabular}{c||c|c|c}
\hline
Feat. & YourTTS & YourTTS $+L_{\mathsf{IC}}$ & YourTTS $+L_{\mathsf{IC}}+L_{\mathsf{SC}}$ \\ \hline \hline
Org.& $0.4011\pm0.0919$ & $0.9026\pm0.0197$ & $0.8390\pm0.0285$ \\ \hline 
$+10^\circ$ & $0.3959\pm0.0922$ & $0.8978\pm0.0204$ & $0.8349\pm0.0291$ \\ \hline
$+20^\circ$ & $0.3804\pm0.0925$ & $0.8832\pm0.0225$ & $0.8220\pm0.0309$ \\ \hline
$+30^\circ$ & $0.3556\pm0.0937$ & $0.8559\pm0.0267$ & $0.7973\pm0.0346$ \\ \hline
$+40^\circ$ & $0.3213\pm0.0949$ & $0.8114\pm0.0345$ & $0.7555\pm0.0418$ \\ \hline
\end{tabular}
}
\label{tab:noise}
\end{table}

\section{Discussion}

\subsection{About DeepFake Detection}
\label{sec:deepfake}
A pertinent question arising from any method generating synthetic or manipulated voice audio is whether existing defense mechanisms, particularly deepfake detectors, can identify such samples. Our attack generates audio that, while mimicking the target speaker's voice identity extracted from an embedding, constitutes speech content potentially never uttered by the actual speaker in that sequence. This places the generated audio in the realm of synthetic speech, which deepfake detection models aim to expose.

To investigate this, we evaluated the audio samples generated by our attack method using an anti-spoofing model. The deepfake detection system used in our evaluation is based on the framework and pre-trained components from the Self-Supervised Learning (SSL)-based anti-spoofing approach. Specifically, the detector~\cite{tak2022automatic} utilizes a powerful pre-trained SSL audio model as a feature extractor. On top of these SSL features, an AASIST back-end classifier is applied to generate prediction scores for the anti-spoofing task. The model is trained using the standard ASVspoof 2019 Logical Access (LA)~\cite{todisco2019asvspoof} dataset, which includes a variety of spoofed audio types relevant to deepfake scenarios. For consistency and reproducibility, we utilize the publicly available pre-trained weights associated with this architecture provided by the authors.~\footnote{\url{https://github.com/TakHemlata/SSL_Anti-spoofing}}

This model outputs a score reflecting the likelihood that an input audio sample is spoofed. Classification decisions are based on comparing this score against a predefined threshold, $\tau_{DF}$. We adopted the standard practice of setting $\tau_{DF}$ to the value corresponding to the Equal Error Rate (EER) operating point, determined from publicly available performance results for this model configuration. According to these results, the detector achieves an EER of 2.85\% on the ASVspoof 2021 deepfake (DF) track evaluation set~\cite{yamagishi2021asvspoof}. Input audio samples producing scores exceeding this EER threshold were consequently classified as deepfakes.

We conducted a comparative test using three distinct sets of audio derived from the VoxCeleb1 testset identities: (1) original, genuine speech samples from VoxCeleb1; (2) synthetic samples generated by the baseline YourTTS~\cite{casanova2022yourtts} using the corresponding speaker embeddings; and (3) synthetic samples generated by our proposed attack method. The deepfake detector was tasked with assessing the authenticity or `spoofiness' of samples from each set.

\begin{table}[t]
\caption{Results (\%) of the Deepfake detection~\cite{tak2022automatic} of audio samples classified as `Spoof' or `Bonafide' for genuine VoxCeleb1 audio, audio generated by YourTTS, and audio generated by our proposed method (`Ours').}
\begin{tabular}{c|c|c|c}
\hline
Decision & VoxCeleb1 Testset & YourTTS & Ours \\ \hline \hline
\multicolumn{1}{c|}{Spoof} & 8.05\% & 55.90\% & 12.55\% \\ \hline
\multicolumn{1}{c|}{Bonafide} & 91.95\% & 44.10\% & 87.45\% \\ \hline
\end{tabular}
\label{tab:deepfake}
\vspace{-3mm}
\end{table}

The results in Table~\ref{tab:deepfake} revealed a clear trend in the detector's assessment. Audio generated by YourTTS was most frequently identified as spoofed or assigned the highest spoofiness scores. Genuine VoxCeleb1 audio, as expected, received the lowest spoofiness scores. Crucially, the audio generated by our method received scores indicating significantly lower spoofiness compared to YourTTS, placing it much closer to the distribution of genuine audio. This suggests that our generated samples are considerably harder for this type of detector to distinguish from real speech compared to those from the  YourTTS synthesis pipeline.

We attribute this increased resilience against detection to a potentially advantageous side-effect of our training strategy, specifically the fine-tuning of the vocoder module. Many deepfake detection systems are trained on large datasets containing audio generated by commonly used, pre-trained vocoders. Consequently, they learn to identify characteristic artifacts or patterns inherent in the outputs of these standard vocoders. Our method, however, incorporates fine-tuning of vocoder parameters as part of its training process. We hypothesize that this fine-tuning process alters the typical acoustic artifacts produced by the baseline vocoder, effectively removing or modifying the very cues the deepfake detector relies upon. As a result, the audio produced by our fine-tuned system lacks the expected signs of synthesis, leading to lower spoofiness scores. This finding underscores that the specific mechanism of synthetic speech generation, particularly modifications to the vocoder, can significantly impact detectability, posing challenges for current artifact-based deepfake detection approaches and highlighting the need for detection methods robust to such variations.

\subsection{Cross-Lingual Attack Evaluation}
\label{subsec:cross-lingual}

Our primary evaluations focused on target speaker verification models predominantly trained on English datasets such as Voxceleb1,2 and VoxBlink2. A natural question arises regarding the generalizability of our proposed attack across languages with differing acoustic and linguistic properties. This is especially pertinent given that prior studies in areas like speech anti-spoofing have shown performance degradation when training and evaluation languages differ~\cite{liu2024towards}. To explore this, we conducted an auxiliary experiment targeting a speaker verification model trained on a linguistically distinct language: Chinese.

We chose Chinese due to the availability of large-scale datasets like CNCeleb~\cite{fan2020cn}, as well as the presence of pre-trained models such as those provided by the WeSpeaker toolkit~\cite{wang2023wespeaker}\footnote{\url{https://github.com/wenet-e2e/wespeaker}}. Furthermore, the substantial linguistic distance between Chinese and English—given differences in phonology, tonal structure, and syntax—makes this a rigorous test case for assessing the cross-lingual transferability of our attack. This experiment was not the central focus of our study but serves as an exploratory evaluation of the language-agnostic potential of targeting speaker embeddings.

The target system in this experiment was a ResNet34-based speaker verification model trained exclusively on the CNCeleb dataset. CNCeleb contains a diverse range of Chinese conversational speech across various conditions, making it a robust benchmark for Chinese SV research.

To evaluate our attack under mismatched linguistic conditions, we assessed this Chinese-trained model using the English VoxCeleb1 testset. Although unconventional for standard SV benchmarking, this cross-lingual evaluation setup allowed us to (1) derive consistent operating thresholds for evaluation, and (2) simulate realistic mismatched-language attack scenarios where attacker queries originate from a different language domain.

We first computed the baseline performance of the model on the VoxCeleb1 testset to determine operating points. The evaluation yielded an Equal Error Rate (EER) of 7.73\% at a threshold of 0.4246, and a minimum Detection Cost Function (minDCF) of 0.6428 at a threshold of 0.6484.

\begin{table}[t]
\caption{Cross-lingual attack performance against a ResNet34 model trained on CNCeleb. Thresholds and baseline metrics were derived using the ASR was computed under these thresholds.}
\centering
\resizebox{0.8\linewidth}{!}{
\begin{tabular}{c|c|c|c|c}
\hline
\multicolumn{2}{c|}{\textbf{Evaluation}} & \textbf{Threshold} & \textbf{ASR(\%)} & \textbf{FRR(\%)} \\ \hline \hline 
EER & 7.73 & 0.4246 & 60.24 & (= EER) \\ \hline
minDCF & 0.6428 & 0.6484 & 0.06 & 60.26 \\ \hline 
\end{tabular}}
\label{tab:cross_lingual_details}
\end{table}

Using the attack methodology Ours-SP, we evaluated the success of our attack under these two operating thresholds. At the EER threshold (0.4246), the attack achieved a success rate (ASR) of 60.24\%, indicating a clear ability to deceive the system despite the mismatch in language between model training and attack data. This result suggests that our attack, which targets the speaker embedding space, is capable of exploiting voice characteristics that are not strictly language-dependent.

In contrast, when evaluated under the stricter minDCF threshold of 0.6484—which corresponds to a high-precision, low false acceptance setting—the ASR dropped to just 0.06\%. This substantial drop is expected, as the minDCF threshold is tuned to reduce false positives and imposes a significantly stricter acceptance criterion. The corresponding False Rejection Rate (FRR) at this threshold was 60.26\%, highlighting the challenge of evaluating a Chinese-trained model with out-of-domain English data.

Taken together, these findings provide two key insights.
First, despite the linguistic mismatch, the attack exhibits a notable degree of cross-lingual transferability, successfully deceiving a Chinese-trained model using embeddings derived from English audio.
This reinforces the idea that speaker embeddings capture speaker discriminative characteristics that are not strictly tied to linguistic content.

Second, the choice of operating threshold plays a critical role in determining the observed attack success.
In this experiment, operating thresholds were derived from English VoxCeleb1 data, which constitutes an out-of-domain setting for the Chinese-trained target model $L_{\mathsf{CN}}$.
As a result, the minDCF threshold reflects a conservative operating point when applied to English inputs.
In practice, speaker verification models trained on Chinese are typically calibrated using in-domain Chinese data
(e.g., CNCeleb), and thresholds derived under such conditions would more accurately reflect the system’s operational vulnerability.
This exploratory evaluation therefore highlights the sensitivity of cross-lingual attack assessment to threshold calibration, and motivates further investigation under fully in-domain settings.

\subsection{Limitation of Evaluation on Commercial APIs}
\label{subsec:commercial}
SRSs have been actively deployed in practice through commercial APIs, and we fully acknowledge the value of evaluating against them. However, by early 2025, major providers such as Azure and Amazon had enforced strict Responsible AI policies, restricting hosted endpoints to prevent potential misuse. To ensure strict compliance with research ethics, we pursued official authorization for academic testing rather than bypassing these restrictions. We contacted smaller vendors—specifically PicoVoice in June 2025 and Phonexia in January 2026—but were unable to obtain the necessary approval.

Despite the lack of direct API access, we anticipate our findings to generalize to commercial systems. It is widely understood that modern commercial SRSs are built upon deep learning architectures~\cite{azure_speaker} similar to those we evaluated. Furthermore, since our proposed attack is score-based, it does not require internal gradient access. Even if commercial APIs return only normalized confidence scores~\cite{azure_speaker, aws_transcribe}, these can be effectively utilized for adversarial perturbations through inverse transformations or directly employed by NES~\cite{zhou2022voice, chen2021real} (Natural Evolution Strategies)-based algorithms without loss of attack efficacy.
To compensate for commercial-grade robustness, we adopted publicly available models provided by Alibaba’s 3D-Speaker research group and the NVIDIA model.

\end{document}